\documentclass[final,12pt,5p,twocolumn]{elsarticle}

\usepackage{amssymb}

\usepackage{ulem}

\usepackage{xcolor}

\usepackage{float}

\usepackage{tabularx,lipsum,graphicx,tabularray}  

\usepackage[hyphens]{url}
\usepackage[colorlinks=true, urlcolor=blue, pdfborder={0 0 0}]{hyperref}
\hypersetup{
  colorlinks   = true, 
  urlcolor     = blue, 
  linkcolor    = blue, 
  citecolor   = red 
}
\hypersetup{breaklinks=true}
\usepackage{subcaption}
\usepackage{booktabs}
\usepackage{multirow}
\usepackage{tabularx}	
\usepackage{color,colortbl}

\usepackage[final]{changes}

\usepackage{algorithmic}
\usepackage{algorithm}
\usepackage{amsmath}

\usepackage{graphicx}
\usepackage{tablefootnote} 
\biboptions{sort&compress}

\makeatletter
\newcommand{\ion}[2]{%
  #1$\;$%
  \if b\expandafter\@car\f@series\relax\@nil
    \begingroup 
      \sbox0{\rmfamily\mdseries\textsc{v}}%
      \resizebox{!}{\ht0}{\rmfamily\@Roman{#2}}%
    \endgroup
  \else
    \textsc{\rmfamily\@roman{#2}}%
  \fi
}
\makeatother

\makeatletter
\def\@author#1{\g@addto@macro\elsauthors{\normalsize%
    \def\baselinestretch{1}%
    \upshape\authorsep#1\unskip\textsuperscript{%
      \ifx\@fnmark\@empty\else\unskip\sep\@fnmark\let\sep=,\fi
      \ifx\@corref\@empty\else\unskip\sep\@corref\let\sep=,\fi
      }%
    \def\authorsep{\unskip,\space}%
    \global\let\@fnmark\@empty
    \global\let\@corref\@empty  
    \global\let\sep\@empty}%
    \@eadauthor={#1}
}
\makeatother

\journal{Blockchain: Research and Applications}

\begin{document}

\begin{frontmatter}



\title{Backtesting Framework for Concentrated Liquidity Market Makers on Uniswap V3 Decentralized Exchange 
}

%

\author[Vega,Ch]{Andrey Urusov}
\ead{{Corresponding author*}{tapwi93@gmail.com}}
\affiliation[Vega]{organization={Vega Institute Foundation},
            city={Moscow},
            country={Russia}}

\affiliation[Ch]{organization={Faculty of Applied Mathematics, Physics and Information Technologies, Chuvash State University},
            city={Cheboksary},
            country={Russia}}

\author[HSE-LSA,Vega]{Rostislav Berezovskiy}
\ead{rostislavberezovskiy@gmail.com}
\affiliation[HSE-LSA]{organization={International Laboratory of Stochastic Analysis and its Applications, HSE University},
            city={Moscow},
            country={Russia}}

\author[Sk,HSE]{Yury Yanovich}
\ead{y.yanovich@skoltech.ru}
\affiliation[Sk]{organization={Skolkovo Institute of Science and Technology},
            city={Moscow},
            country={Russia}}
\affiliation[HSE]{organization={Faculty of Computer Science, HSE University},
            city={Moscow},
            country={Russia}}

\begin{abstract}
    Decentralized finance (DeFi) has revolutionized the financial landscape, with protocols like Uniswap offering innovative automated market-making mechanisms. This article explores the development of a backtesting framework specifically tailored for concentrated liquidity market makers (CLMM). The focus is on leveraging the liquidity distribution approximated using a parametric model, to estimate the rewards within liquidity pools.
    The article details the design, implementation, and insights derived from this novel approach to backtesting within the context of Uniswap V3. The developed backtester was successfully utilized to assess reward levels across several pools using historical data from 2023 (pools Uniswap v3 for pairs of altcoins, stablecoins and USDC/ETH with different fee levels). Moreover, the error in modeling the level of rewards for the period under review for each pool was less than 1\%. This demonstrated the effectiveness of the backtester in quantifying liquidity pool rewards and its potential in estimating LP's revenues as part of the pool rewards, as focus of our next research.
    The backtester serves as a tool to simulate trading strategies and liquidity provision scenarios, providing a quantitative assessment of potential returns for liquidity providers (LP). By incorporating statistical tools to mirror CLMM pool liquidity dynamics, this framework can be further leveraged for strategy enhancement and risk evaluation for LPs operating within decentralized exchanges.
\end{abstract}




\end{frontmatter}


\section{Introduction}\label{sec:intro}

Following the introduction of smart contract systems with arbitrary logic in 2014 \cite{Buterin2014}, the emergence of decentralized exchanges (DEXs) began to disrupt traditional financial systems~\cite{Lin2019,Huang2024}. This shift attracted millions of users to DeFi projects \cite{Samreen2023}, offering a way to bypass the constraints of conventional financial systems. DeFi enabled users to participate in digital markets without censorship, lengthy procedures, or intermediary involvement, granting them unprecedented freedom to conduct various financial operations, particularly cryptocurrency trading.

Prior to 2018, most cryptocurrency trading platforms relied on Centralized Exchanges (CEX), introducing inconveniences for users due to third-party intervention, such as service providers or banks \cite{Lee2023}. This prompted a race among companies to propose market maker mechanisms that could replace traditional order book methods and eliminate third-party interference, leading to the rise of DEXs \cite{DosSantos2022}.

In 2018, Uniswap Labs introduced the Uniswap V1 protocol, revolutionizing DEXs \cite{Adams2018}. This protocol facilitated the exchange of ETH and ERC20 tokens \cite{Vogelsteller2015,Davydov2019a,Angelo2020} on the Ethereum blockchain in a decentralized, censorship-resistant, and secure manner without intermediaries. Uniswap's innovative automated market makers (AMM) in form of constant function market makers (CFMM) concept marked a pivotal shift from CEXs to DEXs by replacing traditional order book mechanisms.

As of 2024, Uniswap operates on Ethereum and dozens of other public blockchain platforms. The smart contract source code is openly accessible, along with transaction histories. Uniswap has evolved through three protocol versions. Uniswap V1 utilized exchange contracts to interact with pools holding ETH and ERC20 tokens during trades. Liquidity providers (LP) supplied equal amounts of ETH and ERC20 tokens to these pools in exchange for special ERC20 liquidity tokens. 

Uniswap V2 introduced ERC20-ERC20 Pools using wrapped ETH (WETH), eliminating the need to convert tokens to ETH for transactions between ERC20 tokens \cite{Adams2020}. Uniswap V2 also incorporated several enhancements, including a highly decentralized on-chain Oracle system. Price determination occurred at the start of each block based on previous transactions, making price manipulation challenging for attackers. Flash swaps allowed users to withdraw ERC20 tokens from Uniswap pools without upfront costs, with fees payable upon completion. 

In both Uniswap V1 and V2, liquidity was evenly distributed along the curve within any positive price. However, challenges arose in stablecoin pairs like DAI/USDC due to most transactions occurring within a narrow price range around one, where LPs expected higher volumes and fees. This situation renders the liquidity outside of this range idle for two main reasons: firstly, traders would avoid transactions outside this range as other platforms may offer more favorable terms, especially given the stability of the two coins. Secondly, any price fluctuations would create arbitrage opportunities, prompting trades that would quickly restore prices back to the specified range. 

Uniswap V3 addressed this challenge by incentivizing LPs to bear the risk of impermanent loss. Uniswap V3 requires LPs depositing assets into liquidity pools to define the price range within which they wish to allocate their assets \cite{Adams2021}, known as concentrated liquidity (CL). This setup ensures that LPs can earn fees only when prices remain within their chosen range. The concentration of liquidity within a defined interval is referred to as a position. 

Uniswap V3 also grants LPs the flexibility to hold multiple positions within a single pool.
Smaller (narrower) price ranges lead to higher liquidity concentration and greater rewards, but there is a risk of price breaking out of the range, leaving the provider unrewarded. In such cases, liquidity reallocation can help, but it comes at a cost. While the CL provides LPs a flexible instrument to balance their profit and risk, management instruments are still necessary.  

In this study, we tackle the challenge of creating a tool to model rewards when placing liquidity in a market maker pool with concentrated liquidity \cite{Heimbach2022}. The aim of this research is to develop a backtesting model to simulate the processes within the liquidity pool to evaluate the model rewards over a specified time period.

When calculating rewards, the goal is broad: rewards can be modeled for the entire liquidity pool, for individual LPs considering a specific strategy or without a strategy as a standalone entity or as part of the pool \cite{Fan2023}. We aim to develop a versatile and flexible tool that, when properly configured, can model rewards for each possible scenario with a straightforward architecture.

The study models state changes of a hypothetical CLMM pool for a token pair, influenced by price movements within the pool (contract price) against changes in the market value of the cryptocurrency token. The pool's rewards depend on its liquidity profile and the model (historical) prices within that pool \cite{Ghazzawi2022}. Rewards for a specific LP can be calculated as a share of the total liquidity pool rewards, depending on the liquidity placement strategy of the LP within the active pool price ranges where the contract price moves through (in this case, the LP liquidity is considered to be a part of the total pool liquidity).


The research contributes to LP backtesting strategies in DEX by the following steps:
\begin{itemize}
    \item
        Develop a methodology for backtesting liquidity provision in a CFMM.
    \item 
        Enhance CFMM backtesting by leveraging GPU acceleration for faster computation.
    \item 
        Showcase the practicality of CFMM backtesting using actual Uniswap pool data.
\end{itemize}

The rest of the paper is organized as follows. The essential information on liquidity provision on DEX is outlined in Section \ref{section:background}: in this section, we delve into the essential concepts and mechanics of providing liquidity on DEXs. This includes an overview of Uniswap protocols, the dynamics of liquidity pools, and the roles of LPs. We also cover the mathematical underpinnings necessary for understanding liquidity provision,  fee structures, and the impact of price fluctuations within the pool, as well as the phenomenon of impermanent loss. Section \ref{section:RelatedWork} considers the related work: here, we examine the existing research and methodologies, highlighting previous studies and their findings. This section contextualizes our work within the broader academic and practical landscape, demonstrating the gaps our research aims to fill and the innovations we introduce. Section \ref{section:ProposedSolution} is the core of our article, where we introduce the backtester designed specifically for evaluating rewards in liquidity pools. We provide an in-depth explanation of the backtester's architecture, including its multi-dimensional NumPy (CuPy) array structure that represents the pool states according to the current pool price and the liquidity distribution configuration of a specific LP. We detail the step-by-step process, from setting the ranges boundaries and the number of this ranges (buckets) to defining the pool price series and implementing the $\tau$-strategy for liquidity placement. The use of the $\tau$-strategy in the current study will be addressed in the theoretical example of the negative impact of impermanent loss (IL). 
Using the developed backtester, we conduct a numerical evaluation of the liquidity pools rewards on Uniswap v2 and v3 in Section \ref{section:Experiments}. This section presents the methodology for the assessment and general assumptions, the data used (including historical price data), and the results obtained. We analyze the backtester accuracy in predicting reward levels, discuss the implications of the results, and compare the tool performance in different pools. Finally, Section \ref{section:Conclusions} summarizes the key findings, reflecting on the backtester effectiveness its and potential applications. We also outline the next steps for our research, including the plans to refine the backtester further, and extend our analysis to other DEX platforms and token pairs.

\section{Background}
\label{section:background}

\subsection{Concentrated Liquidity Market Makers}

The research focuses on decentralized exchanges (DEX) and concentrated liquidity market makers~(CLMM), particularly examining Uniswap v2 and v3. Both versions operate as constant function market makers (CFMM), with a specific emphasis on constant product market makers (CPMM).

The constant product is an invariant function set during the liquidity pool initialization. In Uniswap v2, this function is the product of reserves $x$ and $y$ of tokens $A$ and $B$, for example, a cryptocurrency and a stablecoin, traded in the pool. The constant can change when new liquidity in tokens $A$ and $B$ is added to or withdrawn from the liquidity pool in proportion, maintaining the pool price. Trader actions impact the contractual price but not the constant invariant. The price of token $A$ relative to token $B$ is calculated as $y/x$.

\begin{figure*}[!t]
\centering
\includegraphics[width=0.9\textwidth]{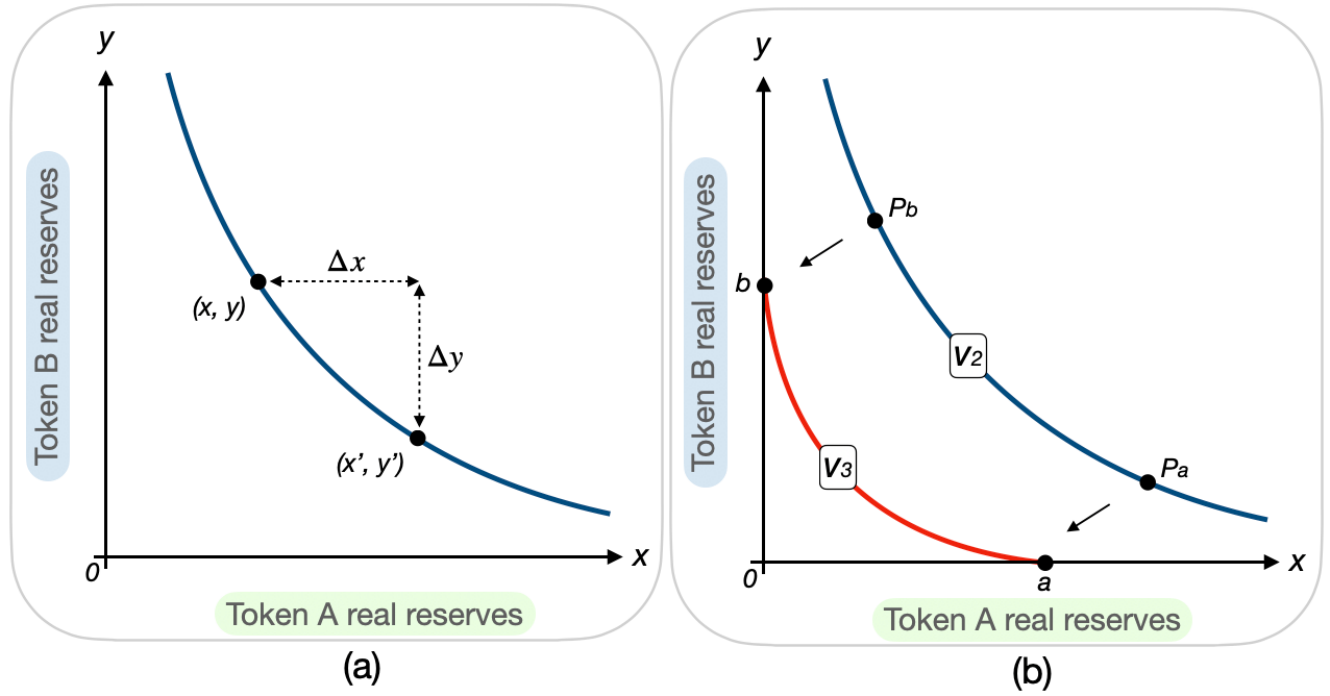}
\smallskip
\caption{Reserve curve of Uniswap v2 (a) and Uniswap v3 (b)}
\label{figure:CFMM}
\end{figure*}

\begin{figure*}[!t]
\includegraphics[width=\textwidth]{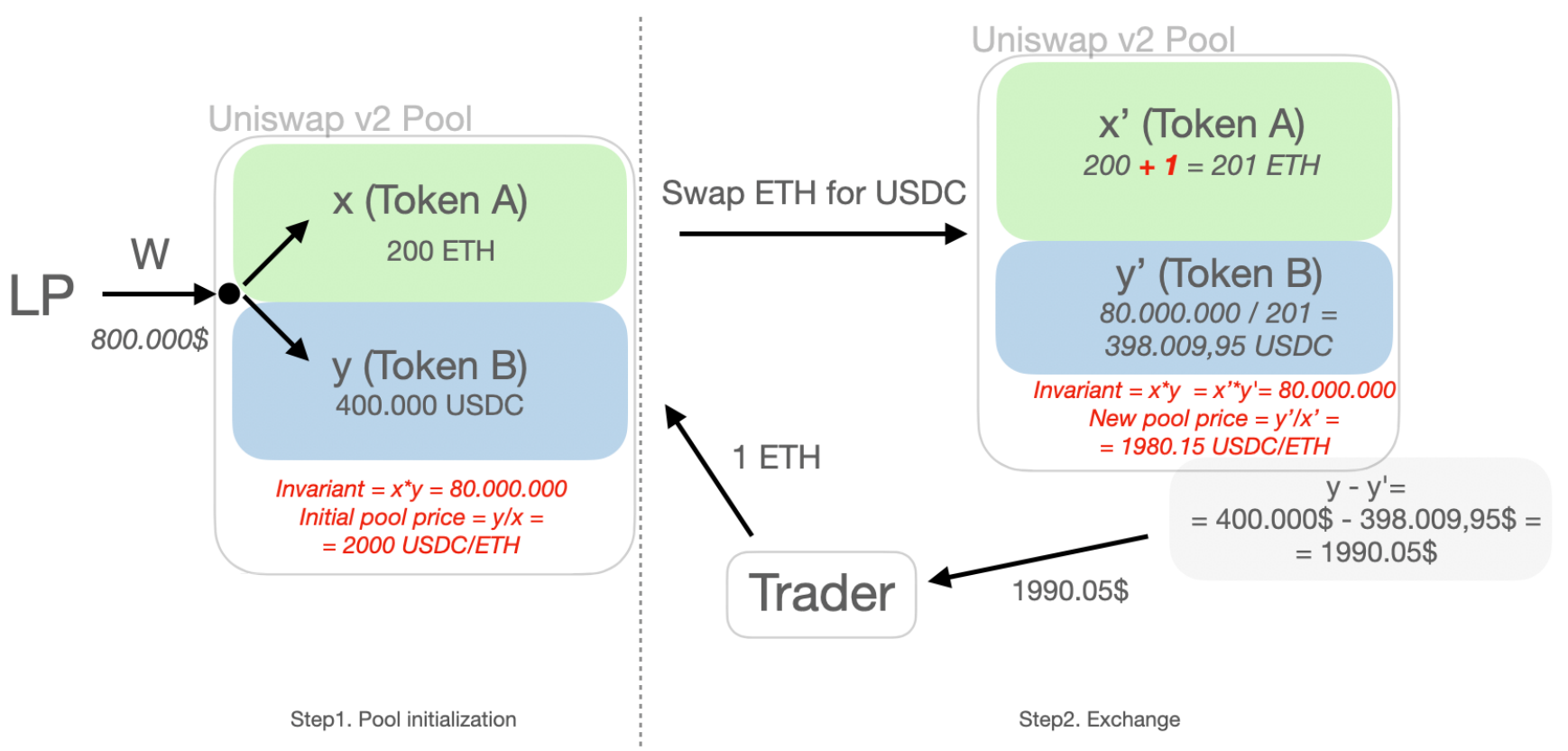}
\smallskip
\caption{The principle of exchange in Uniswap v2}
\label{figure:UniP}
\end{figure*}

Uniswap v2's reserve curve forms a hyperbola (Figure \ref{figure:CFMM} (a)), with $x$ and $y$ reserves on the axes. Traders can exchange $\Delta x$ units of token $A$ for $\Delta y$ units of token $B$, shifting the curve point from $(x, y)$ to $(x', y')$. The product $x'y'$ remains constant, while the contract price $P=y'/x'$ changes (Figure \ref{figure:UniP}). As the trade volume nears the available pool volume for one token, the prices can fluctuate significantly, discouraging complete liquidity extraction. Liquidity provider (LP) offers liquidity across all possible pool prices, distributing capital to areas where prices will not reach -- a drawback of Uniswap v2.

In the case of Uniswap v3 the LPs independently select a specific price range $[P_a, P_b]$ in which they want to provide liquidity. As the contract price changes, similar to Uniswap v2, the provider's ratio of token reserves $A$ and $B$ changes. However, within the price range, the price no longer changes infinitely in both directions. When the price crosses the upper or lower boundary of the range, the LP retains a fixed amount of tokens of one type: tokens $B$ if the price of token $A$ rises or tokens $A$ if the price of token $A$ falls. This effect is achieved by actually shifting the Uniswap v2 reserves curve through an affine transformation, intersecting the axes at points corresponding to the token reserves $A$ and $B$ at prices $P_a$, $P_b$ (Figure \ref{figure:CFMM} (b)). This shifted curve is determined by an invariant function. The intersection points $a$ and $b$ can be calculated by setting $x$ or $y$ to zero when reaching the price corresponding to the range boundary; we will discuss this in more detail below. Here, $x$ and $y$ are the actual reserves of token $A$ and $B$, respectively, and the values in parentheses are the so-called virtual reserves $x$ and $y$ (Formula \ref{eqn:CPMM}). The pool price in Uniswap v3 is determined by the ratio of these virtual reserves.

At this juncture, the outlined theory provides a solid foundation to proceed with further descriptions without delving deeper into protocol variances like those seen in Uniswap. It is important to recognize that while many statements above may seem intuitive, each is grounded in strict mathematical principles \cite{Elsts2021}, although elaborating on this falls beyond the scope of the current discussion.

\subsection{Liquidity Provision Mathematics}

Let us formally introduce some fundamental concepts of liquidity provision. These concepts will be further utilized in developing the architecture of the backtester. LPs offer their capital in one or several price ranges, which typically involves a pair of tokens, $A$ and $B$. Drawing inspiration from \cite{Fan2023}, we will define a function known as the liquidity state function. This function aims to calculate the reserves of each token based on the allocated liquidity ($L$) in a specific range and the current pool price. Previously, we explored the foundational invariant function for Uniswap V3, let us revisit it:
\begin{equation}
    \label{eqn:CPMM}
    \left(x + \frac{L}{\sqrt{P_b}}\right)\left(y + L\sqrt{P_a}\right)=L^2.
\end{equation}

In the formula (\ref{eqn:CPMM}) we deal with the upper $P_b$ and lower $P_a$ bounds of the range in which the LP places its capital $W$, consisting of real reserves $x$ and $y$ (for example: the LP provides its capital $W$ of $8000$ in the form of token $B$ (USDC) in quantity $y = 4000$ and token $A$ (ETH) at a rate of $1:2000$ in quantity $x = 2$, then we have the expression $W = y + x \cdot \texttt{price}$: here $x$ and $y$ are real reserves) and a fixed constant $L$ or range liquidity, which we have mentioned several times before. To understand what $L$ means, let us first learn to calculate this value using the results from \cite{Elsts2021} and then further define the liquidity state function $\mathcal{V}$.

\begin{figure}[!h]
\centering
\includegraphics[width=0.5\textwidth]{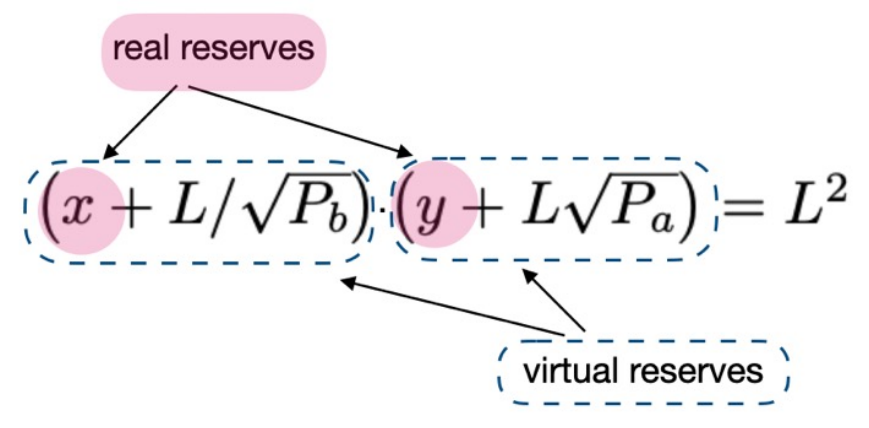}
\smallskip
\caption{Uniswap v3 Invariant Function}
\label{figure:CP}
\end{figure}

The invariant function of Uniswap v3 explicitly differs from the Uniswap v2 function by the presence of so-called virtual reserves (Figure \ref{figure:CP}). Their nature is such that, in addition to real reserves, they contain elements that facilitate transformation to transition to concentrated LP capital deployment (Figure \ref{figure:CFMM} (b)).
Let the LP place its capital in the price range from $P_a$ to $P_b$; in this case, there are three possible scenarios for the current pool price $P$ relative to the range boundaries: $P \leq P_a$, $P_b > P > P_a$, $P \geq P_b$. Let us consider each case in more detail by calculating the value of $L$.

\begin{figure}[!b]
\centering
\includegraphics[width=0.45\textwidth]{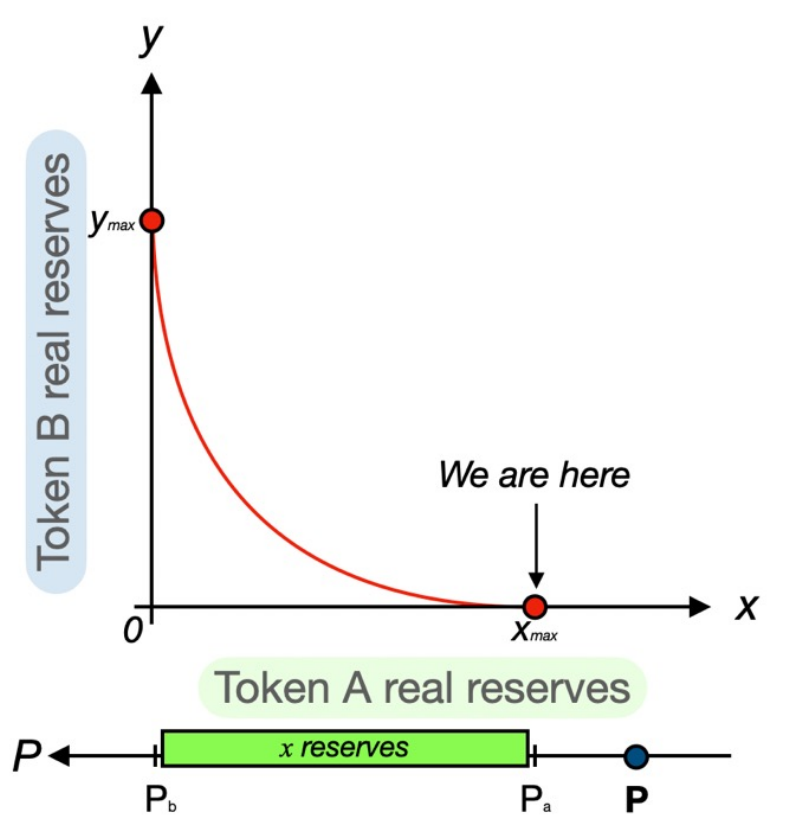}
\smallskip
\caption{Case $P \leq P_a$}
\label{figure:C1}
\end{figure}

\subsubsection{$P \leq P_a$}\label{subsubsection:P_a}

The observed price in the pool is below the lower range boundary (Figure \ref{figure:C1}). When the current price $P$ reaches the level of $P_a$, the maximum possible value $x_{\max}$ of real reserves of token $A$ in the current range is fixed, while the reserves of token $B$ are zero.

Suppose initially the contract price is within the range (Figure \ref{figure:C3}), and then it drops below $P_a$, what does this mean? This happens because the traders have brought in a lot of token $A$ into the pool, wishing to exchange them for tokens $B$, completely depleting the reserves of the latter in the current range and shifting the price to the adjacent one. In fact, this means a drop in the price of token $A$ in terms of token $B$. At this point, the LP who provided liquidity in the current range no longer receives rewards, but even more dangerously, all of their capital is now fully converted into token $A$, and its price is falling. It is important to remember that liquidity $L$ is calculated and fixed at the moment when the LP allocates their capital to the chosen range, and the given example already assumes the existence of $L$. 

Consider now that the LP may also choose to provide liquidity in anticipation of the price increasing, such that at the moment of providing liquidity, $P \leq P_a$, and also the calculation and fixation of LP’s liquidity occur. Then, once the price rises and will be within the specified range, the LP will begin earning fees. In this case, we have the following:
\begin{gather*}
    \left(x + \frac{L}{\sqrt{P_b}}\right) L\sqrt{P_a}=L^2 \\
    x = \frac{L}{\sqrt{P_a}} - \frac{L}{\sqrt{P_b}} = L\frac{\sqrt{P_b} - \sqrt{P_a}}{\sqrt{P_a}\sqrt{P_b}}.
\end{gather*}

The value of liquidity when initializing such a position is equal to:
\begin{equation}
    \label{eqn:liquidity}
    L = x \frac{\sqrt{P_a}\sqrt{P_b}}{\sqrt{P_b} - \sqrt{P_a}}.
\end{equation}

\subsubsection{$P \geq P_b$}

\begin{figure}[!b]
\centering
\includegraphics[width=0.45\textwidth]{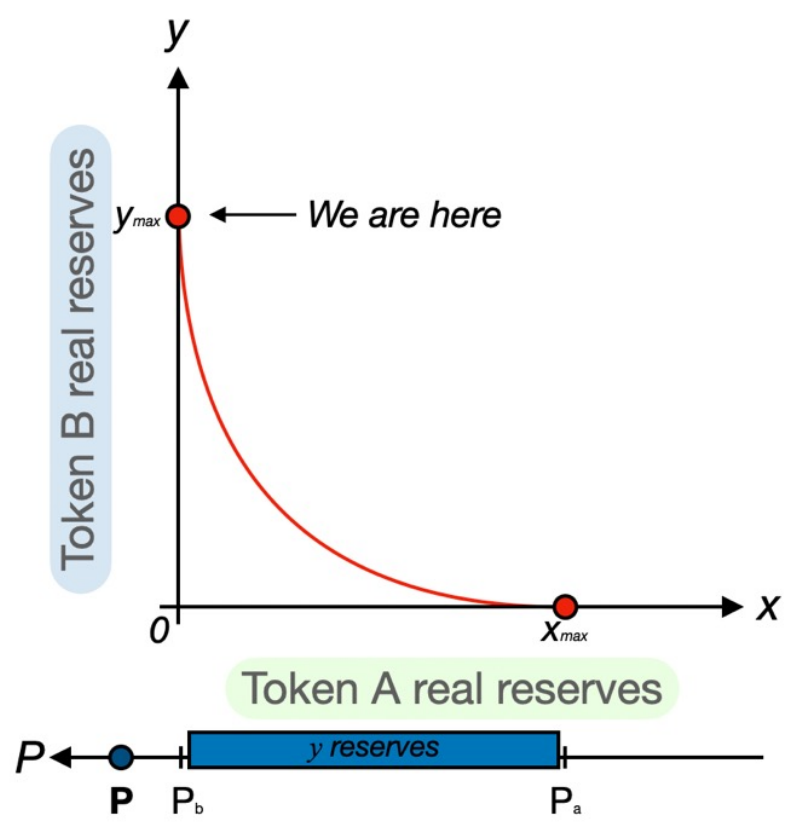}
\smallskip
\caption{Case $P \geq P_b$}
\label{figure:C2}
\end{figure}

The observed price in the pool is above the upper range boundary (Figure \ref{figure:C2}). When the current price $P$ reaches $P_b$, the maximum possible value $y_{\max}$ of the token B real reserves in the current range is locked, while the reserves of token A are zero.

This scenario is completely opposite to the one described in \ref{subsubsection:P_a}. If we assume that initially the contract price is within the range (Figure \ref{figure:C3}), and then it rises above $P_b$, this means that the traders have brought many tokens B into the pool completely depleting the token $A$ reserves and shifting the price to the adjacent range. However, unlike \ref{subsubsection:P_a}, now the token $A$ price grows in terms of token $B$. At this moment, the LP who provided liquidity in the current range no longer receives rewards, as all their capital is now fully converted into token $B$. Similar to the \ref{subsubsection:P_a}, this case assumes a known fixed constant $L$. 

Consider now that the LP decided to place their liquidity in the range lower that the current pool price, expecting a further decrease in the value of token $A$. Let's calculate the value of $L$ for this case; liquidity of the range is calculated upon capital placement and does not change further. In this case, we have the following:
\begin{gather*}
    \frac{L}{\sqrt{P_b}}\left(y + L\sqrt{P_a}\right) =L^2 \\
    y = L(\sqrt{P_b} - \sqrt{P_a}).
\end{gather*}

The liquidity value at the initialization of such a position is equal to:
\begin{eqnarray}
    \label{eqn:L}
    L = \frac{y}{\sqrt{P_b} - \sqrt{P_a}}.
\end{eqnarray}

\subsubsection{$P_a < P < P_b$}

\begin{figure}[!b]
\centering
\includegraphics[width=0.45\textwidth]{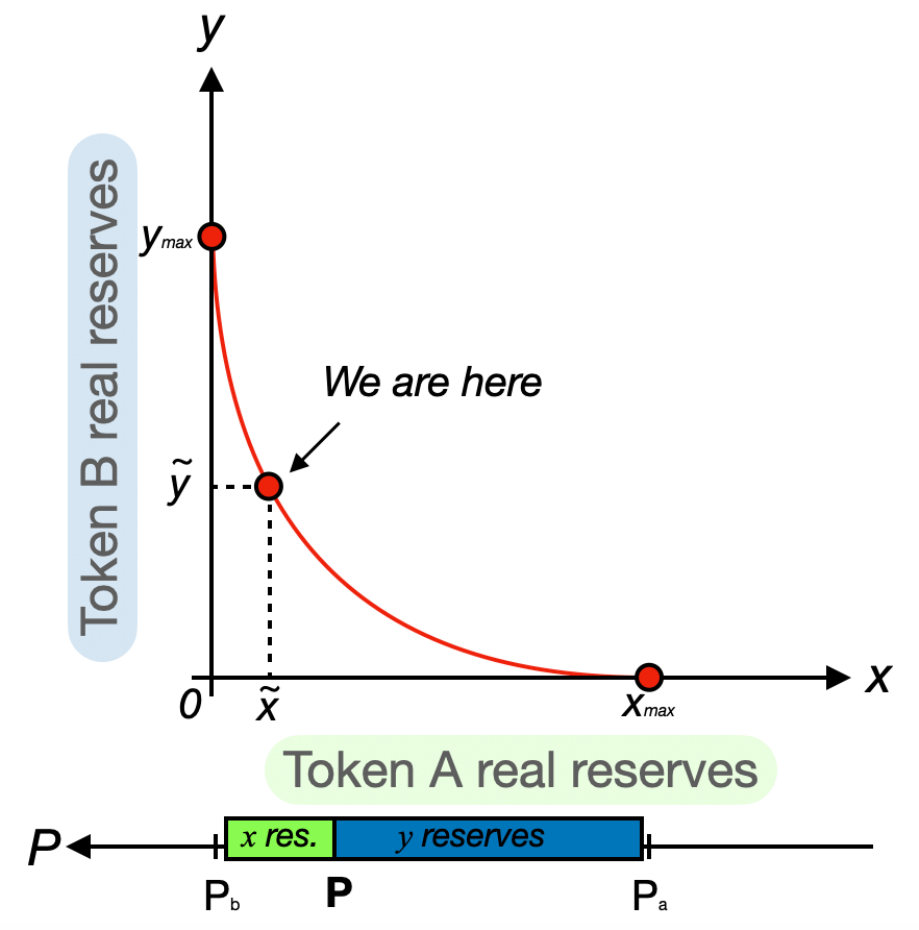}
\smallskip
\caption{Case $P_b > P > P_a$}
\label{figure:C3}
\end{figure}

Consider the example from Figure \ref{figure:UniP} applied to Uniswap v3. The observed pool price $P$ is within the range boundaries $(P_a, P_b)$ (Figure \ref{figure:C3}). Now the token reserves $A$ and $B$ are both non-zero, and the LP allocates capital $W$ nominated in a combination of both tokens. To calculate the liquidity $L$ and token amounts for current price we'll use the result from \cite{Elsts2021}, and also an additional condition that we introduce ourselves.

Part 1: according to \cite{Elsts2021}, when the price is in the range $(P_a, P_b)$, both assets should contribute equally to the liquidity. In other words, the liquidity $Lx$ provided in terms of $x$ (token $A$) on one side of the range $(P, P_b)$ should be equal to the liquidity $Ly$ provided in terms of $y$ (token $B$) on the other side of the range $(P_a, P)$. By mentally dividing $(P_a, P_b)$ into two sub-ranges where the liquidity is nominated only in $x$ or only $y$ correspondingly, we obtain an expression $Lx(P, P_b) = Ly(P_a, P)$:
\begin{equation}
    \label{eqn:P_c}
    x \frac{\sqrt{P}\sqrt{P_b}}{\sqrt{P_b} - \sqrt{P}} = \frac{y}{\sqrt{P} - \sqrt{P_a}}.
\end{equation}

Part 2: to find the actual reserves $x$ and $y$ that ensure the condition (\ref{eqn:P_c}) is met, we need a second equation. We can write the initial capital $W$ in terms of the current price $P$ and the token reserves:
\begin{equation}
    \label{eqn:W}
    y + x P = W.
\end{equation}

Introducing notations for the left and right sides of (\ref{eqn:P_c}):
\begin{gather*}
    x_L = \frac{\sqrt{P}\sqrt{P_b}}{\sqrt{P_b} - \sqrt{P}} \\
    y_L = \frac{1}{\sqrt{P} - \sqrt{P_a}}.
\end{gather*}

Next, using these notations, we solve the system of equations (\ref{eqn:P_c}) and (\ref{eqn:W}) with respect to $x$ or $y$ since determining $L$ only requires calculating either the left or right side of (\ref{eqn:P_c}):
\begin{eqnarray}
\label{eqn:actualReserves}
\begin{cases}
    y = W \left(1 - \frac{y_L P}{x_L + P y_L}\right) \\
    x = W y_L / \left(x_L + P y_L\right).
\end{cases}
\end{eqnarray}

We have obtained the expression (\ref{eqn:actualReserves}) for the actual reserves $x$ and $y$ calculation given known boundaries, current price, and the allocated capital amount. By substituting either $x$ or $y$ from (\ref{eqn:actualReserves}) into the left or right side of (\ref{eqn:W}) respectively (both sides will be equal), we calculate and fix the value of liquidity $L$ for the range in the case of $P_a < P < P_b$.

\subsection{Liquidity State Function}
After learning how to calculate the liquidity $L$ for all possible price values, let us return to the liquidity state function $\mathcal{V}(L_i, P, B_i)$. Now we know all the input data. Leveraging the results from \cite{Fan2023}, we state that the LP provides liquidity $L_i$ in a specific segment $B_i = [P_{a_i}, P_{b_i}]$, called $B_i$-liquidity, by sending a set of tokens $A$ and $B$ into the contract. The token set required to issue $L_i$ units of $B_i$-liquidity is determined by the liquidity value function and represents a tuple $\mathcal{V}(L_i, P, B_i)$, where the first component is the quantity in token $A$ and the second is the quantity in token $B$. For $P_a < P_b$, we use notations $\Delta_{P_b,P_a}^x = \frac{1}{\sqrt{P_a}} - \frac{1}{\sqrt{P_b}}$ and $\Delta_{P_a,P_b}^y = \sqrt{P_b} - \sqrt{P_a}$. 
For the contract price $P$, bucket $B_i = [P_{a_i}, P_{b_i}]$, and quantity of allocated liquidity $L_i > 0$ for the bucket, the liquidity state function is defined as $\mathcal{V}(L_i, P, B_i) = (x, y)$ \cite{Fan2022}:
\begin{flalign}
\label{eqn:liquidityStateFunction}
\mathcal{V} = 
\begin{cases}
    \left(L_i\Delta_{P_{b_i},P_{a_i}}^x, 0\right), & \text{} P \leq P_{a_i} \\
    \left(L_i\Delta_{P_{b_i},P}^x, L_i\Delta_{P_{a_i},P}^y\right), & \text{} P_{b_i} > P > P_{a_i} \\
    \left(0, L_i\Delta_{P_{a_i},P_{b_i}}^y\right), & \text{}  P \geq P_{b_i}
\end{cases}.
\end{flalign}
and determines the sets of tokens $A$ and $B$ respectively for $L_i$ units of $B_i$-liquidity.

\subsection{Impermanent Loss}

\begin{figure*}[h]
\centering
\includegraphics[width=\textwidth]{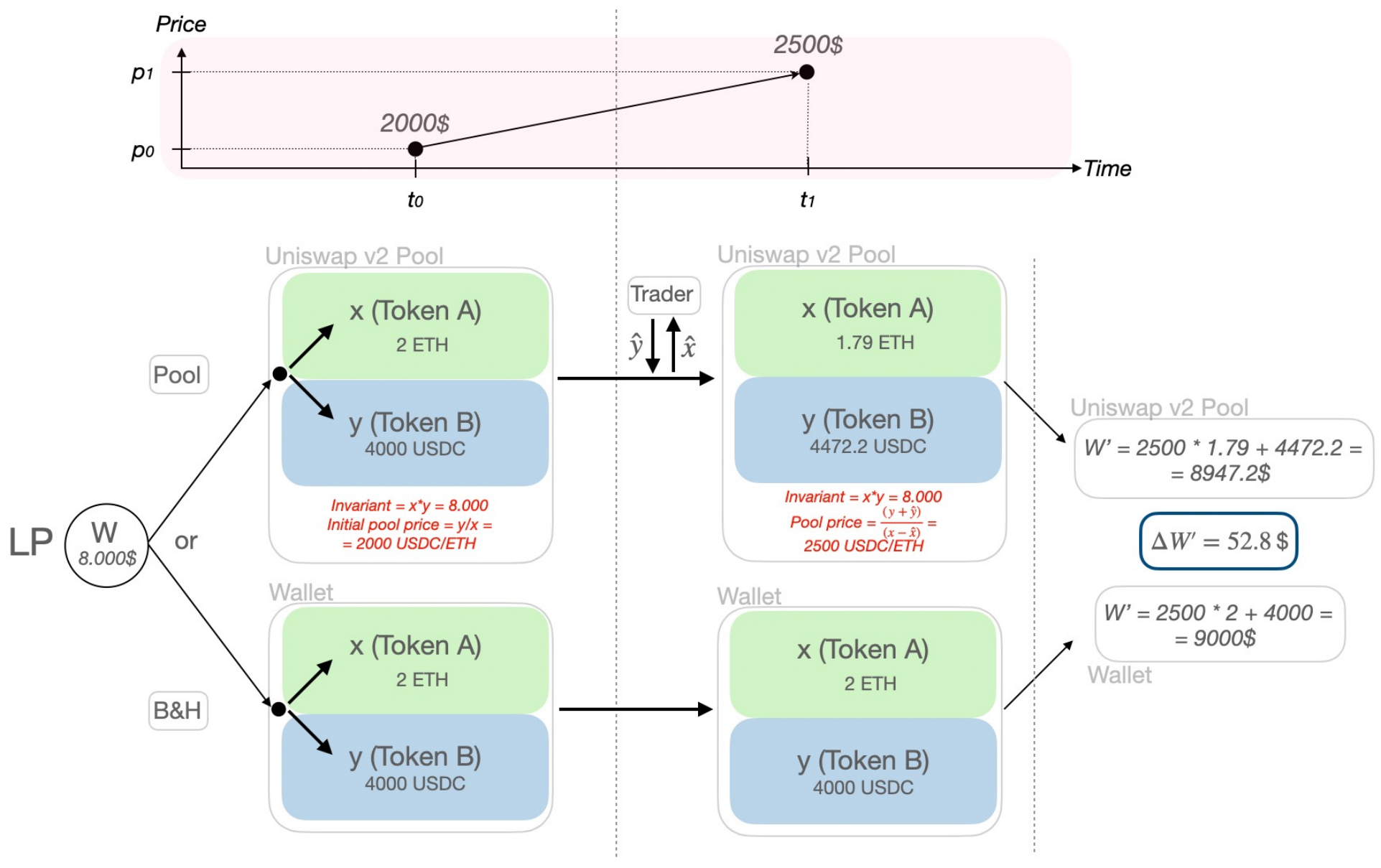}
\caption{Impermanent Loss occurrence example}
\label{figure:IL}
\end{figure*}

Impermanent Loss (IL) refers to the variation in asset value within a liquidity pool caused by price fluctuations between two assets on a decentralized exchange (DEX). Detailed technical information on IL can be found in \cite{Milionis2022}. In this context, we will demonstrate the concept using a hypothetical example and later validate it with the backtester on artificial data in Section \ref{subsection:ILEonAD}. Let us delve deeper into how these losses manifest and why they occur specifically on DEXs. Take the USDC/ETH pair and a hypothetical price scenario on Uniswap v2 as an example (Figure \ref{figure:IL}). For additional technical information, please refer to \cite{Milionis2022}.

At the initial time $t_0$, the LP divides their capital $W$ into two tokens based on the current price of the pair, $p_0$ ($\$2000/ETH$). They can either add tokens to the liquidity pool at the current pool price $p_0$ (for simplicity, let us assume they initialize the pool at $p_0$ and are the sole LP) or simply hold them in their wallet following a buy \& hold (B\&H) strategy.

At the next time instance, the fair market price of the pair becomes $p_1$ ($\$2500/ETH$). In the case of holding tokens in the wallet, the reserves of both tokens remain unchanged, and considering the market value of ETH, their current capital amounts to $\$9000$. Simultaneously, to establish the fair market price within the pool, a trader needs to execute an arbitrage transaction within the pool. To achieve this, the trader must input a volume $\hat{y}$ of token B, receiving in return a volume $\hat{x}  = x - \frac{L^2}{y + \hat{y}}$ of token A, thereby setting the pool price at $p_{1_{Pool}} = \frac{y + \hat{y}}{x - \hat{x}}$. Due to the isolated nature of the liquidity pool with a fixed invariant, we calculate $\Delta W'$, the difference in final capital that the LP misses out on by providing liquidity to the pool compared to simply holding tokens in a wallet (B\&H). However, it is not all bleak. The LP rewards for supplying liquidity to pools are designed to offset impermanent loss, making liquidity provision in pools more profitable and preferable for LPs than just holding tokens under a B\&H strategy.

Speaking in terms of impermanent loss concerning Uniswap v3, while the nature of this phenomenon remains the same, the negative impact may be greater due to fundamental protocol differences from Uniswap v2. When the current pool price exceeds the fixed bounds of the placement range, LPs face a situation where one token's reserves hit zero, converting all LP reserves into the second token. These non-zero reserves consist of tokens that are not very advantageous for LPs. If the current pool price surpasses the upper bound (Figure~\ref{figure:C2}), token A's reserves, which are currently appreciating in value, become zero or decrease as the contract price rises and breaches the liquidity placement range, gradually leaving LPs with diminishing exposure to the appreciating asset in the market, fixing their capital level upon exiting above the upper liquidity placement boundary. Conversely, if the current pool price drops below the lower bound (Figure~\ref{figure:C3}), the LPs are left with only reserves of the depreciating token, and their capital level can plummet as low as possible (until the LP position is closed), depending on the contract price. This presents a classic market risk scenario but is no longer directly related to impermanent loss.

\section{Related Work}
\label{section:RelatedWork}

The properties of DEXs and specifically CLMMs gave been studied in a number of papers. The basic idea behind the CLMM is described in the pioneer works of Hayden Adams \cite{Adams2018}, \cite{Adams2020}, and \cite{Adams2021} where the authors provide the ideas behind the automated market makers, introduce the concept of concentrated liquidity, and describe the details of the Uniswap DEX technical implementation in the Ethereum blockchain. 

The paper \cite{Adams2021} contains the basic CLMM invariant but lacks details for liquidity operations. To look at the DEX pool from the LPs point of view and understand deeply the maths behind the liquidity provision and withdrawal, as well as tokens transformations within and outside the price range one should refer to \cite{Elsts2021}. 

Basic understating of the liquidity properties and impermanent loss may be found in \cite{Aigner2021} where the authors derive the expressions for impermanent loss in the case of Uniswap V2 and Uniswap V3, and include very instructive illustrations. The LPs capital value properties have been studied more deeply in \cite{Milionis2022} and \cite{loesch2021impermanent}. The properties of impermanent loss and real market data have been studied in \cite{loesch2021impermanent} where the infamous conclusion is made about LPs quite often loosing money in the liquidity pools. The paper \cite{Milionis2022} studies the concept of impermanent loss and the properties of passive liquidity provision, their results include ideas to hedge out the market risk in the liquidity pool based on the concept of loss versus rebalancing. The authors position their results as a “Black-Scholes formula for AMMs”.

Another worth mentioning paper is \cite{echenim2023thorough} which adds the trading fees accumulated by the LPs into consideration. The paper continues the attempts to understand the liquidity math behind DEX pools which was started in \cite{Elsts2021} and derives very important estimates for the total accumulated amount of fees in the CLMM pool with the assumption that the price follows a Geometric Brownian motion, but it does not take into account the trading volumes which are not directly derived from the price process in general case.

The trading volume is crucial for the purpose liquidity provision and especially for active liquidity provision. The case of a static price range has been studied in the papers described above, but it is more interesting to understand what the LP can do in a dynamic setting. This is studied in \cite{Bar-On2024} for where the authors derive a liquidity provision strategy in a non-stochastic case, and in \cite{Fan2023} where the authors consider a special class of $\tau$-reset liquidity provision strategies. They will be considered in detail further in current paper. 

One should be able to compare the liquidity provision strategies among the wide range of different approaches, taking into consideration not only the accumulated fees but also the impermanent loss. In this context is is very important to mention the paper \cite{milionis2023flair} which develops the idea behind loss versus rebalancing and introduces a metric for LP competitiveness. 

Since current work focuses on the trading volumes and accumulated fees, and develops a backtesting solution for the CLMMs it is important to be aware of numerous backtesters solutions available in the DeFi space. These include \href{https://github.com/DefiLab-xyz}{DeFiLab}, \href{https://github.com/Bella-DeFinTech/uniswap-v3-simulator}{Tuner}, \href{https://github.com/mellow-finance/mellow-strategy-sdk}{Mellow Finance}, \href{https://docs.revert.finance/revert/technical-docs/backtester}{Revert finance}, \href{https://www.gamma.xyz/}{Gamma strategies}, \href{https://github.com/UniverseFinance/UniCode2021-UF-Backtesting}{Universe Finance}, \href{https://github.com/zelos-alpha/demeter}{Demeter} which provide different types of instruments facilitating the LP strategy analysis. Such kind of analysis is supported by the \href{https://www.uniswapfoundation.org/liquidity}{Uniswap Foundation} and is also develop by numerous analytics and active liquidity provision solutions like \href{https://www.charm.fi/}{Charm finance}, \href{https://github.com/credmark}{Credmark}, \href{https://www.visor.finance/}{Visor}, \href{https://aloe.capital/}{Aloe}, \href{https://flipsidecrypto.xyz/}{Flipside}, \href{https://github.com/unbound-finance/}{Unbound}, \href{https://uniswap-simulator.vercel.app/}{Vercel}. Current work adds to the research topic and introduces a comprehensive CLMM backtesting solution which is applicable to both real world and simulated data and may support not only LP strategy choise but also the new pool launch decision making process. 

\section{Proposed Solution}
\label{section:ProposedSolution}

All technical details and descriptions are provided below. An IPython notebook with the code containing the backtesting tool configurations and the raw data \replaced{is available on Github \cite{UrusovGithub2024}.}


The developed backtester is a semi-manual tool that allows a DeFi researcher to see every stage of modeling in detail. Standard Python libraries were used for development, and all main structures were implemented using CuPy, which is a GPU-accelerated counterpart of NumPy for computations. The backtester algorithm incorporates a $\tau$-reset strategy studied in \cite{Fan2023}.

\begin{figure*}[h]
\centering
\includegraphics[width=\textwidth]{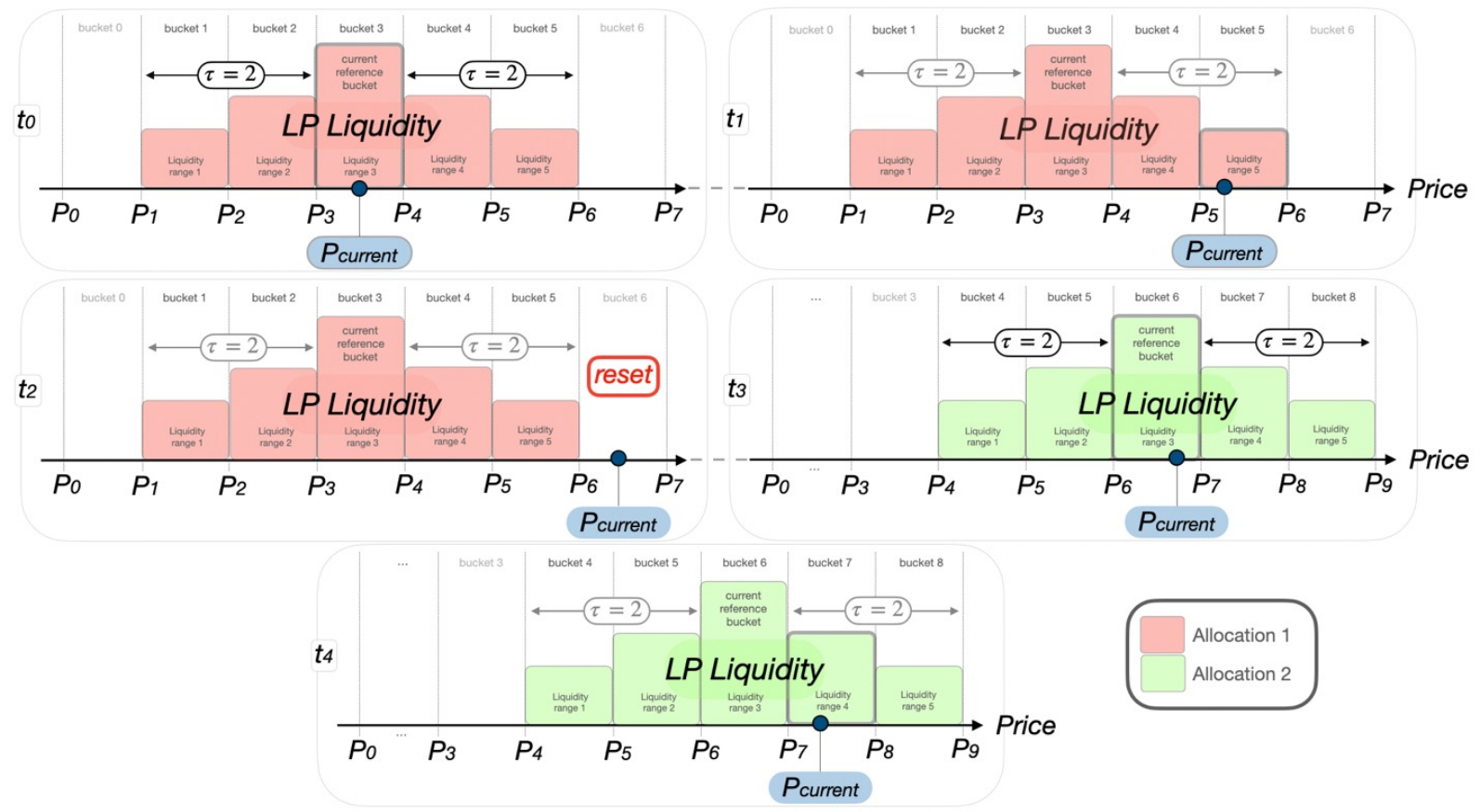}
\caption{Example of $\tau$-reset strategy for $\tau$=2}
\label{figure:M1}
\end{figure*}

The essence of the strategy involves dynamically placing liquidity based on the current value of the contract price. At the initial time $t=0$, the LP selects a fixed range length dividing the price series into buckets of fixed length $\beta = \{B_1,...,B_n\}$. The benchmark bucket $\boldsymbol{s}$ (current reference bucket) is the bucket containing the current contract price, and $\tau$ is a parameter that determines the number of buckets to the left and right of $\boldsymbol{s}$ where LP will place liquidity along with $\boldsymbol{s}$. In this setup the upper boundary of each bucket is the lower boundary of the next bucket, except for the last one (similarly with the lower boundary, except for the first bucket). Subsequently, the fixed liquidity placement in each bucket is maintained until the current contract price moves more than $\tau$ buckets to the left or right ($t=2$) from the benchmark bucket $\boldsymbol{s}$, after which liquidity is redistributed similar to $t=0$, and so forth (Figure \ref{figure:M1}).

\begin{figure*}[h]
\centering
\includegraphics[width=0.9\textwidth]{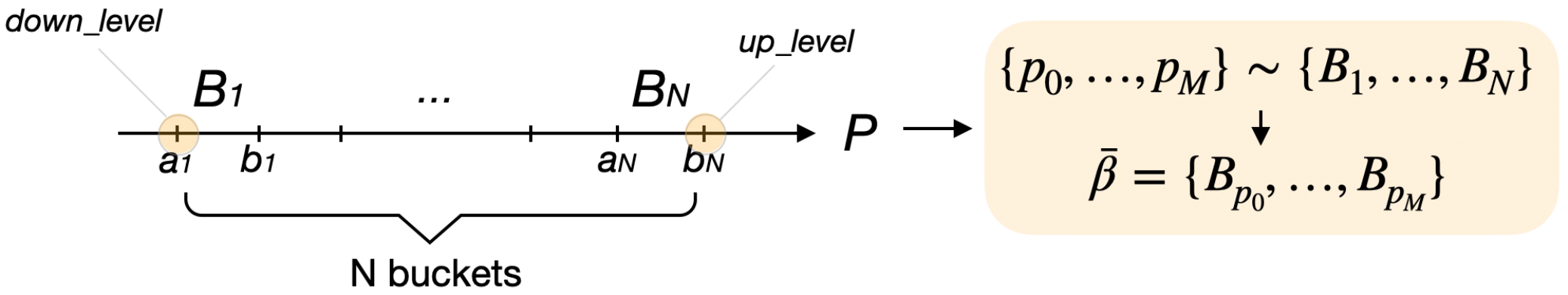}
\caption{The initial bucketing process}
\label{figure:M2}
\end{figure*}

We start by taking a historical price series $P$=$\{p_0,p_1,...p_M\}$, which will either be the actual price series of the modeled pool or quotes from a centralized exchange, considered as the market prices to which the hypothetical pool prices aspire, as will be explained in more detail below. Next, we set the upper and lower boundaries of our partitioning $\beta$ (bucket $B_n$ and $B_1$, respectively) and the parameter $N$ for the number of buckets (Figure \ref{figure:M2}), determining which bucket each price in the set $P$ belongs to, forming $\bar{\beta}$.

\begin{figure*}[h]
\centering
\includegraphics[width=\textwidth]{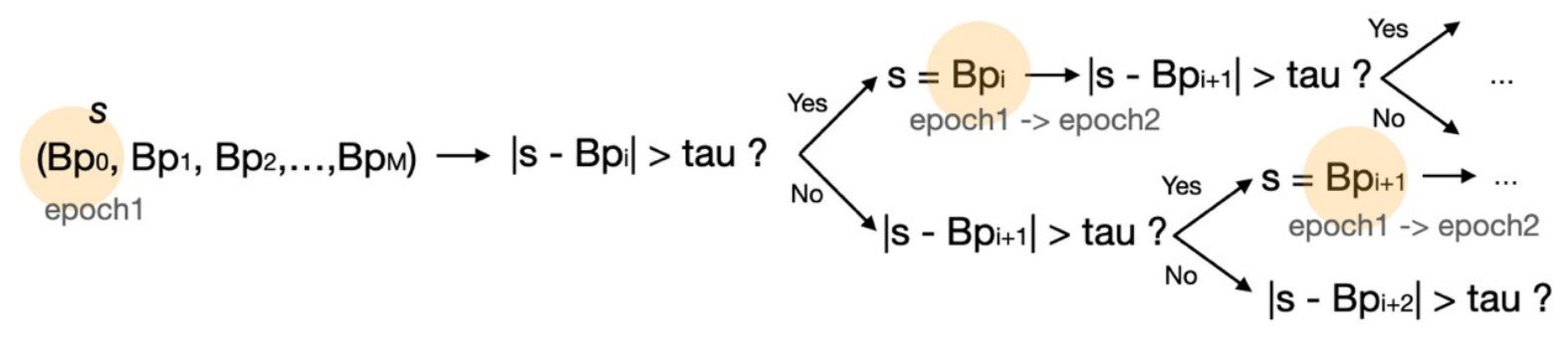}
\caption{Reference bucket $\boldsymbol{s}$ determination and the epochs change in the $\tau$-strategy}
\label{figure:M3}
\end{figure*}

After the initial partitioning into buckets, associating each price with a specific bucket, and forming $\bar{\beta}$, we will determine and denote as $\boldsymbol{s}$ the benchmark bucket where the very first price in the set $p_0$ is located. Next, we will define the moments of transition between epochs (periods of fixation for each benchmark bucket), along with their corresponding prices and buckets. Recall the rule: the bucket containing the price  $p_0$ is recognized as the benchmark bucket $\boldsymbol{s}$. Then, if the bucket number of price $p_i$ ($i$ $\in$ $[1,M]$) exceeds the bucket number of 
$\boldsymbol{s}$ by more than $\tau$, the bucket of price $p_i$ is recognized as the benchmark (here liquidity is reallocated), and the process continues until the next $\boldsymbol{s}$-bucket is found, repeating this until reaching price $p_M$ (Figure~\ref{figure:M3}). The moments of updating the benchmark bucket corresponding to specific prices serve as epoch boundaries: for a new epoch, they constitute the first price, and for the previous epoch, they constitute the last (except for $p_0$ and $p_M$). Prices between these boundaries belong to a single epoch.

\begin{figure*}[h]
\centering
\includegraphics[width=0.85\textwidth]{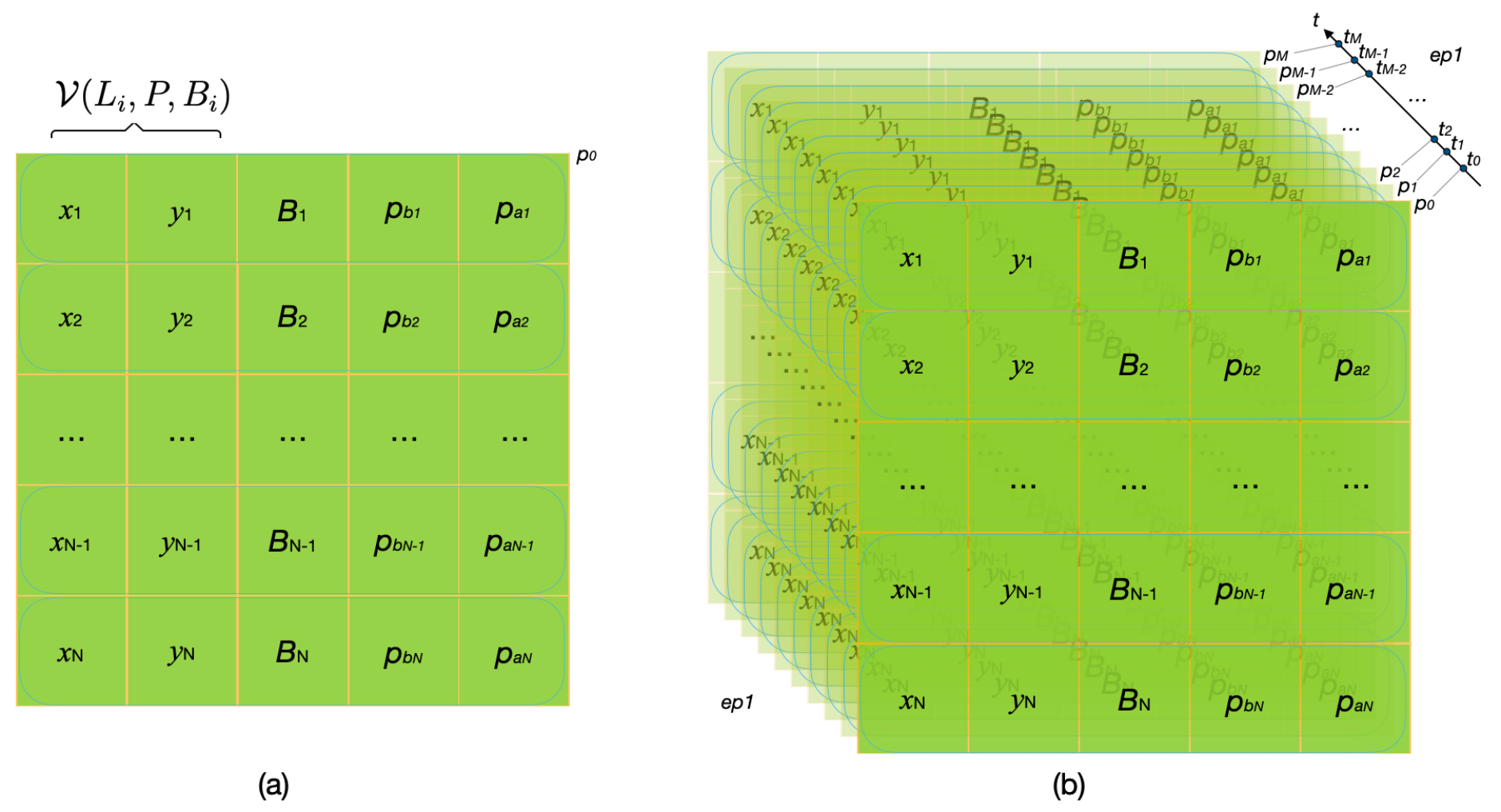}
\caption{(a) Pool state sheet structure for $p_0$.\\
(b) A multidimensional contract states array for a pool prices set of length $M$ for one epoch}
\label{figure:M4}
\end{figure*}

\begin{figure*}[!h]
\centering
\includegraphics[width=0.7\textwidth]{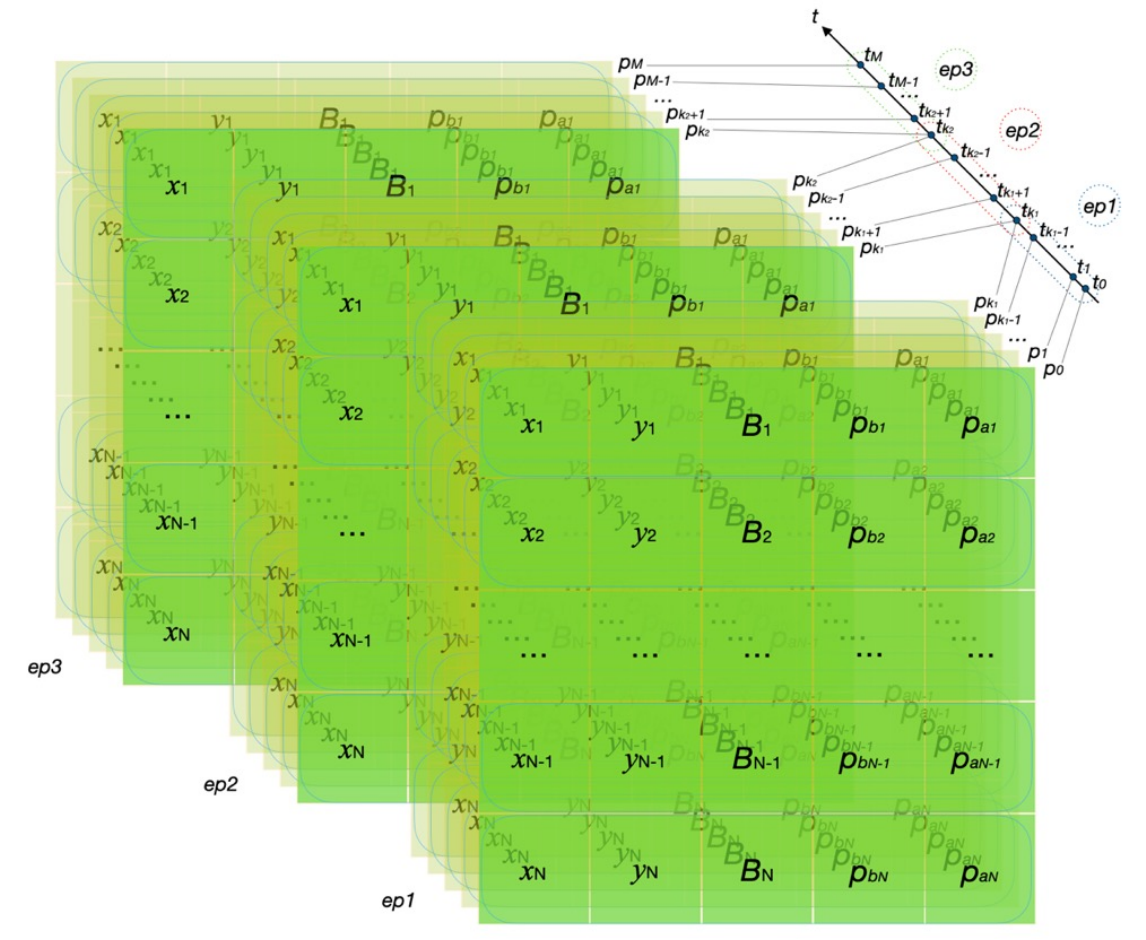}
\caption{Structure of a multidimensional array of pool states for the case of three epochs}
\label{figure:M5}
\end{figure*}

The backtester architecture is based on representing a three-dimensional NumPy (CuPy) pool states array according to the current pool price and the liquidity range configuration specific to each LP. Here are the main steps:

\begin{enumerate}
    \item Define the upper and lower boundaries of $\beta$ partitioning $P_{b_N}$ and $P_{a_1}$ (theoretical boundaries of liquidity provision ranges);
    \item Determine the number of partitions $N$ (number of equal-length buckets);
    \item Define the price set $P$=$\{p_0,p_1,...p_M\}$ used to model the pool states. These can be actual prices of a specific pool or, in their absence, the quotes of a specific pair from a centralized exchange, considered fair, assuming that the pool price will converge towards them. Below, we will discuss this issue in detail, including the assumptions and parameters to use in each case;
    \item Set the parameter $\tau$ for the $\tau$-strategy;
    \item Forming $\bar{\beta}$, initiate the search for benchmark buckets and epoch change moments, described above and illustrated in Figure \ref{figure:M2};
    \item At the beginning of the first epoch, liquidity is allocated for each bucket where the LP has placed capital at the price $p_0$ (Figure \ref{figure:M4} (a)). Then, using the liquidity state function $\mathcal{V}(L_i, P, B_i)$, the liquidity range states are calculated for each price of the current epoch (Figure \ref{figure:M4} (b)). This process is repeated for all epochs, fixing liquidity for the first price of each epoch and modeling pool (contract) range states (Figure \ref{figure:M5});
    \item For each epoch, the LP can apply different values of the $\tau$ parameter and the proportions (strategy) in which they allocate their capital in the ranges.
\end{enumerate}

These formed structures allow a detailed understanding of the reasons that led to the various states of the modeled pool, upon which the final LP reward depends.

It is worth noting again that we are not modeling flow of arbitrage and non-arbitrage swap transactions that change the pool price; instead, we model the pool state corresponding to the historical price, thereby ignoring non-arbitrage transactions that do not bring the current pool price closer to the market price. In theory, this provides a more conservative estimate of pool trading volumes and LP rewards.

\begin{figure*}[h]
\centering
\includegraphics[width=0.9\textwidth]{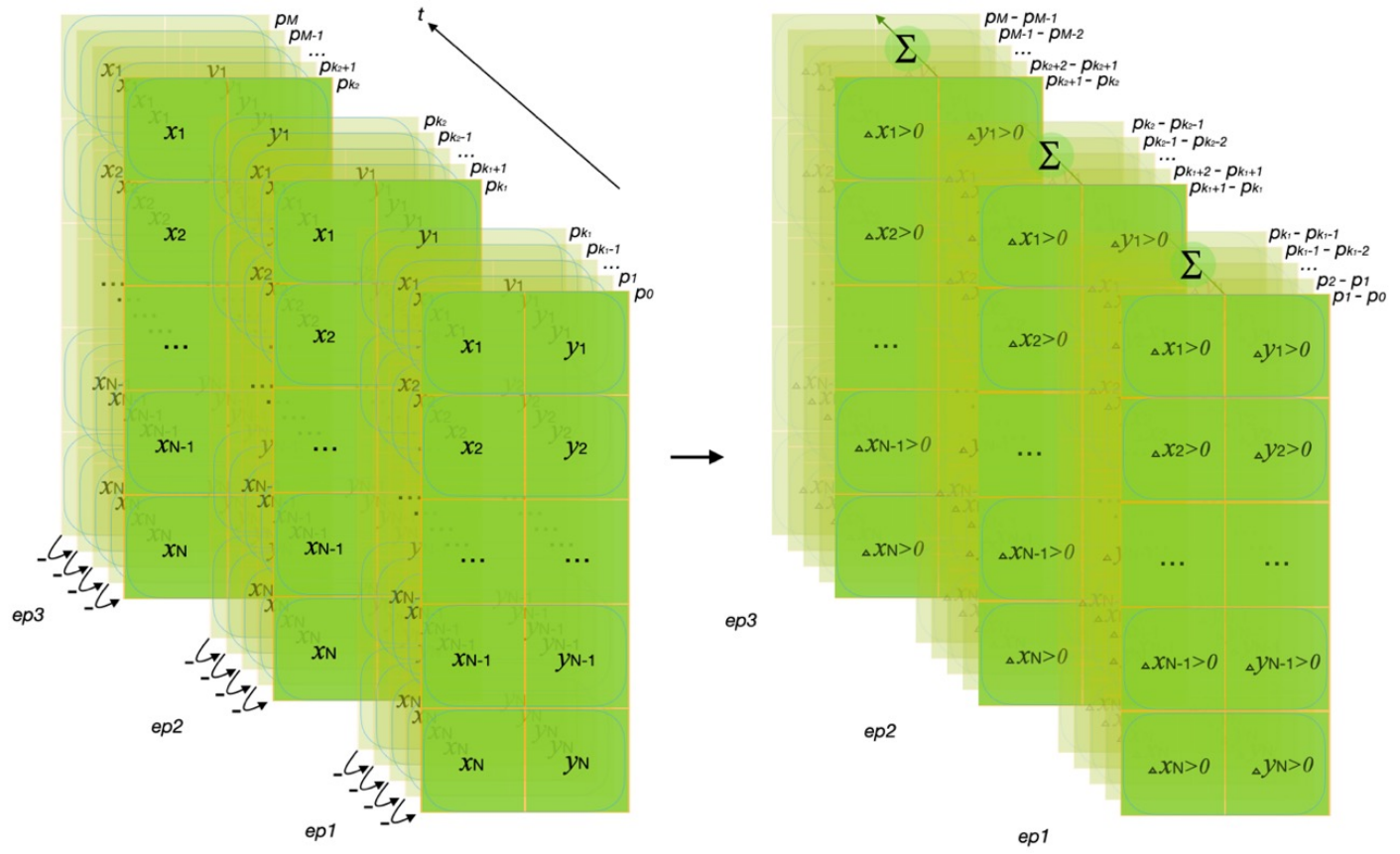}
\caption{Architecture of the reward calculation block}
\label{figure:M6}
\end{figure*}

\begin{figure*}[h]
\centering
\includegraphics[width=0.9\textwidth]{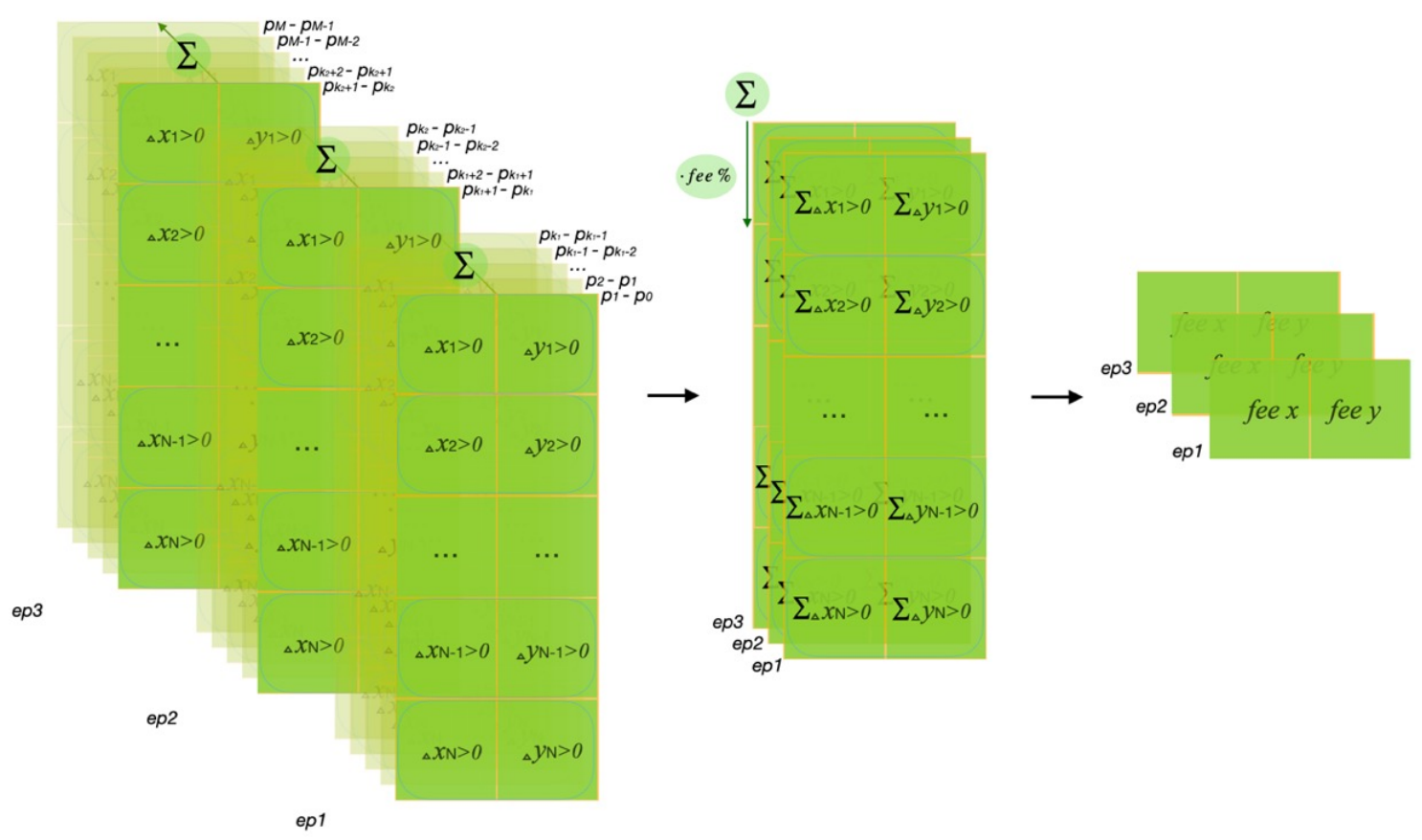}
\caption{Architecture of the reward calculation block}
\label{figure:M7}
\end{figure*}

Now let's consider the reward calculation algorithm implemented in the backtester. The underlying idea is quite simple: suppose an LP has provided liquidity in certain ranges at price $p_0$ at time $t_0$, and at time $t_1$, the pool price becomes $p_1$. How do we calculate the LP's earned reward? To do this, we need to know the reward level in the pool under consideration (fee \%) and the transactions volume that passed through the pool, shifting the pool price from $p_0$ to $p_1$. In this theoretical scenario, the pool's liquidity consists entirely of one LP's liquidity. 
Let's set the actual reward level of the specific modeled pool as the value of fee \% and calculate the transaction volume through the difference in the modeled pool's states. Let's consider the approach in a slightly more general sense, using the example of three epochs (Figure \ref{figure:M6} and \ref{figure:M7}). Epoch changes occur at the moments when the pool prices reach $p_{k_1}$ and $p_{k_2}$ (where $t_M > t_{k_2} > t_{k_1} > t_0$).

Knowing the liquidity states of each bucket at every moment in time (at each price), we can determine the volume of tokens $A$ ($x$ reserve) and tokens $B$ ($y$ reserve) that should enter each range (bucket) to change the pool price to the desired level. To do this, we calculate the differences between the states of our multidimensional array, moving from the last state to the first in each epoch. Then we consider only the positive differences, which correspond to the volumes of tokens contributed by traders to the pool for exchange: these are the volumes where the LPs earn rewards as the pool price moves through the LP's liquidity range.

Next, we will find the sum of all positive differences for each range within each epoch. We will adjust the contributed amounts of each token type within each epoch by multiplying them by the pool's reward rate, and will determine the final sizes of rewards for each epoch in terms of tokens $A$ and $B$. To fix the sum of rewards in terms of token $B$ at the end of each epoch, we simply multiply the rewards sum in token $A$ by the final pool price in each epoch. 

\begin{figure*}[h]
\centering
\includegraphics[width=0.8\textwidth]{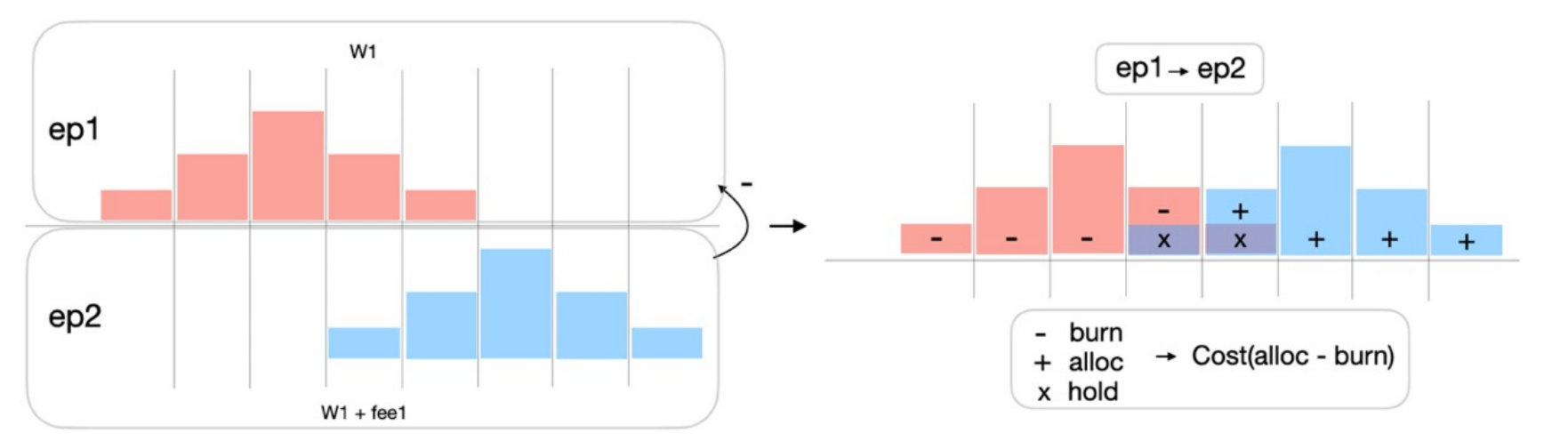}
\caption{Architecture of the block for calculating commissions when redistributing liquidity}
\label{figure:M8}
\end{figure*}

Now, let's talk about accounting for gas fees. When providing liquidity, LPs pay gas fees in two cases: when they add their liquidity in specific ranges and when they bern it.
Within the framework of the strategy we are considering, the LP places its liquidity at the initial moment, reallocates it at the epoch transition points, and fully withdraws it at the last moment (after reaching the final model pool price). The gas fee calculation will be based on the following rules:

\begin{enumerate}
    \item LP pays for each range where liquidity is deposited or withdrawn, regardless of the exchange of the placed liquidity;
    \item The cost for placing liquidity in one range is fixed at 430 000 gas units;
    \item The cost for withdrawing (burning) liquidity in one range is fixed at 215 000 gas units;
    \item During liquidity reallocation between epochs, the number of ranges in which liquidity placement and withdrawal must occur (fully or partially) is determined based on the liquidity placement strategy in both epochs (Figure \ref{figure:M8});
    \item The gas price for modeling the primary model is fixed at 100 Gwei (an elevated level to ensure transaction inclusion in the current block, although this level is significantly higher than the one after the Dencun Ethereum upgrade). 
\end{enumerate}

In addition to the possible backtester versions mentioned above, there are several settings that make it a truly flexible tool, including: the ability to set the desired level of the strategy's $\tau$ parameter, partition boundaries $\beta$, range lengths for liquidity placement through the bucket count parameter $\beta$; the ability to perform strategies backtesting with rewards reinvestment for past epochs, without their inclusion, or with capital fixation at the beginning of each epoch at a certain level; the ability to generate a random liquidity placement strategy with a known $\tau$ or manually define this distribution, including the overall pool operation simulation; the ability to use a single price series for modeling or multiple price series, for example, synthetic trajectories obtained from a parametric model; the ability to limit the capital turnover in the pool at a known level of the modeled pool.

\section{Pool Modeling on Real Data}
\label{section:Experiments}

\begin{figure*}[h]
\centering
\includegraphics[width=\textwidth]{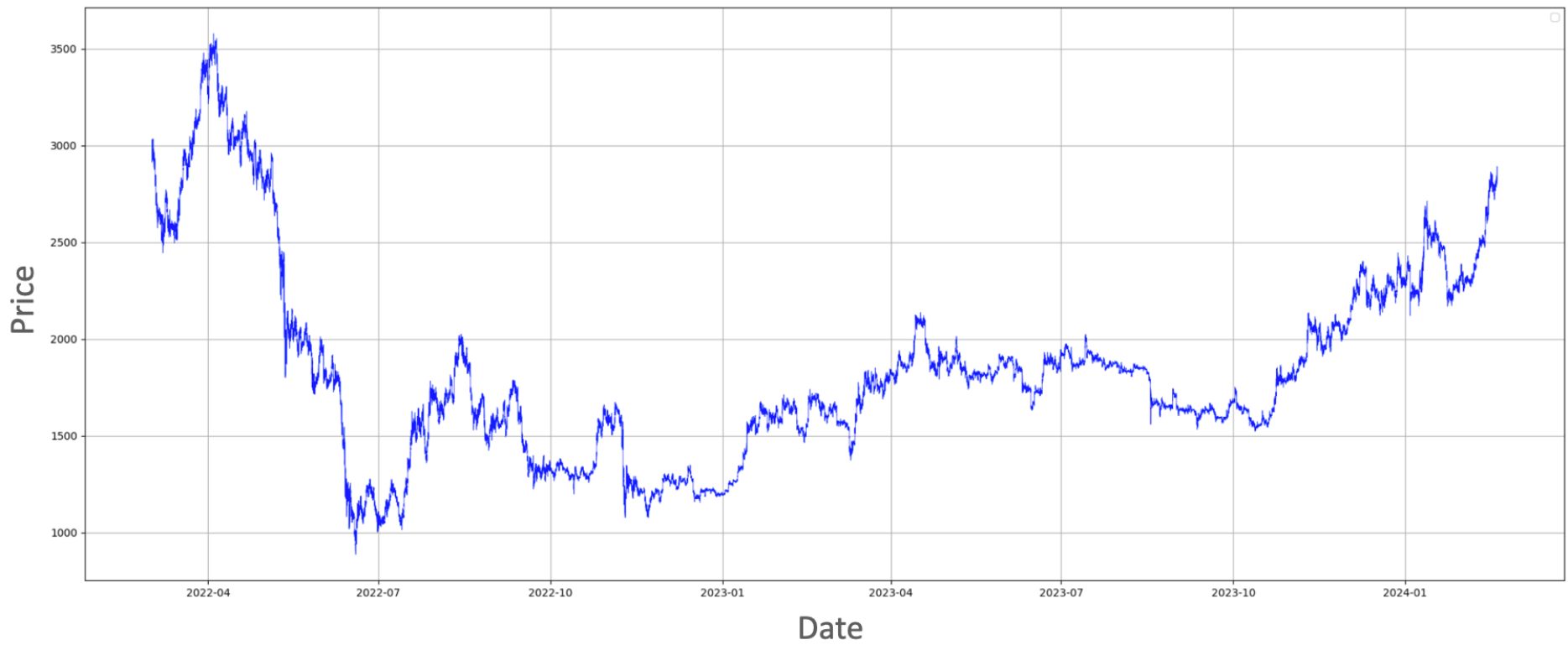}
\caption{Price dynamics for the USDC/ETH pair from March 2022 to February 2024}
\label{figure:BS}
\end{figure*}

In this section, we will examine the backtester operation results. We need to ensure the correct functioning of the developed tool, so we will simulate several examples using real data: Uniswap v2 pool, Uniswap v3 pool with a single range, and Uniswap v3 pool\footnote{Contract 0x8ad599c3A0ff1De082011EFDDc58f1908eb6e6D8} for the USDC/ETH pair at 0.3\% with starting capital $W$= \$1m. Pool modeling will be conducted without applying the $\tau$-reset strategy; the entire considered period will be treated as a single epoch. Initially, we will use minute-level quotes from Binance for the period from March 2022 to February 2024 (Figure \ref{figure:BS}).

\subsection{Uniswap v2}\label{subsection:Uv2}
\begin{figure*}[!h]
\centering
\includegraphics[width=\textwidth]{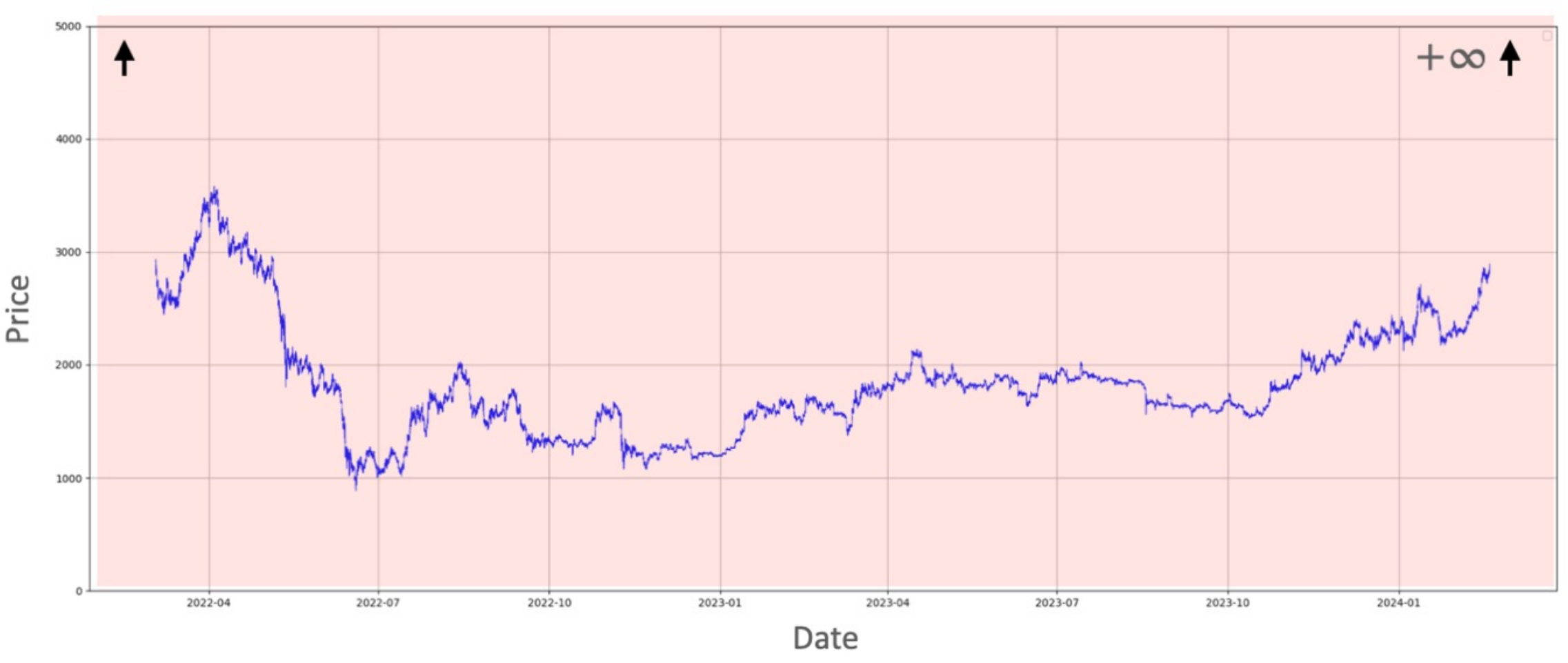}
\caption{Range of liquidity allocation in Uniswap v2}
\label{figure:BS1}
\end{figure*}

To configure this pool, it is sufficient to set one range with a lower bound near zero ($0.1^{15}$) and a large value as the upper bound ($10^{15}$), simulating a range from zero to infinity for Uniswap v2. In the Figure \ref{figure:BS1}, the liquidity provision range for the current configuration is highlighted in red.

\begin{figure*}[!h]
\centering
\includegraphics[width=\textwidth]{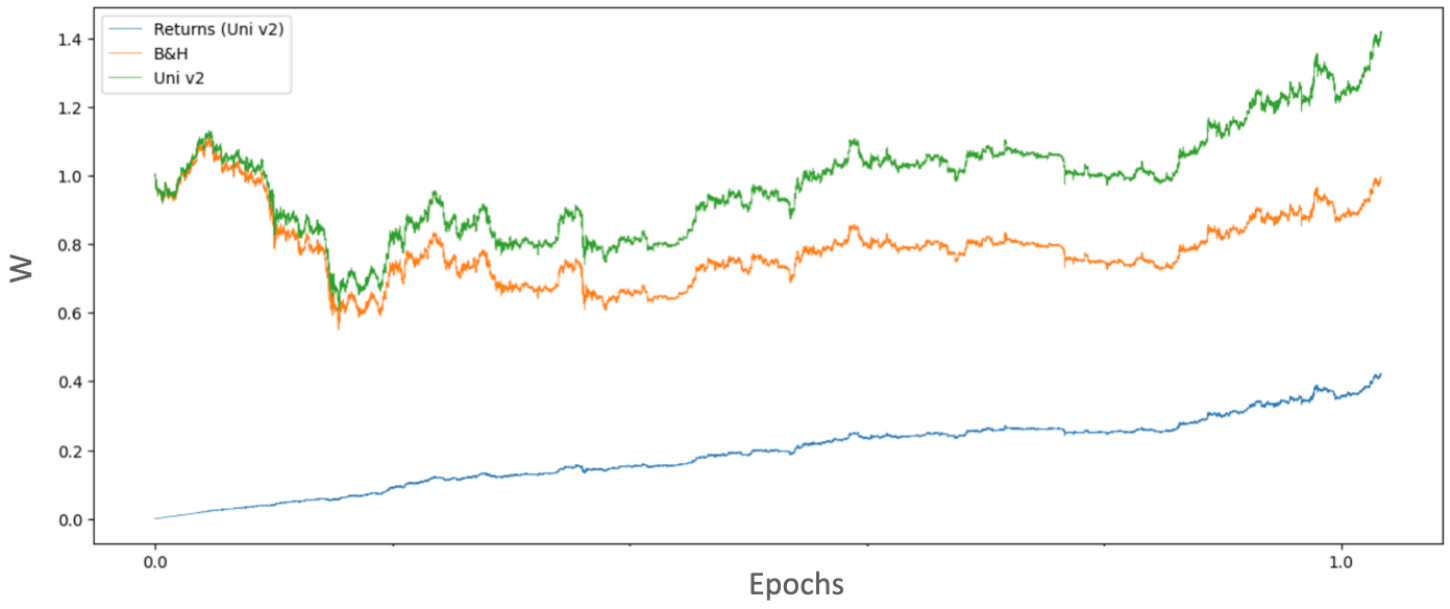}
\caption{Dynamics of LP capital position (Uniswap v2 vs B\&H)}
\label{figure:BS2}
\end{figure*}

The number of epochs with the current configuration will be equal to one, as already mentioned. Results visualization for the entire considered period is shown in Figure \ref{figure:BS2}.

We will use a B\&H strategy as a benchmark. It involves splitting the capital into two tokens, similar to splitting when providing liquidity in the pool, but thereafter, we do nothing with it, mimicking asset holding in a wallet. 

The results of the initial modeling approach: the B\&H strategy shows approximately zero return because the price at the start of the period under consideration is approximately the same as at the end (\$2907.7 vs. \$2881.2). In contrast, providing LP capital in Uniswap v2 via exchange rewards can yield significantly higher profits. In Table \ref{table:t1} are the modeling results.

\begin{table}[h!]
\centering
\caption{Summary Uniswap v2}
 \begin{tabular}{|c | c|} 
 \hline
 Strategy & {Profit rate, \% year} \\ [0.5ex] 
 \hline\hline
 Buy and Hold & -0.24\% \\ 
 \hline
 Uniswap v2 & +21.81\% \\ [1ex] 
 \hline
 \end{tabular}
 \label{table:t1}
\end{table}

It appears that nearly +22\% profit per year for Uniswap v2 is an inflated result, all due to the use of minute-level quotes for USDC/ETH from Binance. Next, we will show why this modeled reward level overestimation occurs and what tools help to combat it. However, at the moment, we are modeling a hypothetical Uniswap v2 pool where the current price changes every minute (i.e., there are 60 trades per hour) according to the available price history. It's important now to demonstrate our backtester's reward level sensitivity to pool configuration, result accuracy will be achieved using backtesting tools, which will be discussed below.

\subsection{Uniswap v3 with a single range}\label{subsection:Uv3one}
\begin{figure*}[!h]
\centering
\includegraphics[width=\textwidth]{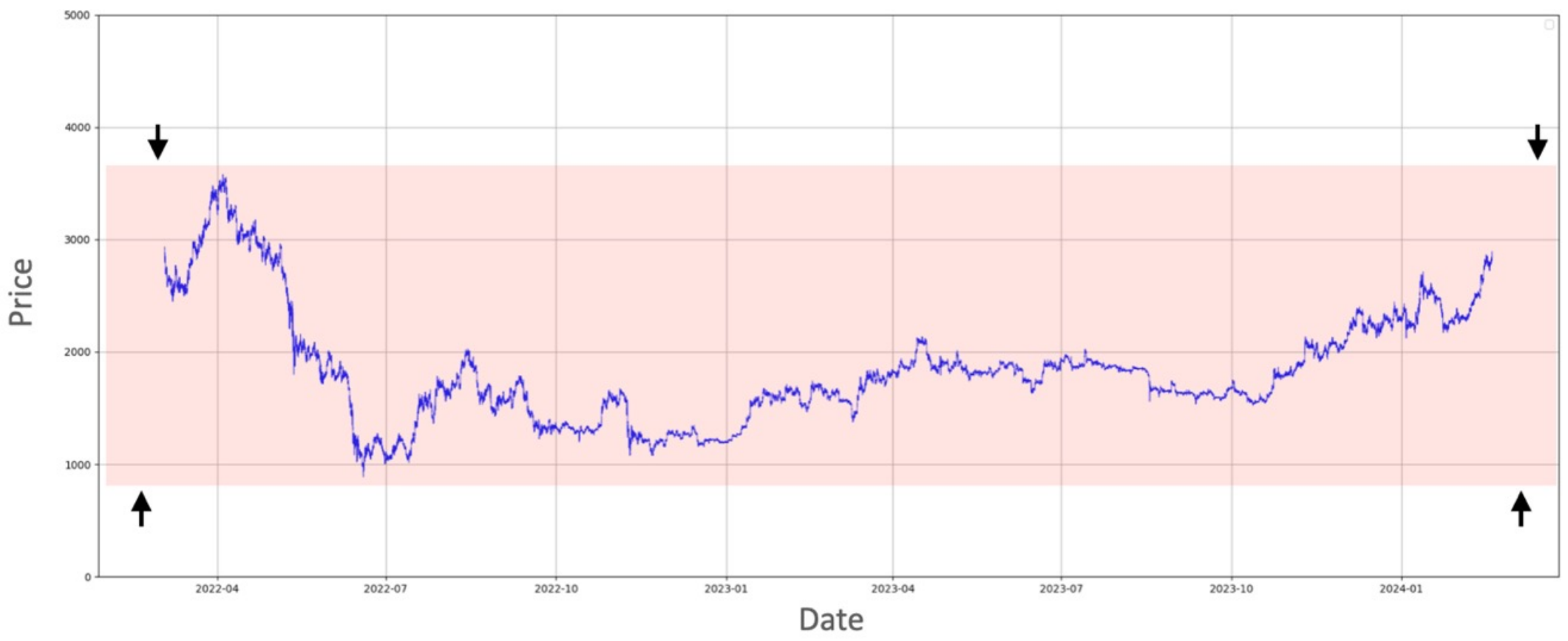}
\caption{Liquidity staking range in Uniswap v3 with one range}
\label{figure:BS3}
\end{figure*}

For this configuration, we also need to set a single range with a lower bound equal to 0.99 * the minimum price over the considered period (\$886.03) and an upper bound of 1.01 * the maximum price over the considered period (\$3579.5). In Figure \ref{figure:BS3}, the liquidity provision range for the current configuration is highlighted in red.

\begin{figure*}[!h]
\centering
\includegraphics[width=\textwidth]{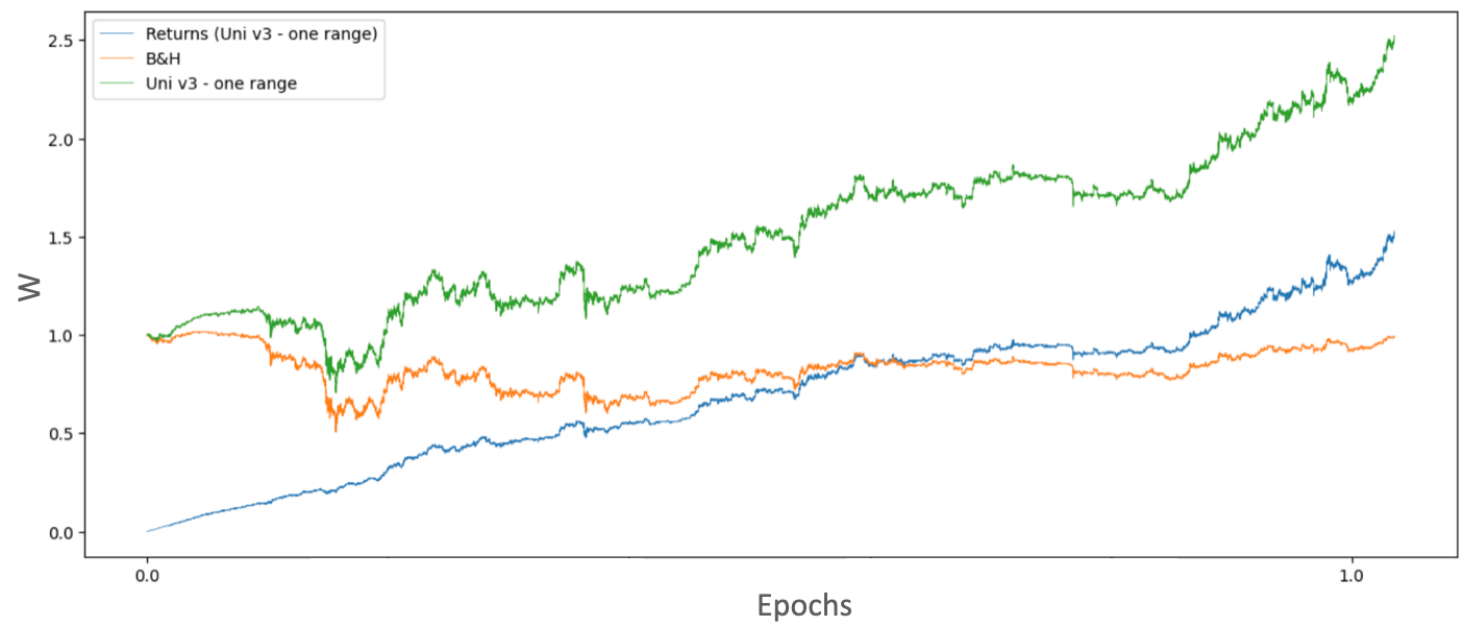}
\caption{LP capital position dynamics (Uniswap v3 one range vs B\&H)}
\label{figure:BS4}
\end{figure*}

The liquidity provision range for LP narrows, corresponding to the transition from Uniswap v2 to Uniswap v3, but still with only one working range containing all quotes from the considered period. The number of epochs in the current configuration will also be 1 because we cannot apply the price exiting logic beyond the specified range. Results visualization for the entire considered period is shown in Figure \ref{figure:BS4}.

The LP reward level significantly increased when transitioning to a narrower range. The B\&H strategy, similar to the first configuration, shows profitability close to zero, although not exactly equal to the first case due to the different splitting of the starting capital into two token sets when providing liquidity (due to different boundaries). In Table \ref{table:t2} are the modeling results:

\begin{table}[h!]
\centering
\caption{Summary Uniswap v3 with one range}
 \begin{tabular}{|c | c|} 
 \hline
 Strategy & {Profit rate, \% year} \\ [0.5ex] 
 \hline\hline
 Buy and Hold & -0.09\% \\ 
 \hline
 Uniswap v3 with one range & +79.51\% \\ [1ex] 
 \hline
 \end{tabular}
 \label{table:t2}
\end{table}

We observe a qualitative increase in the modeled reward level when providing LP liquidity in narrower Uniswap v3 ranges compared to Uniswap v2, this is an expected result due to more efficient capital utilization, which we aimed to verify. 

However, nearly +80\% to the initial capital per year with a non-dynamic Uniswap v3 strategy and +22\% with Uniswap v2 are overly optimistic results. We have demonstrated the sensitivity of our backtester to pool configuration, now it's time to talk about accuracy. To do this, we will configure a real pool where actual reward values are known and delve into the nature of modeling liquidity pool trading volumes.

\subsection{Uniswap v3. Real pool}\label{subsection:Uv3}
\subsubsection{One day}

\begin{figure*}[!h]
\centering
\includegraphics[width=\textwidth]{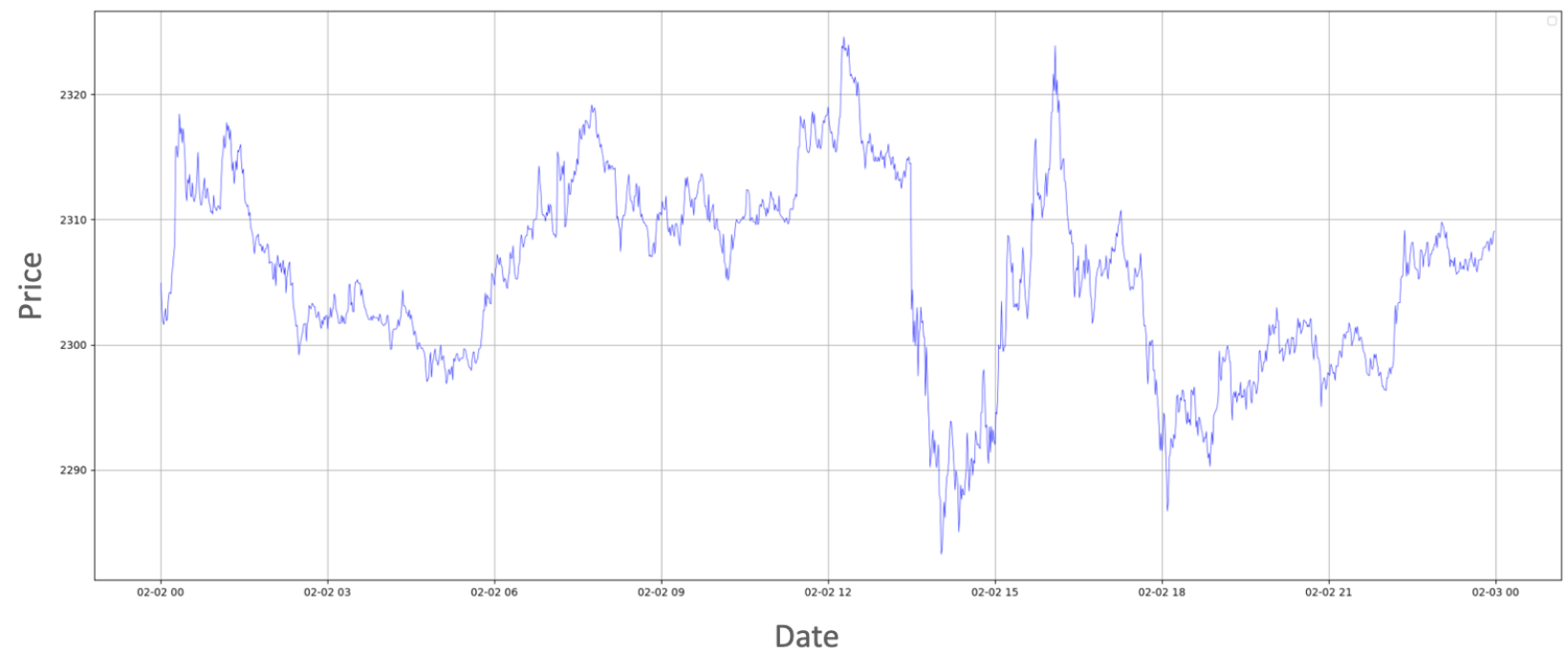}
\caption{The USDC/ETH price dynamics as of 02/02/2024 (minutes, CEX)}
\label{figure:BS5}
\end{figure*}

We'll start with the most obvious assumptions, gradually complicating the approach to obtain an accurate result. First, let's look at the price dynamics for the selected date (Figure \ref{figure:BS5}): the minimum price for the day was \$2283.27, and the maximum was \$2324.58. By examining the liquidity distribution profile (\ref{App13}), we can identify a plateau of liquidity around the current price.

\begin{figure*}[!h]
\centering
\includegraphics[width=0.8\textwidth]{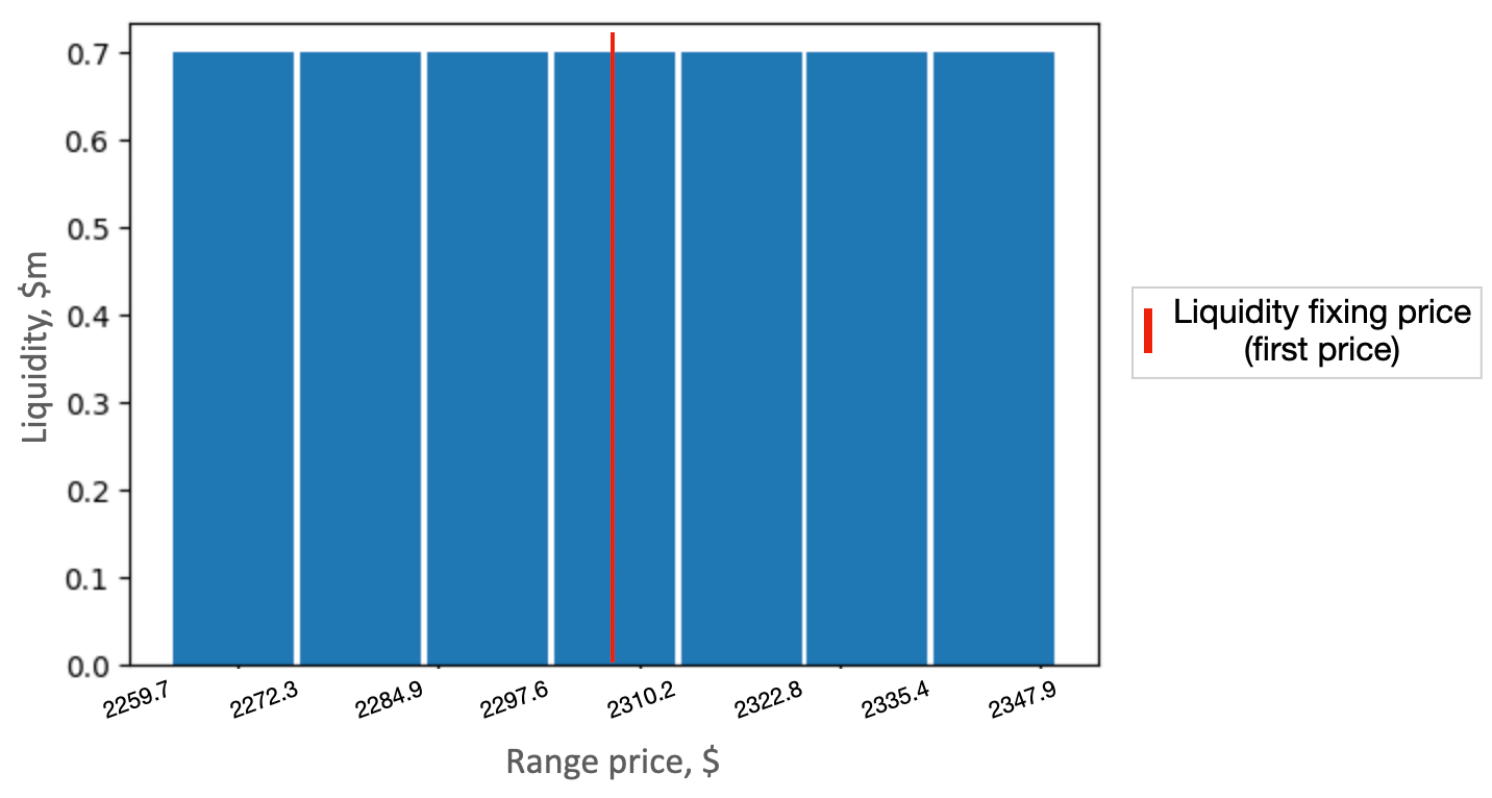}
\smallskip
\caption{Liquidity plateau of the price range under consideration}
\label{figure:BS6}
\end{figure*}

Since the price range for the day is small, it is reasonable to assume that the pool price did not exceed the liquidity plateau. Let's model this described local area of the pool (liquidity plateau), assuming that the width of one plateau range is approximately \$10-\$15, and each plateau range holds an average of \$0.7m with a total TVL of \replaced{$\sim$}{\~}\$65m; the plateau size consists of 7 ranges (Figure \ref{figure:BS6}).

\begin{figure*}[!h]
\centering
\includegraphics[width=0.8\textwidth]{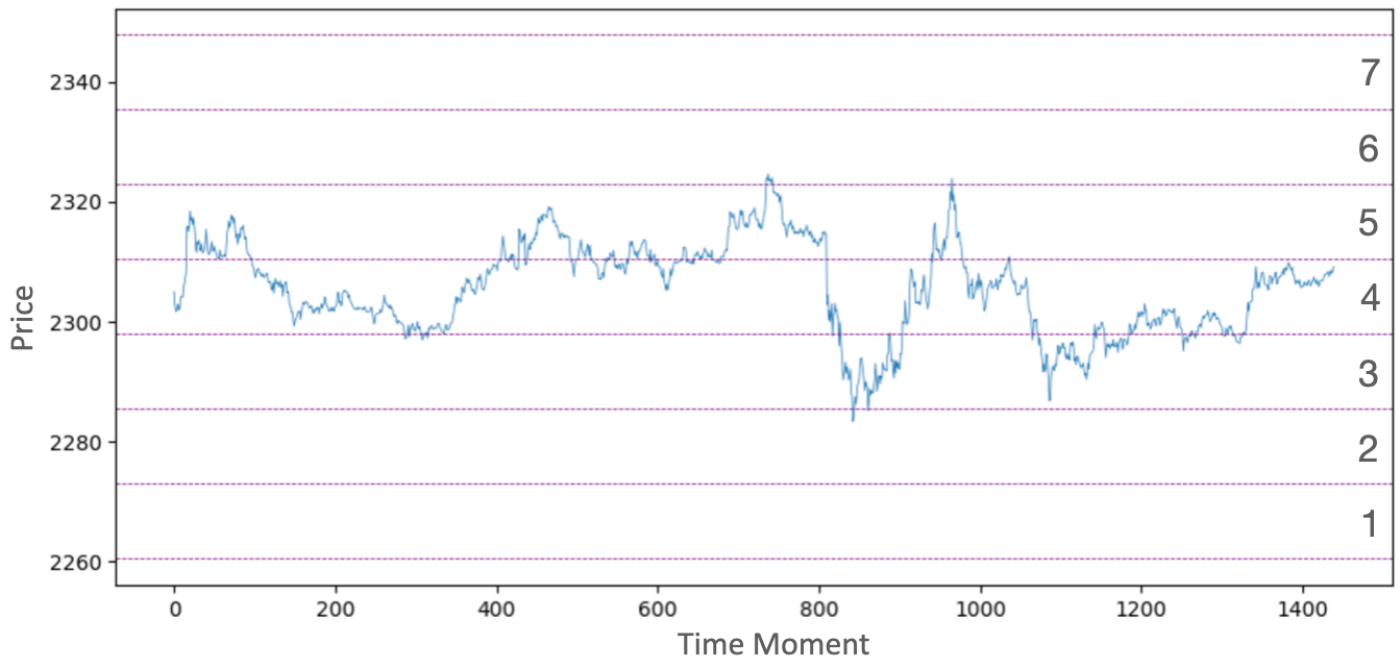}
\caption{Price dynamics relative to liquidity plateau ranges}
\label{figure:BS7}
\end{figure*}

The assumptions used correspond to the structure on 02/02/2024, as provided in \ref{App1}. The number of epochs in the current configuration, similar to 5.1 and 5.2, will also be 1, as we do not apply reset logic when the price exits beyond the specified range boundaries when modeling the pool and its local areas. We will create an additional visualization (Figure \ref{figure:BS7}) to demonstrate how the price moves between the specified price ranges.\\

Modeling also considers costs for liquidity provision at the initial moment and burning at the end, according to \ref{section:ProposedSolution}. The modeling results are presented in the Table \ref{table:t3}.

\begin{table}[h!]
\centering
\caption{Summary Uniswap v3.\\ Liquidity Plateau (minutes, CEX)}
 \begin{tabular}{| l | l | c |} 
 \hline
  \multicolumn{2}{|l|}{Indicator} & Value \\
  \hline\hline
  \multicolumn{2}{|l|}{W} & \$4.9m \\
  \hline
  \multicolumn{2}{|l|}{Cost} & \$1.04k \\
  \hline
  \multirow{2}*{Volume} & model & \$60.7m \\
       & fact & \$5.69m \\
  \hline  
  \multirow{2}*{Fees} & model & \$182.1k \\
       & fact & \$17.06k \\
 \hline
 \end{tabular}
 \label{table:t3}
\end{table}

We see that the predicted reward level is more than 10 times higher than the actual. What could be causing this? It's evident that the reward level directly depends on the volumes passing through the liquidity-provision ranges. Also, assuming a hypothetical pool turnover of 12 times per day doesn't seem plausible. It appears that the modeled transaction volume is too inflated, and the transaction volume is clearly influenced by the transaction frequency.

\begin{figure*}[!h]
\centering
\includegraphics[width=\textwidth]{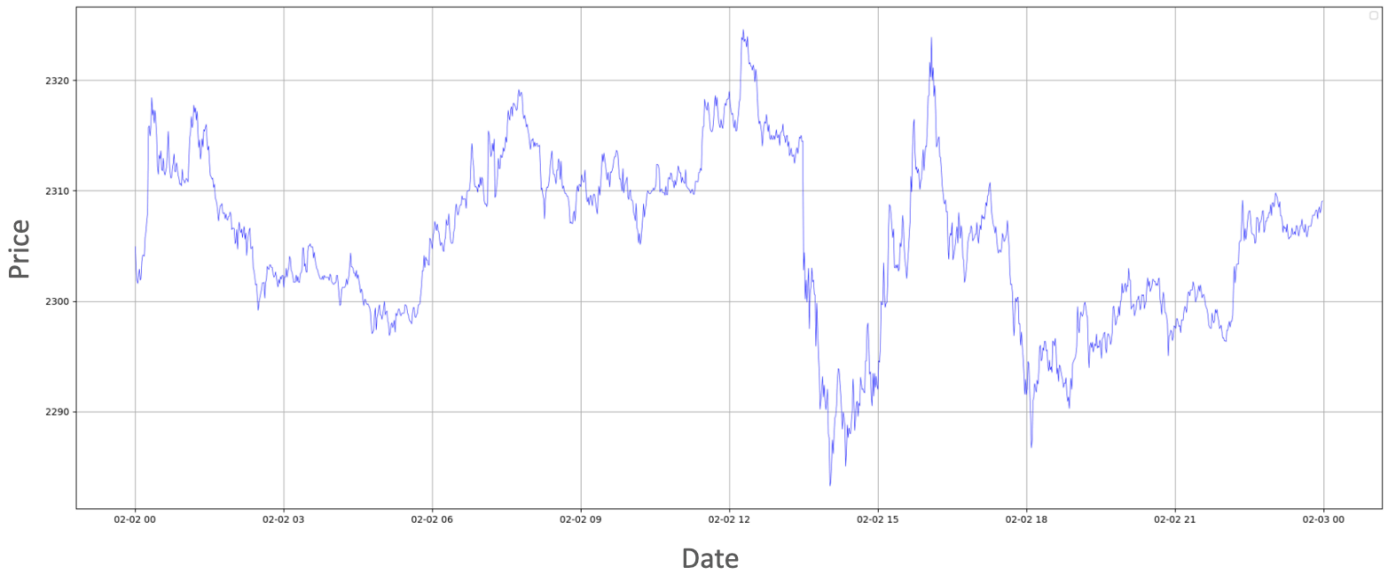}
\caption{The USDC/ETH price dynamics as of 02/02/2024 (seconds, CEX)}
\label{figure:BS8}
\end{figure*}

Let's test the almost obvious hypothesis that the price change frequency in the pool affects the modeled transaction volume. We will increase the transaction frequency by considering seconds-level quotes from Binance (Figure \ref{figure:BS8}) on the same date 02/02/2024.

We see that the profile in Figure \ref{figure:BS8} is identical to Figure \ref{figure:BS5}, but the price change frequency differs significantly. Again, let's model the liquidity plateau under the same parameters, replacing only the minute-level quotes with seconds-level quotes. The modeling results are presented in the Table \ref{table:t4}.

\begin{table}[h!]
\centering
\caption{Summary Uniswap v3.\\ Liquidity Plateau (seconds, CEX)}
 \begin{tabular}{| l | l | c |} 
 \hline
  \multicolumn{2}{|l|}{Indicator} & Value \\
  \hline\hline
  \multicolumn{2}{|l|}{W} & \$4.9m \\
  \hline
  \multicolumn{2}{|l|}{Cost} & \$1.04k \\
  \hline
  \multirow{2}*{Volume} & model & \$165.8m \\
       & fact & \$5.69m \\
  \hline  
  \multirow{2}*{Fees} & model & \$497.4k \\
       & fact & \$17.06k \\
 \hline
 \end{tabular}
 \label{table:t4}
\end{table}

We see that the modeled reward level has become more than 30 times higher than the actual due to increased trading volumes passing through the hypothetical pool, increasing the pool turnover by a factor for threefold. The main reason for the difference between the actual reward level and the modeled one here, as in sections 5.1 and 5.2, lies in the frequency of price changes within the modeled pool (DEX), which differs from the frequency of price changes in the USDC/ETH pair on a centralized exchange (CEX) in general and Binance, in particular. We cannot simply take CEX quotes for the USDC/ETH pair and apply them to model a DEX pool, even if the internal pool prices match the "market" prices from CEX, because the transaction frequency on DEX can differ significantly from that on CEX.

\begin{figure*}[!h]
\centering
\includegraphics[width=\textwidth]{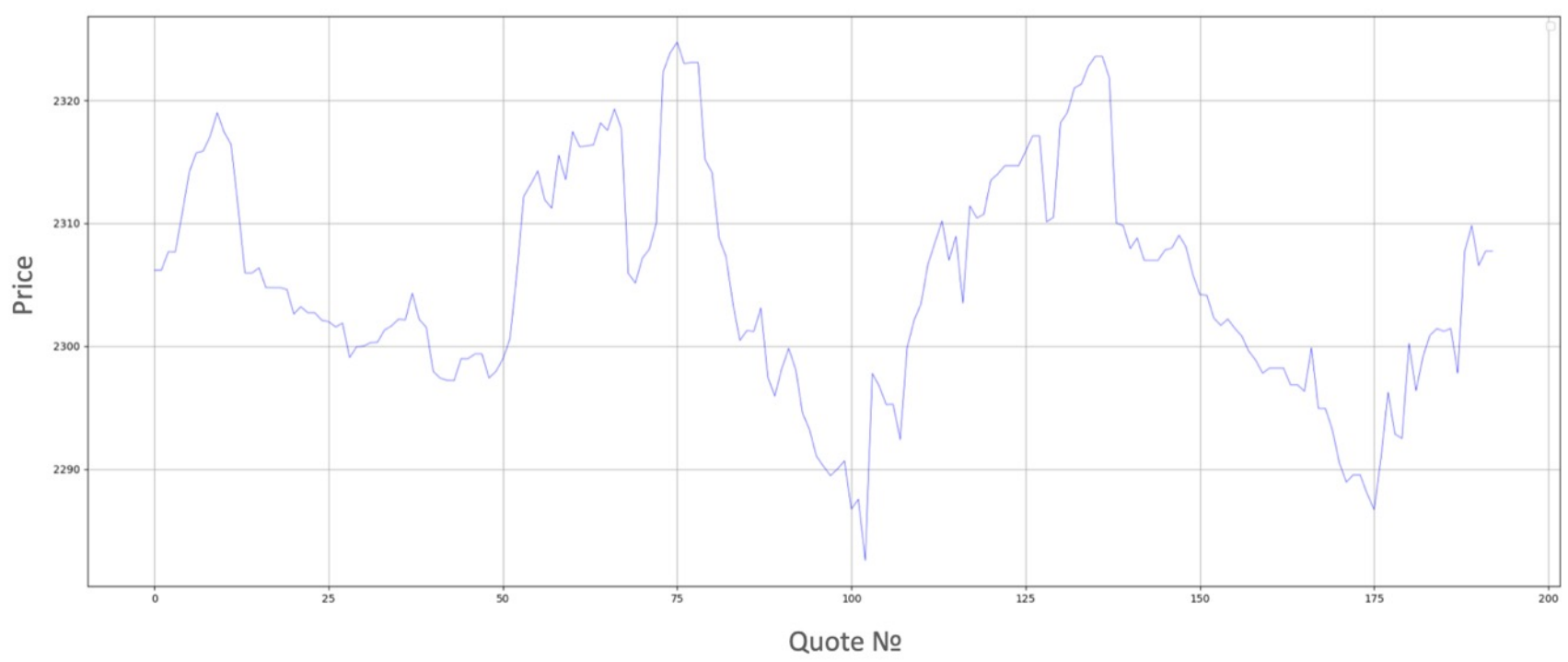}
\caption{Actual USDC/ETH DEX price dynamics as of 02/02/2024}
\label{figure:BS9}
\end{figure*}

Let's consider the actual prices (Figure \ref{figure:BS9}) of the modeled pool on the same date 02/02/2024. We see that the price dynamics in Figure \ref{figure:BS9}, \ref{figure:BS8}, and \ref{figure:BS5} are identical, but the frequency is completely different. Thus, the actual number of swap transactions in the pool on the specified date is 193, which is much lower than the frequency of the previously considered CEX quotes.

Let's model the liquidity plateau under the same parameters as before but using actual quotes. The modeling results are presented in the Table \ref{table:t5}:

\begin{table}[h!]
\centering
\caption{Summary Uniswap v3.\\ Liquidity Plateau (one day, DEX)}
 \begin{tabular}{| l | l | c |} 
 \hline
  \multicolumn{2}{|l|}{Indicator} & Value \\
  \hline\hline
  \multicolumn{2}{|l|}{W} & \$4.9m \\
  \hline
  \multicolumn{2}{|l|}{Cost} & \$1.04k \\
  \hline
  \multirow{2}*{Volume} & model & \$20.5m \\
       & fact & \$5.69m \\
  \hline  
  \multirow{2}*{Fees} & model & \$61.6k \\
       & fact & \$17.06k \\
 \hline
 \end{tabular}
 \label{table:t5}
\end{table}

The results are much closer to reality but still overestimated by several times. The reason for this lies in the constant liquidity level in the ranges of the used liquidity plateau: if we analyze the dynamics of actual prices from Figure \ref{figure:BS9} more closely, we can see sharp price jumps with a rebound to the previous level in the next moment. This is caused by price slippage in the pool due to transitions between ranges with different liquidity levels. In other words, the liquidity plateau concept is not an optimal assumption, which inflates the modeled transaction volume level.

\begin{figure*}[!h]
\centering
\includegraphics[width=\textwidth]{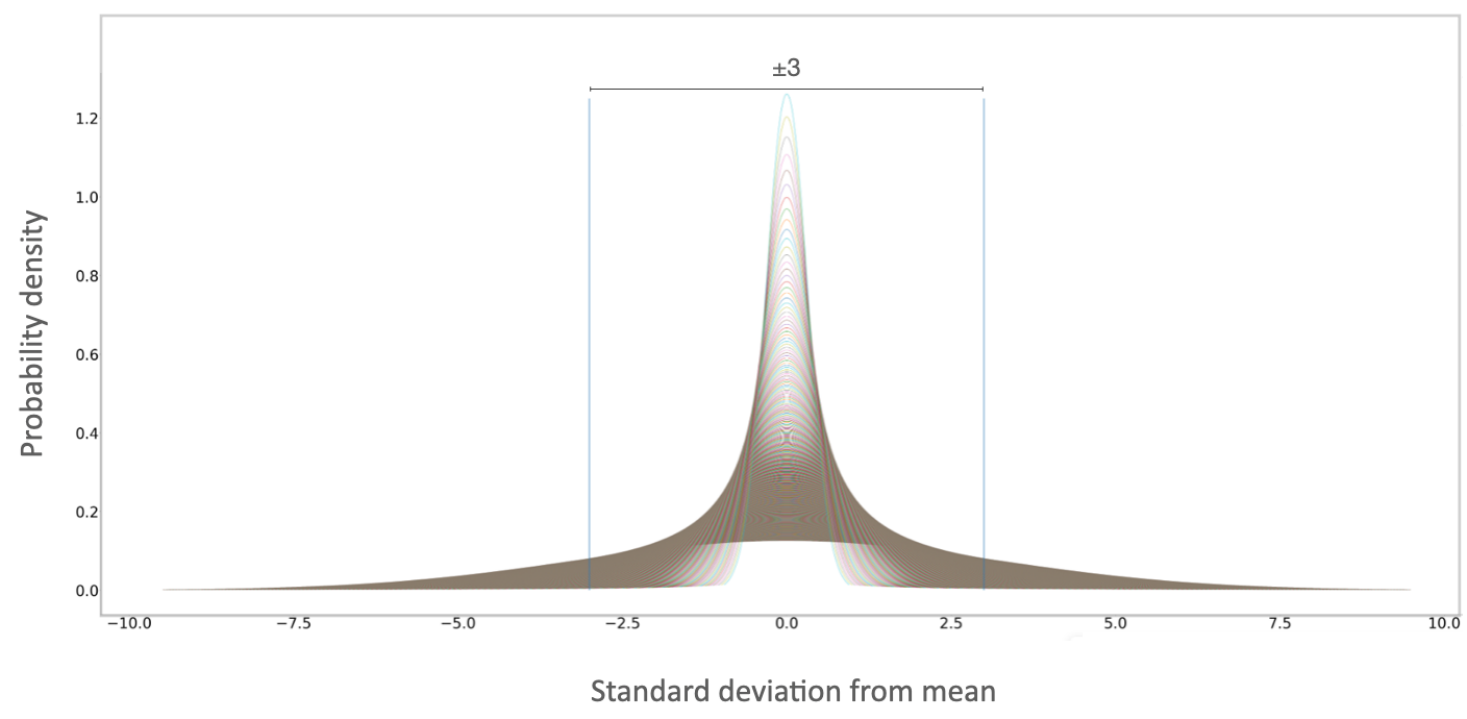}
\caption{Normal distribution density for the shape of the pool's liquidity profile.
We limit the area under consideration to ±3}
\label{figure:BS10}
\end{figure*}

For more accurate modeling, we will not consider a hypothetical liquidity plateau in a local area around the current price but instead we'll model the liquidity profile of the entire pool using a normal distribution (Figure \ref{figure:BS10}) within the defined region with the boundary of the considered distribution area ±3 (analogous to the placement strategy for a single LP in different ranges, generalized to the entire pool). 
\begin{figure*}[!h]
\centering
\includegraphics[width=\textwidth]{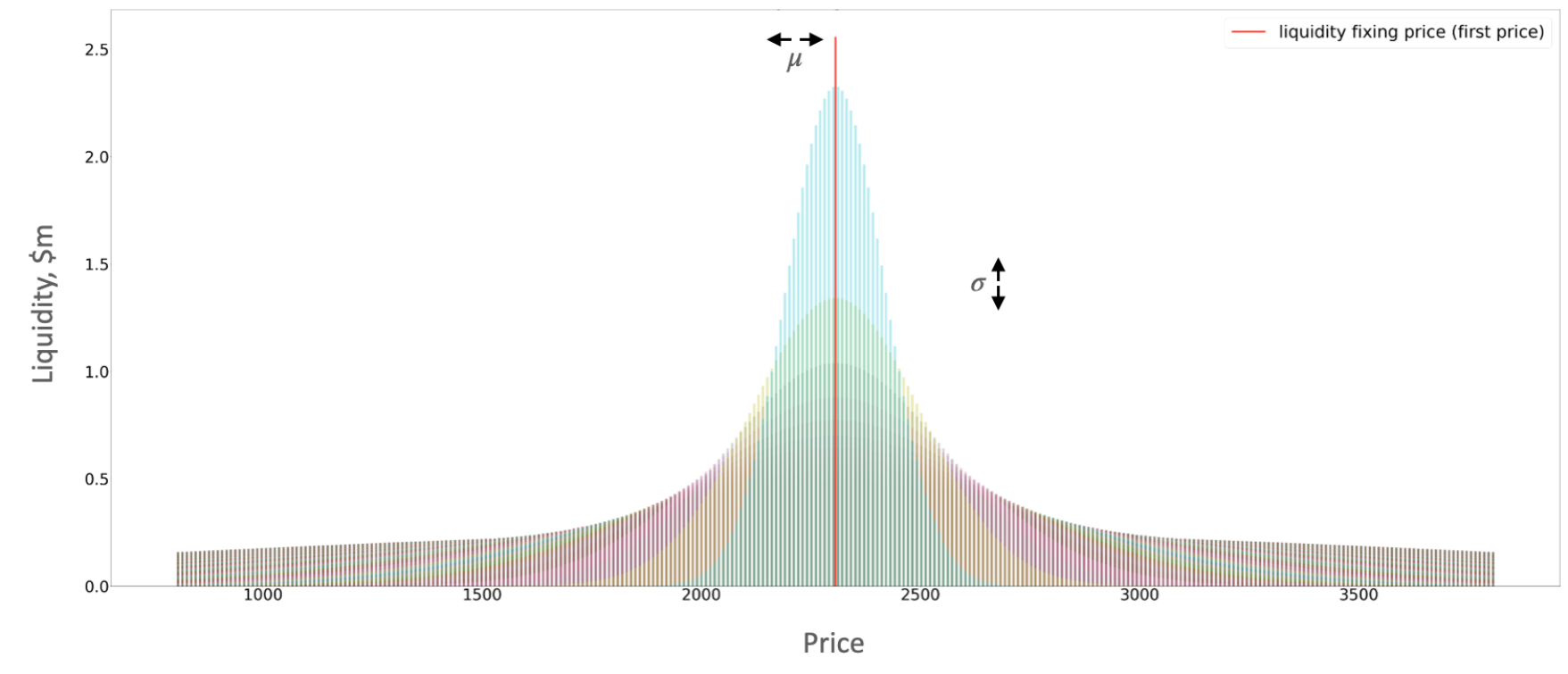}
\caption{Pool liquidity profile shape variants depending on the dispersion}
\label{figure:BS11}
\end{figure*}
We will use normalization of values within the defined area to 1, with variance and expected value ($\mu$) as parameters to approximate an accurate result (Figure \ref{figure:BS11}). We will also use the pool parameters set earlier, including the average pool TVL of \$65m on that date.

\begin{figure*}[!h]
\centering
\includegraphics[width=0.9\textwidth]{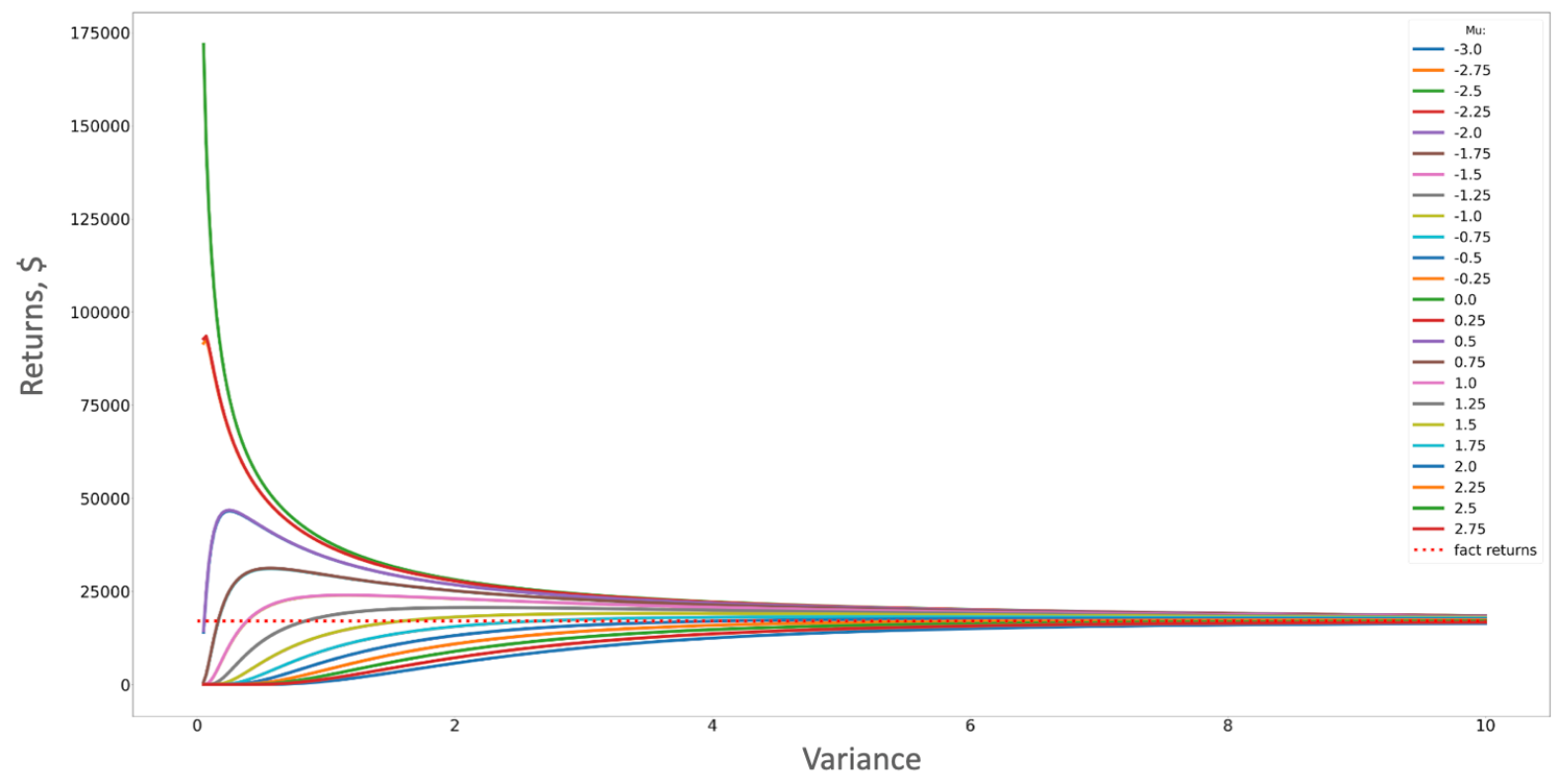}
\caption{The model remuneration level curve for the period under consideration 02/02/2024 depending on the dispersion parameter and mathematical expectation
\label{figure:BS12}}
\end{figure*} 

Using the normal distribution, we have defined a bar form of liquidity distributions in the pool. Let's approach the modeled reward value using the dynamic liquidity curve profile, resulting in the outcome on Figure \ref{figure:BS12}.

\begin{figure*}[!h]
\centering
\includegraphics[width=0.9\textwidth]{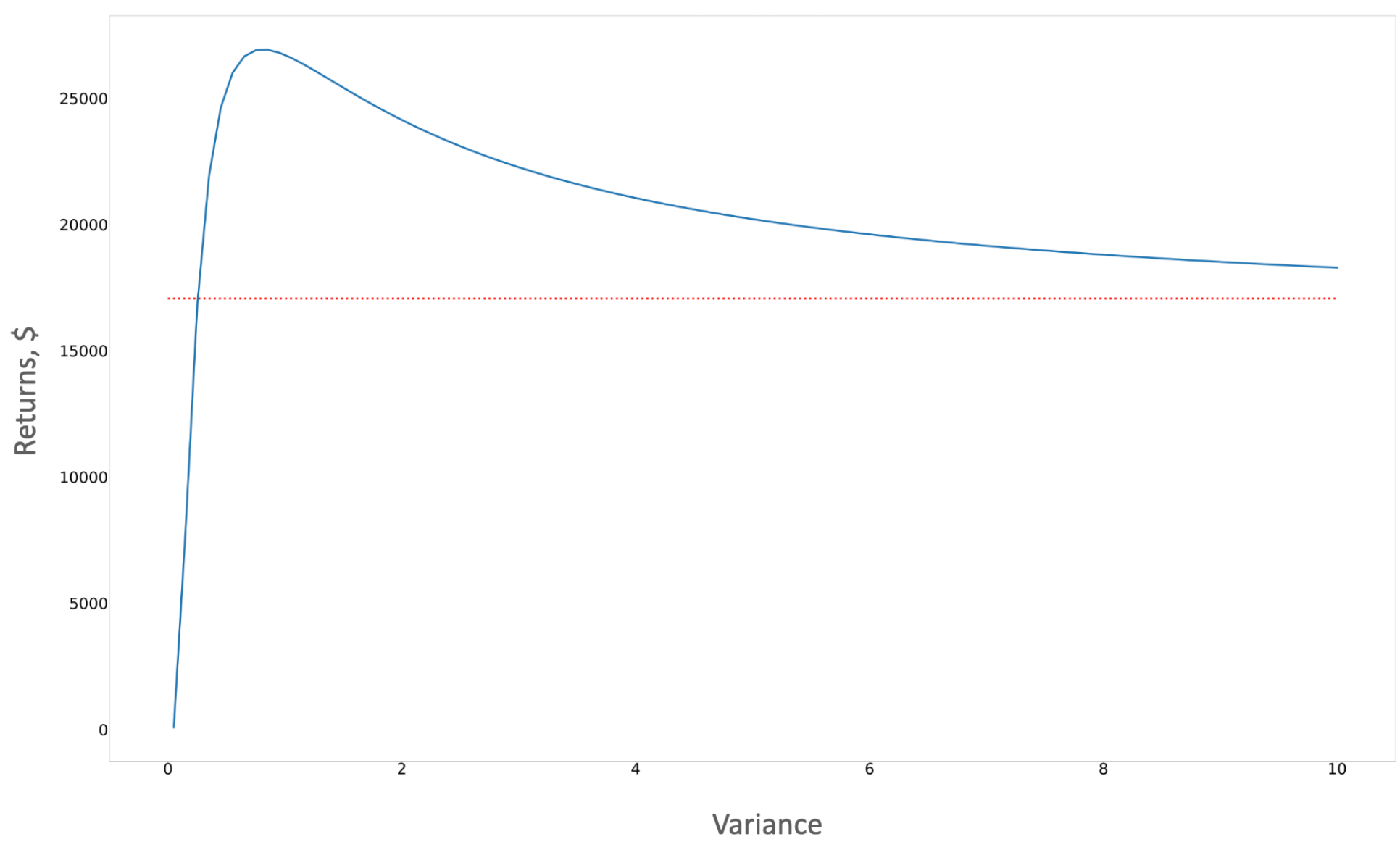}
\caption{The model pool rewards level curve for the period under consideration 02/02/2024 with $\mu$ = 0.875}
\label{figure:BS13}
\end{figure*}

The red dashed line corresponds to the actual reward level (\$17.06k). Depending on the values of two dynamic parameters, we either see an intersection of the reward level curves with the actual one or an aspiration towards it. The aspiration towards the actual result is due to the low level of actual rewards and the previously chosen boundary of the considered distribution area ±3: with high volatility, we get a "flattened" liquidity profile, but the fixed boundaries ±3 do not allow it to "smear" to zero, fixing the minimum reward level at a certain point. In this case, the minimum reward level seems to match the actual, but this is a random coincidence dictated only by the current configuration. For further discussion, we will use the variance level at which the first intersection (Figure \ref{figure:BS13}) of the reward curve with the actual level occurs at the selected value of the $\mu$ parameter.

\begin{table}[h!]
\centering
\caption{Summary Uniswap v3.\\ Liquidity Curve (one day, DEX)}
 \begin{tabular}{| l | l | c |} 
 \hline
  \multicolumn{2}{|l|}{Indicator} & Value \\
  \hline\hline
  \multicolumn{2}{|l|}{$\mu$} & 0.875 \\
  \hline
  \multicolumn{2}{|l|}{Variance} & 0.254 \\
  \hline
  \multicolumn{2}{|l|}{W} & \$65m \\
  \hline
  \multicolumn{2}{|l|}{Cost} & \$44.8k \\
  \hline
  \multirow{2}*{Volume} & model & \$5.67m \\
       & fact & \$5.69m \\
  \hline  
  \multirow{2}*{Fees} & model & \$17.01k \\
       & fact & \$17.06k \\
 \hline
 \end{tabular}
 \label{table:t6}
\end{table}

Thus, the parameters have been obtained to approximate the modeled reward level to the actual one for a specific period (one day). We can apply these parameters for modeling a single day. The modeling results are presented in the Table \ref{table:t6}.

\begin{figure*}[!h]
\centering
\includegraphics[width=\textwidth]{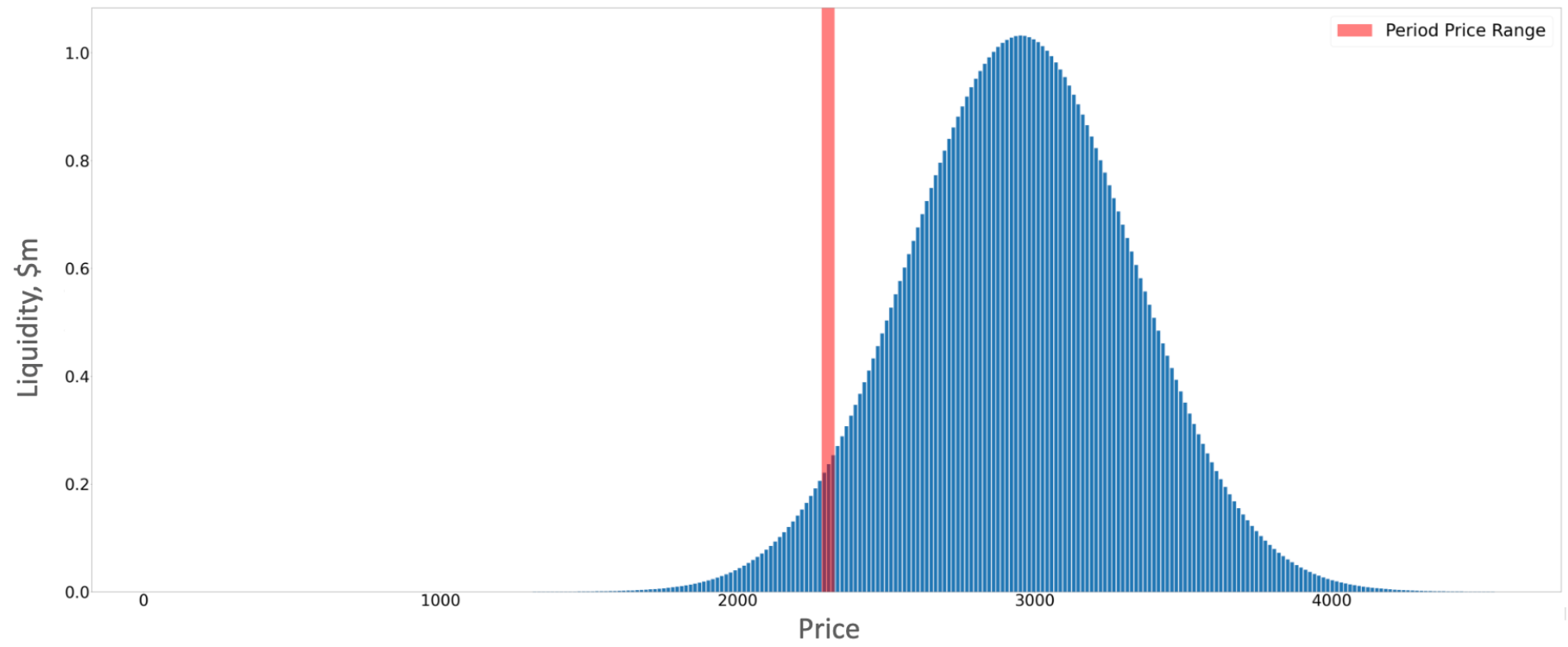}
\caption{Model pool liquidity profile and the price range position for the period under review (one day)}
\label{figure:BS14}
\end{figure*}

The result looks almost perfect, with an error less than 1\%. Let's visualize the price range area for the considered period and its position on the liquidity curve after parameter fitting (Figure \ref{figure:BS14}).

\begin{figure*}[!h]
\centering
\includegraphics[width=\textwidth]{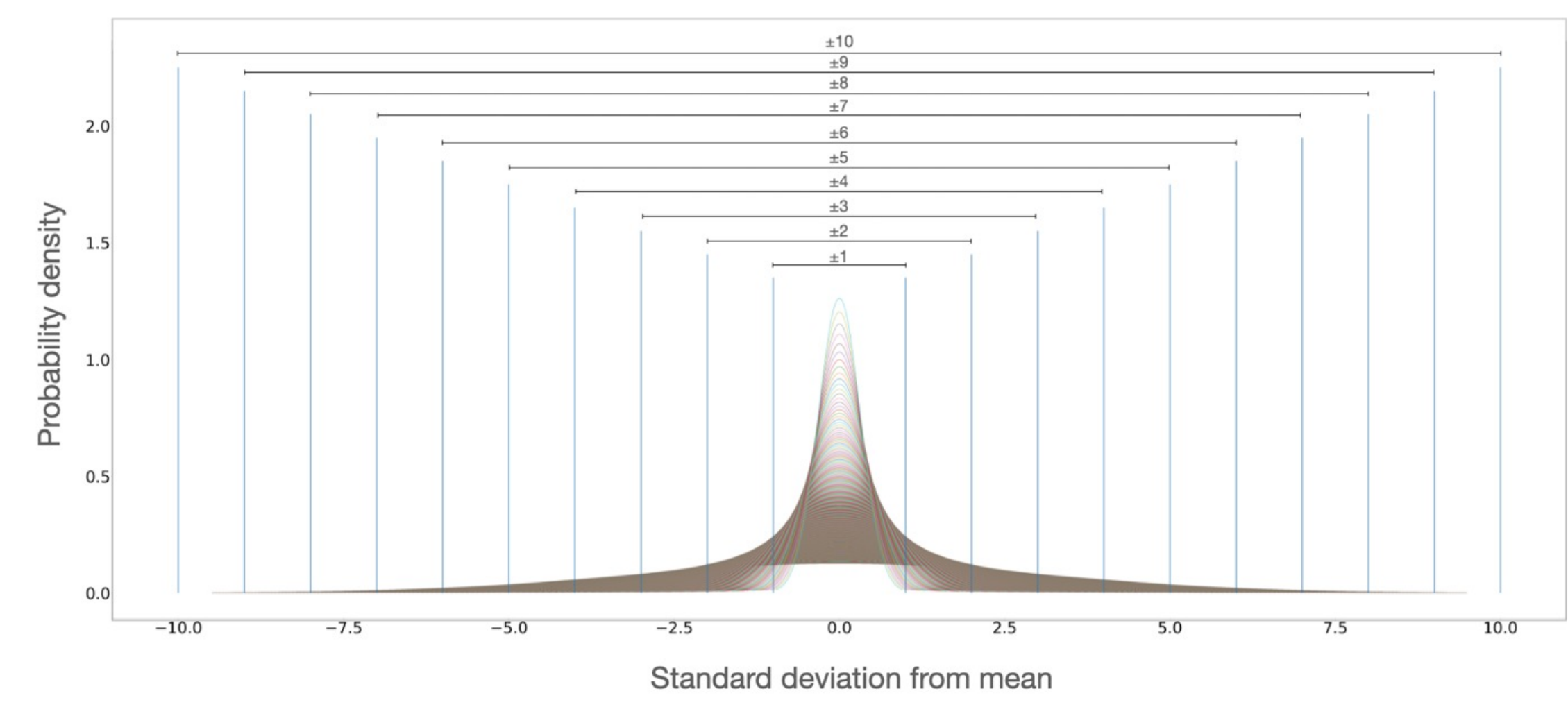}
\caption{Normal distribution density for the pool liquidity profile shape\\ with different boundaries}
\label{figure:BS15}
\end{figure*}

It is important to note that a solution exists regardless of the boundaries of the considered normal distribution area, within which the modeled pool ranges proportions are normalized. To illustrate this, let's consider different boundaries ranges (Figure \ref{figure:BS15}) and demonstrate that the solution will exist for each case for the same period.

\begin{figure*}[!h]
\centering
\includegraphics[width=\textwidth]{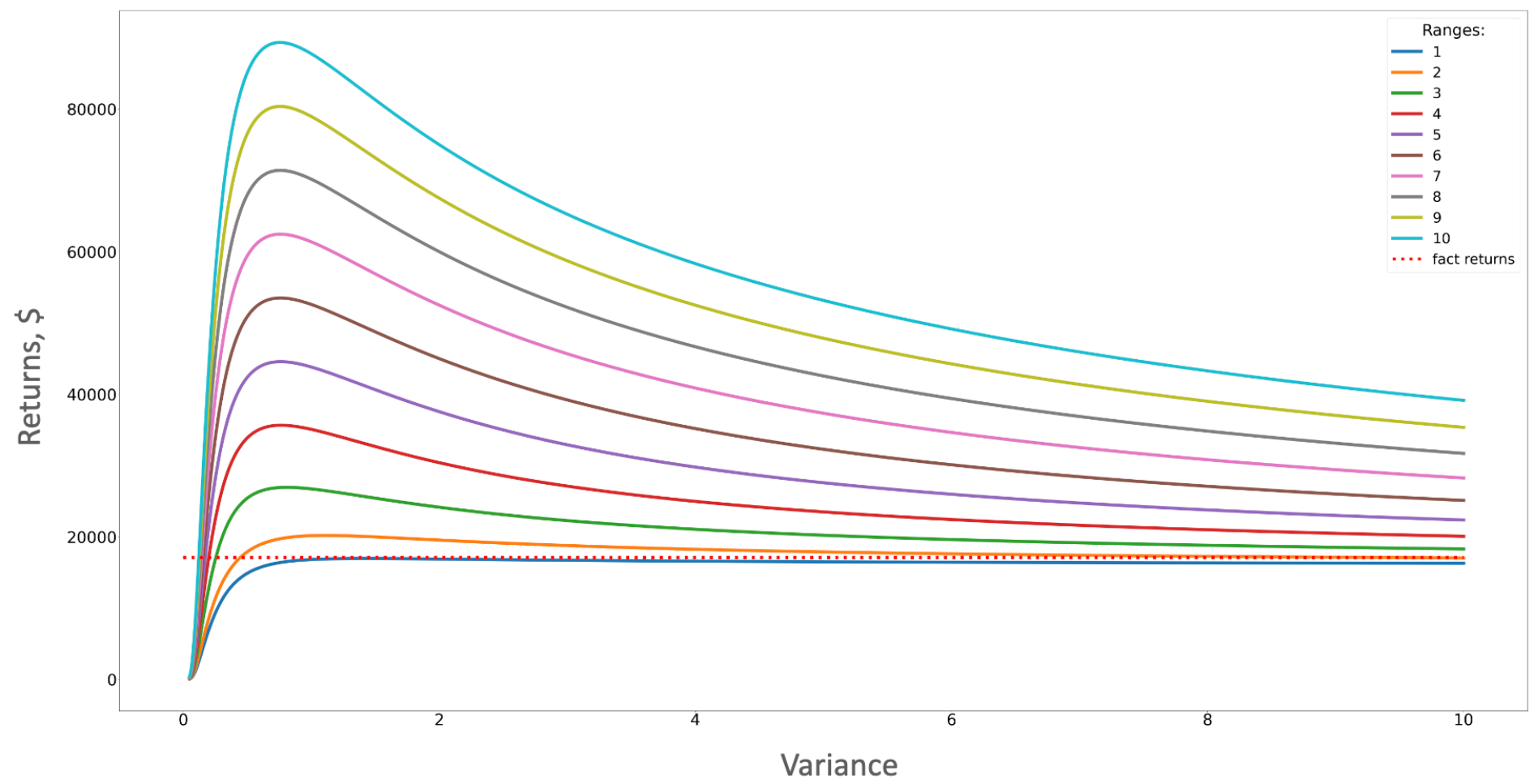}
\caption{The model remuneration level curves for the period under consideration 02/02/2024 \\ with $\mu$ = 0.875 for each area under consideration boundary value}
\label{figure:BS16}
\end{figure*}

Next, with a fixed level of the $\mu$ parameter, we will plot reward curves for each boundary scenario (Figure \ref{figure:BS16}).

We see that except for a very narrow boundary, we can approximate the modeled reward value to the actual one by correctly choosing the variance parameter. The values of the variance for each boundary value, where the actual reward level is achieved, are presented in the Table \ref{table:t7}.

\begin{table}[h!]
\centering
\caption{Summary Uniswap v3. \\ Liquidity Curve (one day, DEX).
 Different ranges}
 \begin{tabular}{|c|c|c|} 
 \hline
 Range & Variance & Model fee\\ [0.5ex] 
 \hline\hline
 1 & 1.42 & \$16.9k \\ 
 \hline
 2 & 0.44 & \$17.03k \\ 
 \hline
 3 & 0.25 & \$17.01k \\ 
 \hline
 4 & 0.2 & \$17.01k \\ 
 \hline
 5 & 0.17 & \$17.39k \\ 
 \hline
 6 & 0.15 & \$16.85k \\ 
 \hline
 7 & 0.14 & \$17.3k \\ 
 \hline
 8 & 0.13 & \$17.05k \\ 
 \hline
 9 & 0.13 & \$17.75k \\ 
 \hline
 10 & 0.12 & \$16.39k \\ 
 \hline
 \end{tabular}
 \label{table:t7}
\end{table}

We observe that regardless of the range boundary values, with the correct variance parameter setting and a fixed $\mu$ parameter, it is possible to approximate the actual reward value for each scenario. The boundary choice essentially sets the modeled rewards limit for the specified period. The modeled liquidity profile visualization for each scenario is presented in \ref{App2}. Next, we will also consider the configuration with a ±3 boundary.

\subsubsection{One year}

\begin{figure*}[!h]
\centering
\includegraphics[width=\textwidth]{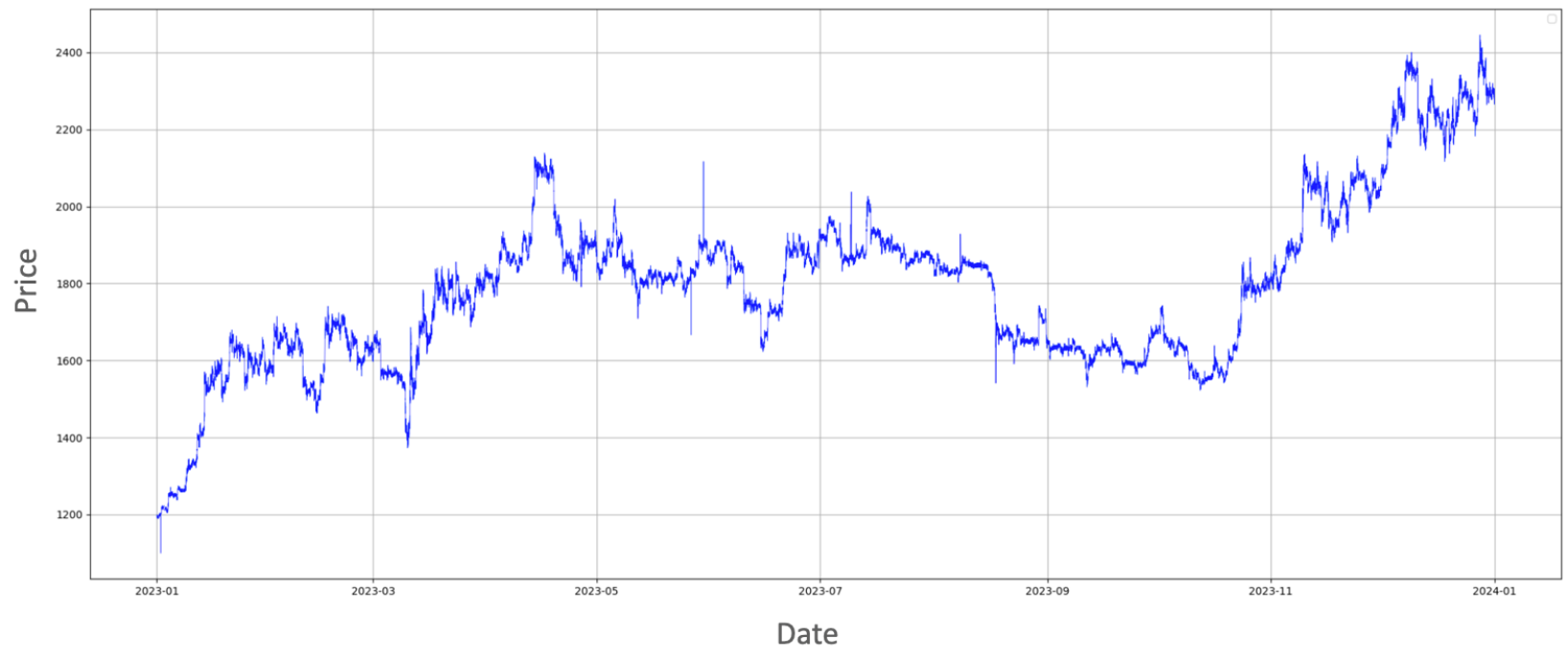}
\caption{Actual price dynamics for the USDC/ETH pair for 2023 in the simulated pool}
\label{figure:BS17}
\end{figure*}

Now let's scale the task up from considering a specific day to a whole year. As the new period, we'll consider the year 2023 (Figure \ref{figure:BS17}) for the same Uniswap v3 pool with the USDC/ETH 0.3\% pair.

\begin{figure*}[!h]
\centering
\includegraphics[width=0.85\textwidth]{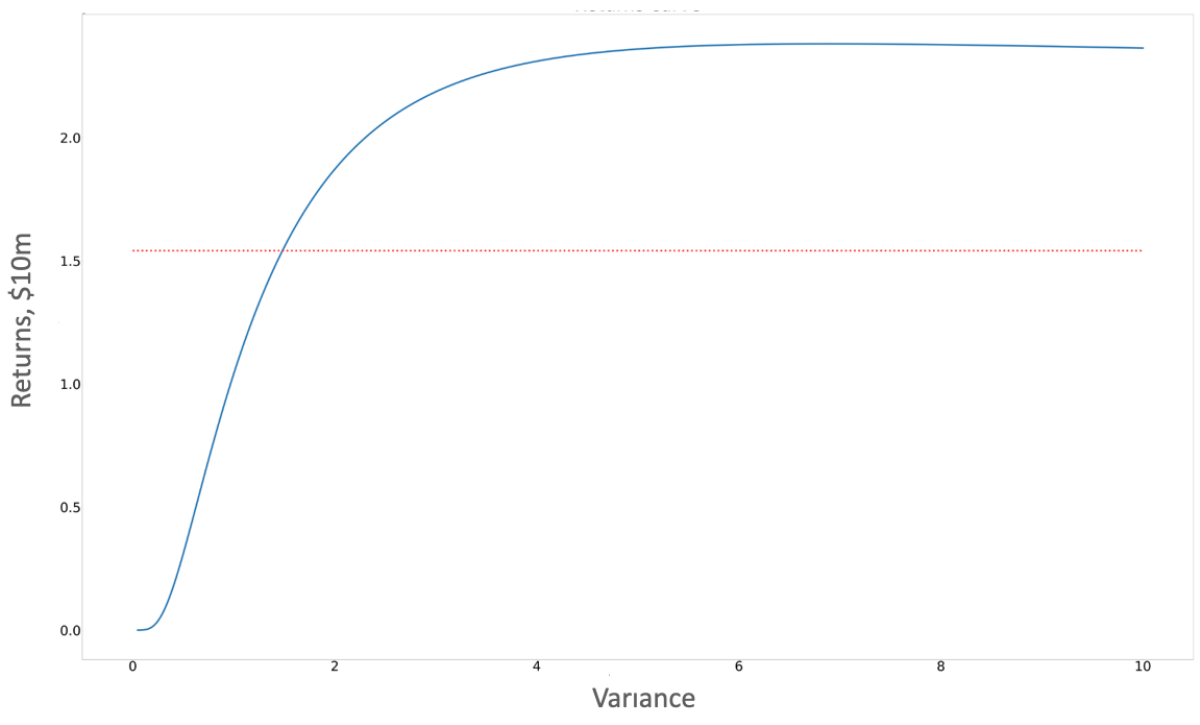}
\caption{Curve of the model level of remuneration. 2023 year, with the parameter $\mu$ = 1.9}
\label{figure:BS18}
\end{figure*}

Similarly, we'll adjust the $\mu$ and variance parameters to closely approximate the modeled reward level to the actual one (Figure \ref{figure:BS18}). 

Having obtained the parameters to approximate the modeled reward level to the actual one, we apply them to model the period. The modeling results are presented in the Table \ref{table:t8}:

\begin{table}[h!]
\centering
\caption{Summary Uniswap v3.\\ Liquidity Curve (2023Y, DEX)}
 \begin{tabular}{| l | l | c |} 
 \hline
  \multicolumn{2}{|l|}{Indicator} & Value \\
  \hline\hline
  \multicolumn{2}{|l|}{$\mu$} & 1.9 \\
  \hline
  \multicolumn{2}{|l|}{Variance} & 1.484 \\
  \hline
  \multicolumn{2}{|l|}{W} & \$76.4m \\
  \hline
  \multicolumn{2}{|l|}{Cost} & \$30.2k \\
  \hline
  \multirow{2}*{Volume} & model & \$5.133b \\
       & fact & \$5.135b \\
  \hline  
  \multirow{2}*{Fees} & model & \$15.42m \\
       & fact & \$15.41m \\
 \hline
 \end{tabular}
 \label{table:t8}
\end{table}

\begin{figure*}[!h]
\centering
\includegraphics[width=0.8\textwidth]{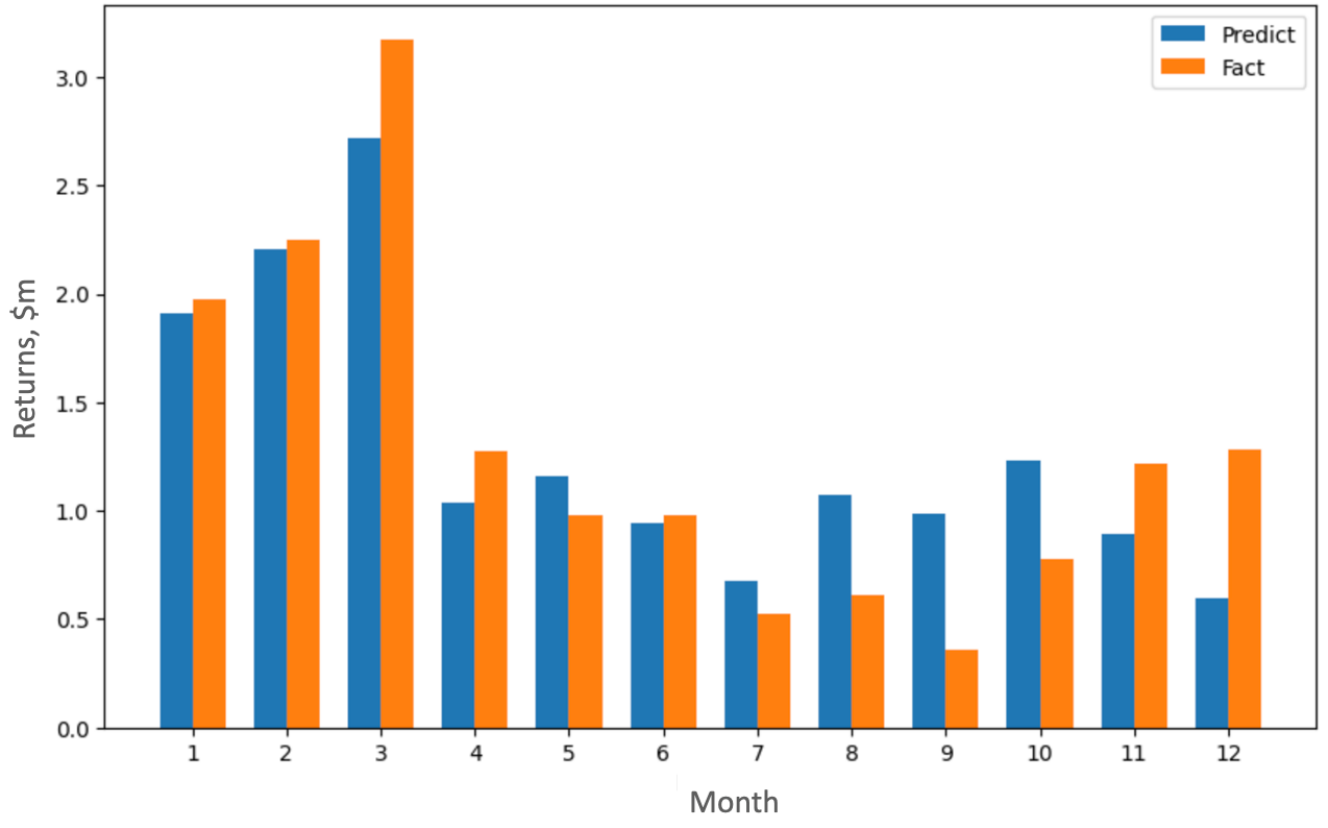}
\caption{The model and fact remuneration distribution\\levels by month for the period 2023}
\label{figure:BS19}
\end{figure*}

The result looks perfect, with an error near 0\%. However, this result is achieved over the entire period. Let's analyze the modeled reward level against the actual one on a monthly basis. In Figure \ref{figure:BS19}, you can see the varying relationship between the actual and predicted levels from month to month.

\begin{figure*}[!h]
\centering
\includegraphics[width=\textwidth]{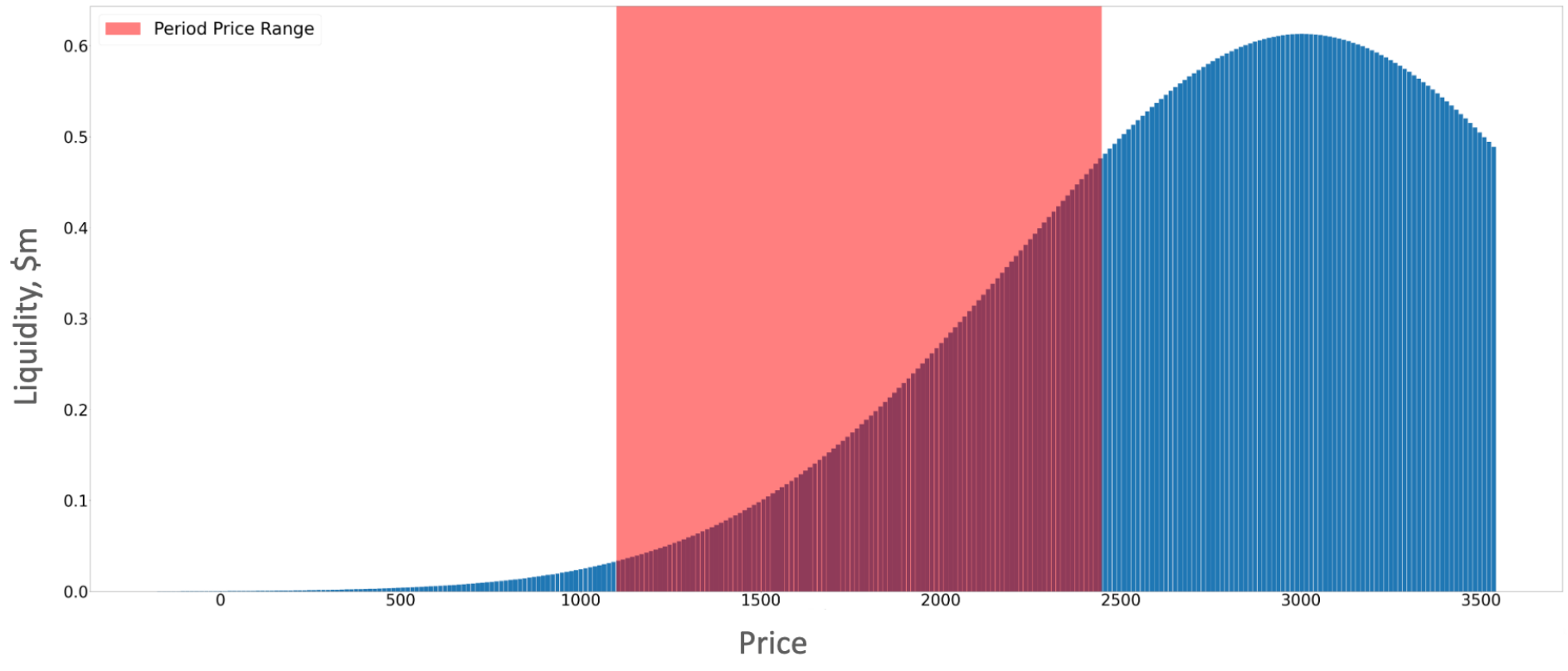}
\caption{Model pool liquidity profile and the price range position for the period under review (one year)}
\label{figure:BS20}
\end{figure*}

This variability is caused by the shape of the unified liquidity curve that we use to approximate the reward level. Towards the end of the period under consideration, there is a steady price increase (Figure \ref{figure:BS17}), but the predicted reward volumes in the last two months are underestimated relative to the actual ones. This is due to the narrowing of available liquidity on the right side (Figure \ref{figure:BS20}) of the modeled liquidity profile that falls within the range of historical prices for the period under review.

By changing parameters, we adjust the liquidity profile and the relationship between the modeled and actual reward levels to the desired level of detail.

\subsubsection{Other Pools}
We apply the developed approach and the backtesting tool to other pools, considering Uniswap v3 pools for pairs of altcoins, stablecoins and USDC/ETH with different fee levels. The results are presented in the Table \ref{table:t9}:

\begin{table*}[h!]
\centering
\caption{Summary, different pools.}
 \begin{tabular}{| l | l | p{0.13\linewidth} | p{0.13\linewidth} | p{0.14\linewidth} | p{0.14\linewidth} | p{0.14\linewidth} |} 
 \hline
  \multicolumn{2}{|l|}{Indicator} & USDC/ETH 0.3\%\tablefootnote{Contract 0x8ad599c3A0ff1De082011EFDDc58f1908eb6e6D8}
  & USDC/ETH 0.05\%\tablefootnote{Contract 0x88e6A0c2dDD26FEEb64F039a2c41296FcB3f5640} & WBTC/ETH 0.3\%\tablefootnote{Contract 0xCBCdF9626bC03E24f779434178A73a0B4bad62eD} & USDC/USDT 0.01\%\tablefootnote{Contract 0x3416cF6C708Da44DB2624D63ea0AAef7113527C6} & TON/ETH 1\%\tablefootnote{Contract 0x4b62Fa30Fea125e43780DC425C2BE5acb4BA743b } \\
  \hline\hline
  \multicolumn{2}{|l|}{Period} & 2023 & Q3 2023 & 2023 & 2023 & 2023 \\
  \hline
  \multicolumn{2}{|l|}{Transactions} & 102.5k & 339k & 64.4k & 198.8k & 18.1k \\
  \hline
  \multicolumn{2}{|l|}{$\mu$} & 1.9 & 0.75 & 1.35 & -0.75 & 2.5 \\
  \hline
  \multicolumn{2}{|l|}{Variance} & 1.484 & 0.145 & 0.359 & 0.194 & 0.198 \\
  \hline
  \multicolumn{2}{|l|}{W (avTVL)} & \$76.4m & \$139.6m & \$283m
  & \$27m & \$5.2m\\
  \hline
  \multicolumn{2}{|l|}{Cost} & \$30.2k & \$30.2k & \$30.2k & \$30.2k & \$30.2k \\
  \hline
  \multirow{2}*{Volume} & model & \$5.138b & \$15.71b & 144.5k BTC & \$18.59b & 33.77m TON\\
       & fact & \$5.135b & \$15.74b & 144.1k BTC & \$18.54b & 33.67m TON\\
  \hline  
  \multirow{3}*{Fees} & model & \$15.42m & \$7.85m & 433.6 BTC & \$1.86m & 337.7k TON \\
       & fact & \$15.41m & \$7.87m & 432.1 BTC & \$1.85m & 336.7k TON \\ \cline{2-7}
       & error & 0.05\% & 0.20\% & 0.37\% & 0.30\% & 0.32\% \\

 \hline
 \end{tabular}
 \label{table:t9}
\end{table*}

We can see that the backtesting tool works with pools of different pair types and calculation periods. The fee level is technically informative and is the same for all pools because in each case, the same number of liquidity ranges creation is simulated at the initial moment and burning is done at the end of the modeling. The ranges sizes and the liquidity level are individual for each case.

Liquidity profiles and current TVL levels for each pool are provided in \ref{App3}.

\subsection{Impermanent Loss Estimation on Artificial Data}
\label{subsection:ILEonAD}

Let us consider an artificial impermanent loss example in Uniswap v3. Using the backtester, we model the following situation: there is a capital of one million USD that an LP  places in a hypothetical pool on the USDC/ETH pair. A $\tau$-reset strategy is implemented with a range length of one USD, $\tau=5$, and liquidity is uniformly distributed across the ranges. In other words, the liquidity is uniformly distributed in a relatively narrow price range, causing liquidity to be frequently dropped and relocated due to the $\tau$-reset strategy. We intentionally do not consider fees and rewards of past epochs to observe what happens to the initial capital solely under the influence of impermanent loss and the protocol's architecture when the $\tau$-reset strategy is applied without much thought. For this exercise, we use minute data from Binance for the year 2023 on the USDC/ETH pair.

\begin{figure*}[h]
\centering
\includegraphics[width=\textwidth]{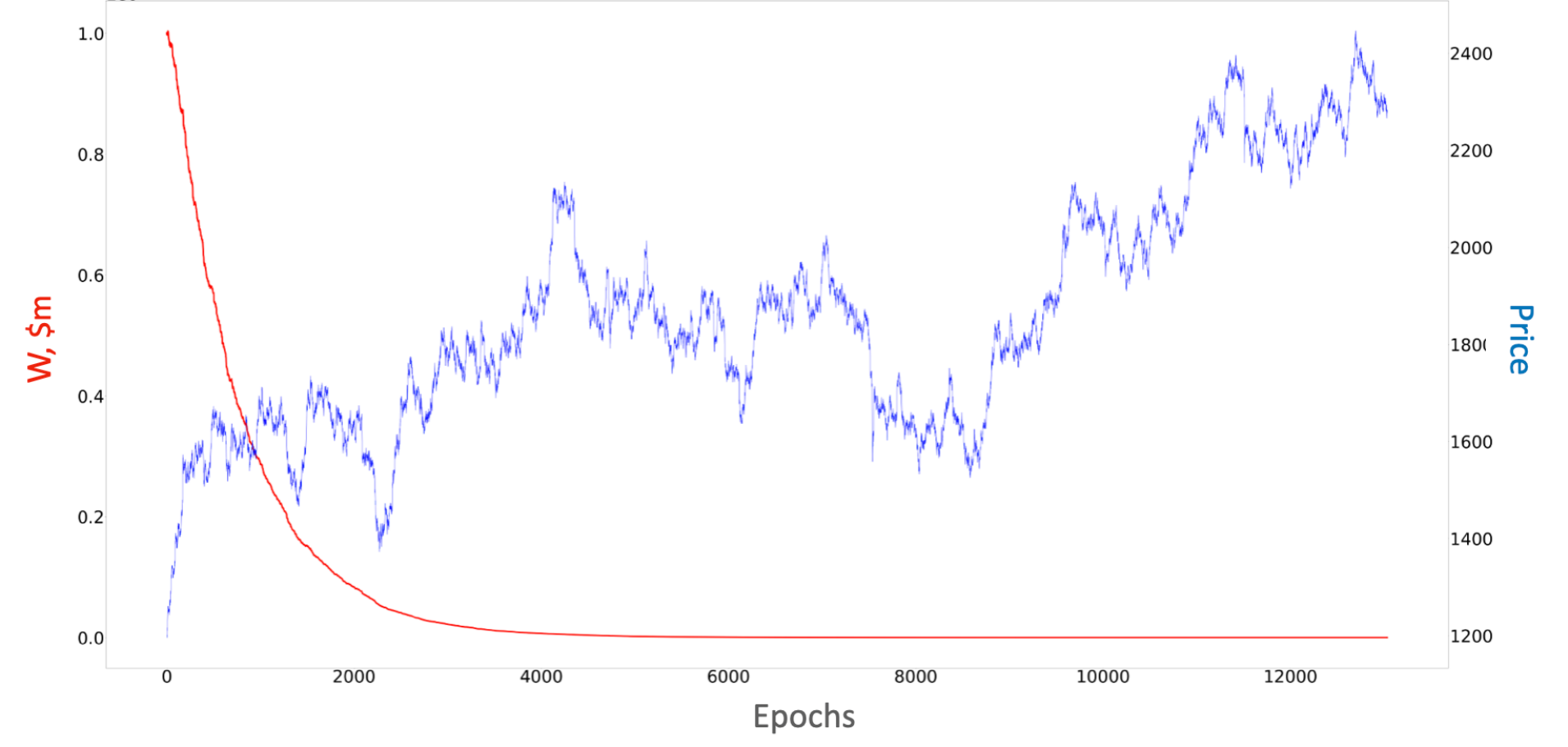}
\caption{Capital decrease due to the features of Uniswap v3 and IL}
\label{figure:IL-Uniswap}
\end{figure*}

We observe a sharp decline in the capital level $W$ even with a steadily rising price throughout 2023. This happens due to relatively narrow liquidity allocation; even with small price changes, we constantly encounter situations where the current pool price drops below the lower liquidity provision boundary, leading to a reduction in capital due to the full token reserves conversion into locally "depressed" ETH. Let us delve deeper into this case by examining a simple example in Figure~\ref{figure:IL-Uniswap}.

At time $t_0$, the LP believes that the pool price will drop below and wants to earn rewards by fixing their liquidity in the range $[P_a, P_b]$ at the current pool price $P_0 > P_b$. At time $t_0$, all actual LP reserves consist only of token $B$ reserves ($y$ reserves).

\begin{figure*}[h]
\centering
\includegraphics[width=\textwidth]{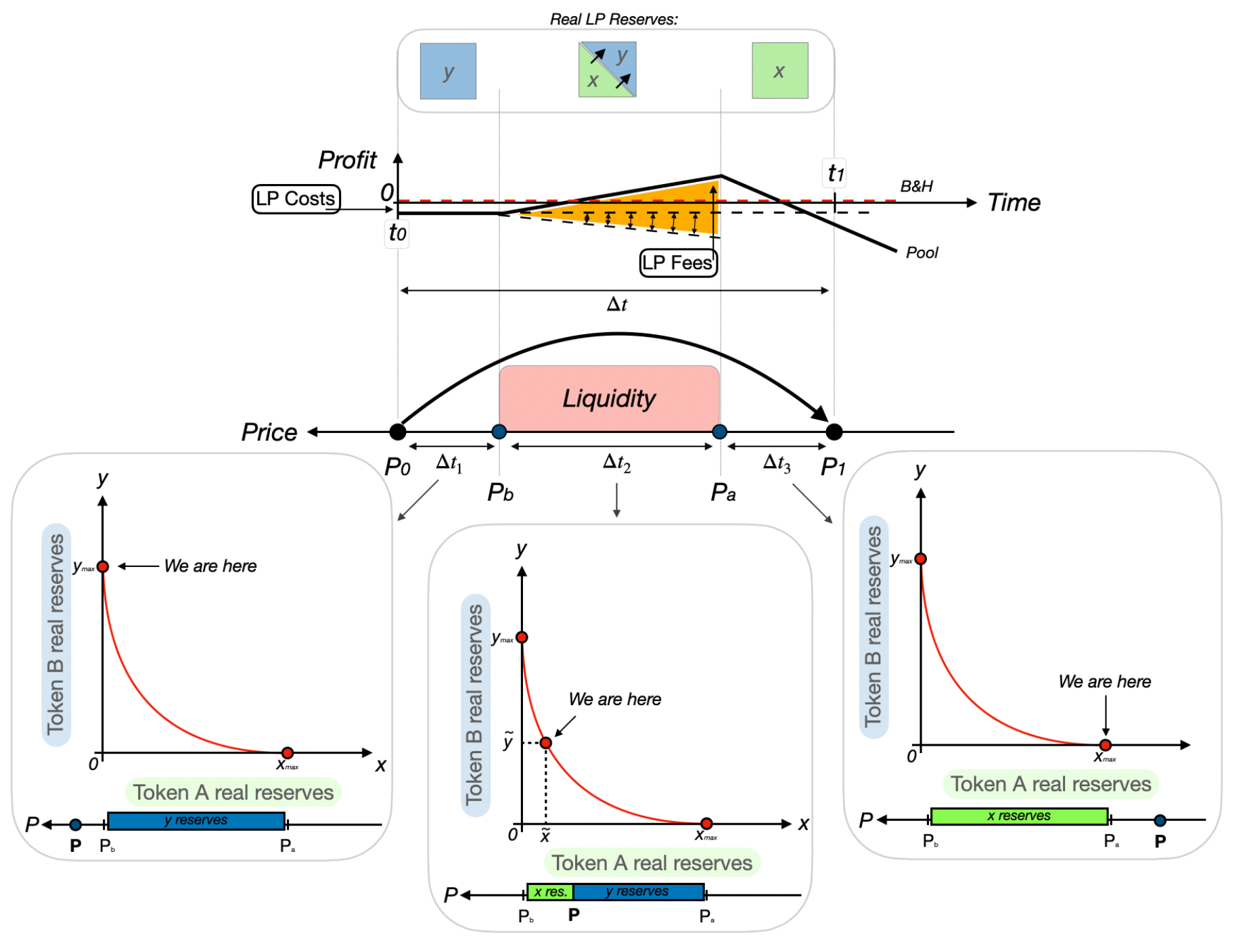}
\caption{IL for $P_1 < P_0$}
\label{figure:IL-Uniswap-1}
\end{figure*}

The alternative B\&H strategy in this case would be to simply hold onto this set of tokens in a wallet, which in our case is some amount of stablecoins in USDC for the USDC/ETH pair. Then, at time $t_1$, the price decreases from $P_0$ to $P_1$, where $P_1 < P_a$. The period $\Delta t$ between these moments can be divided into three sub-periods:
\begin{itemize}
    \item 
        $\Delta t_1$: A sub-period where the pool price drops to the upper range boundary. In this period, the fixed token reserve $y$ remains unchanged, with the only difference from B\&H being that the LP paid a fee for providing liquidity, resulting in their profit consistently below zero.
    \item 
        $\Delta t_2$: A sub-period where the pool price moves through the range where the LP provided liquidity. As the price decreases, the tokens reserves ratio changes. This leads to larger losses for LP over time due to the dynamic increase in the depreciating assets share. By the time we move to $\Delta t_3$, the LP only has reserves of token $A$ $x$ and feels the price drop significantly since placement at $t_0$. Rewards from the exchange during this sub-period are intended to compensate for LP's losses, giving an advantage over the B\&H strategy as depicted in the Figure \ref{figure:IL-Uniswap-1}.
    \item 
        $\Delta t_3$: A sub-period where the pool price falls below the lower range boundary. The LP holds reserves consisting of depreciating assets and incurs losses due to market risk at every moment. By the end of this sub-period, losses from such placement may exceed the profits in terms of exchange rewards and lose out to the B\&H strategy in this case.
\end{itemize}

\begin{figure*}[h]
\centering
\includegraphics[width=\textwidth]{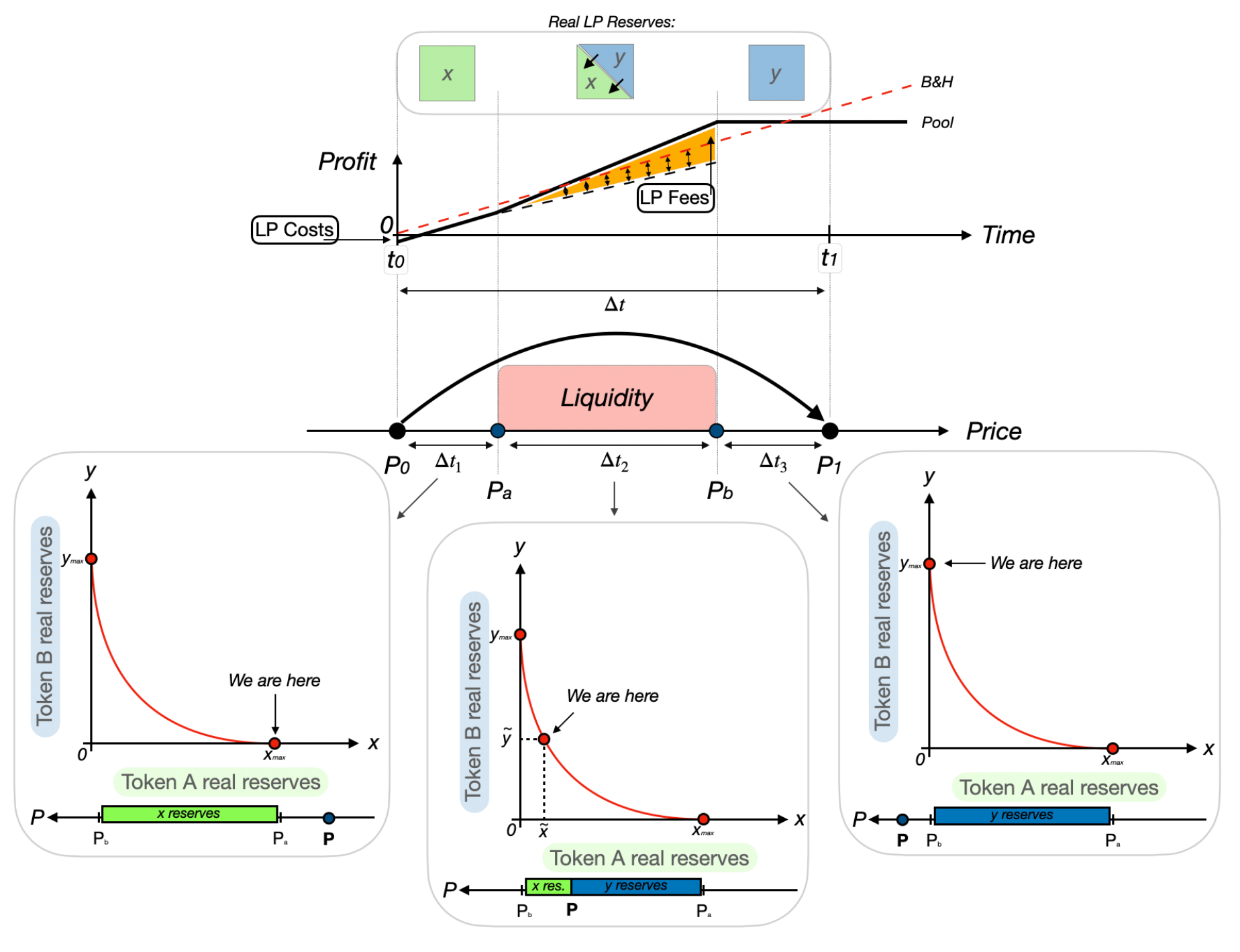}
\caption{IL for $P_1 > P_0$}
\label{figure:IL-Uniswap-2}
\end{figure*}

It is worth noting that with a dynamic $\tau$-reset strategy and similar dynamic pool price range placement, there is a liquidity dump and reduced capital reinvestment in the form of new liquidity into new ranges. This case can be compared to the risk of capital reduction in dynamic delta hedging on a CEX. Now let us consider the reverse case with an increase in the current price (Figure \ref{figure:IL-Uniswap-2}) of a hypothetical pool.

At time $t_0$, the LP believes that the price in the pool will rise and that they will receive rewards by fixing their liquidity in the range $[P_a, P_b]$ at a price of $P_0 < P_a$. At time $t_0$, all actual reserves consist only of Token $A$ ($x$ reserves). In this case, an alternative B\&H strategy would involve fixing a set of cryptocurrency tokens in a wallet, in this case, ETH in the USDC/ETH pair.

Next, at time $t_1$, the price rises from $P_0$ to $P_1$, where $P_1 > P_b$. Similarly, the period $\Delta t$ between the considered time points can be divided into three sub-periods:
\begin{itemize}
    \item
        $\Delta t_1$: A sub-period during which the pool price rises to the lower range boundary. The fixed reserve $x$ of token $A$ remains unchanged, but its market price increases similar to B\&H, accounting for liquidity provision fees, resulting in LP's profit being below zero at the initial time.
    \item 
       $\Delta t_2$: A sub-period during which the pool price passes through the range where LP provided liquidity. As the price increases, the tokens reserves ratio changes accordingly. Due to the dynamic reduction in the growing asset share, the LP starts earning less profit compared to the B\&H strategy at each moment. By the time $\Delta t_3$ is reached, the LP only has reserves of token $B$, and its capital is no longer sensitive to the growth of token A. Exchange rewards in this sub-period are designed to compensate for LP's lag behind B\&H, as seen in the Figure \ref{figure:IL-Uniswap-2}.
    \item 
        $\Delta t_3$: A sub-period during which the pool price exceeds the upper range boundary. The LP has fixed reserves consisting of a stablecoin. Essentially, providing liquidity in this case represents an on-chain short put option. At the end of this sub-period, the LP's fixed capital level may start lagging behind the B\&H strategy due to further growth in token A's price and pool price increase. Therefore, it is important for LP to redistribute liquidity to achieve maximum rewards.
\end{itemize}

This concludes an example of IL occurrence on artificial data.

\section{Conclusions}\label{conclusion}
\label{section:Conclusions}


DeFi has empowered users with unprecedented autonomy in digital markets, particularly in cryptocurrency trading, by eliminating intermediaries. The transition from centralized exchanges (CEX) to decentralized exchanges (DEXs) gained momentum with the introduction of Uniswap V1 protocol by Uniswap Labs in 2018, reshaping DEXs on the Ethereum blockchain. Uniswap's innovative automated market makers (AMM) concept marked a significant departure from traditional order book mechanisms. As Uniswap evolved through three protocol versions, Uniswap V3 emerged to address challenges by introducing concentrated liquidity to incentivize Liquidity Providers and mitigate impermanent loss risks.

This study involved the creation of a custom backtesting framework to evaluate rewards within liquidity pools, featuring a streamlined architecture and minimal required input data tailored specifically for CLMMs. Noteworthy is the backtester's use of a parametric approach to approximate liquidity distribution within pools for precise reward estimation. By analyzing reward levels across various pools using historical data from 2023, including altcoin pairs, stablecoins, and USDC/ETH with different fee structures, the backtester demonstrated a remarkable level of accuracy, validating the effectiveness of the methodology. In addition to the tool's development, the research establishes essential theoretical groundwork for further advancement, covering crucial mathematical aspects and processes within CLMM pools.

The next phase of our research will focus on enhancing the backtesting tool to accurately project LP revenues integrated into pool liquidity, aiming to develop an optimal dynamic strategy model for providing liquidity in CLMM pools.

The backtesting framework offers a means to optimize liquidity provision strategies for individuals or entities participating in DEXs. By forecasting LP revenues accurately and analyzing historical data, users can make informed decisions on maximizing returns while minimizing risks.

The research findings could contribute to the development of risk management tools and strategies for participants in decentralized finance (DeFi) ecosystems \cite{Bertomeu2024,Chaleenutthawut2024}. Understanding how concentrated liquidity impacts rewards and mitigates impermanent loss risks can help users navigate the complexities of DeFi more effectively.

The theoretical foundation established in this research could prove valuable for protocol developers looking to enhance existing decentralized exchange protocols or create new ones. Insights into liquidity distribution, reward estimation, and mathematical processes within CLMM pools could guide the development of more efficient and secure DeFi protocols, such as stablecoins and lending protocols~\cite{Klages-Mundt2020,KlagesMundt2022,Bullmann2019}.




\bibliographystyle{elsarticle-num}
\bibliography{main}
\newpage

\appendix

\onecolumn

\section{Actual Pool Data.}\label{App1}
\subsection{Actual level of pool rewards as of 02/02/2024.}\label{App11}

\begin{figure*}[!h]
\centering
\includegraphics[width=0.8\textwidth]{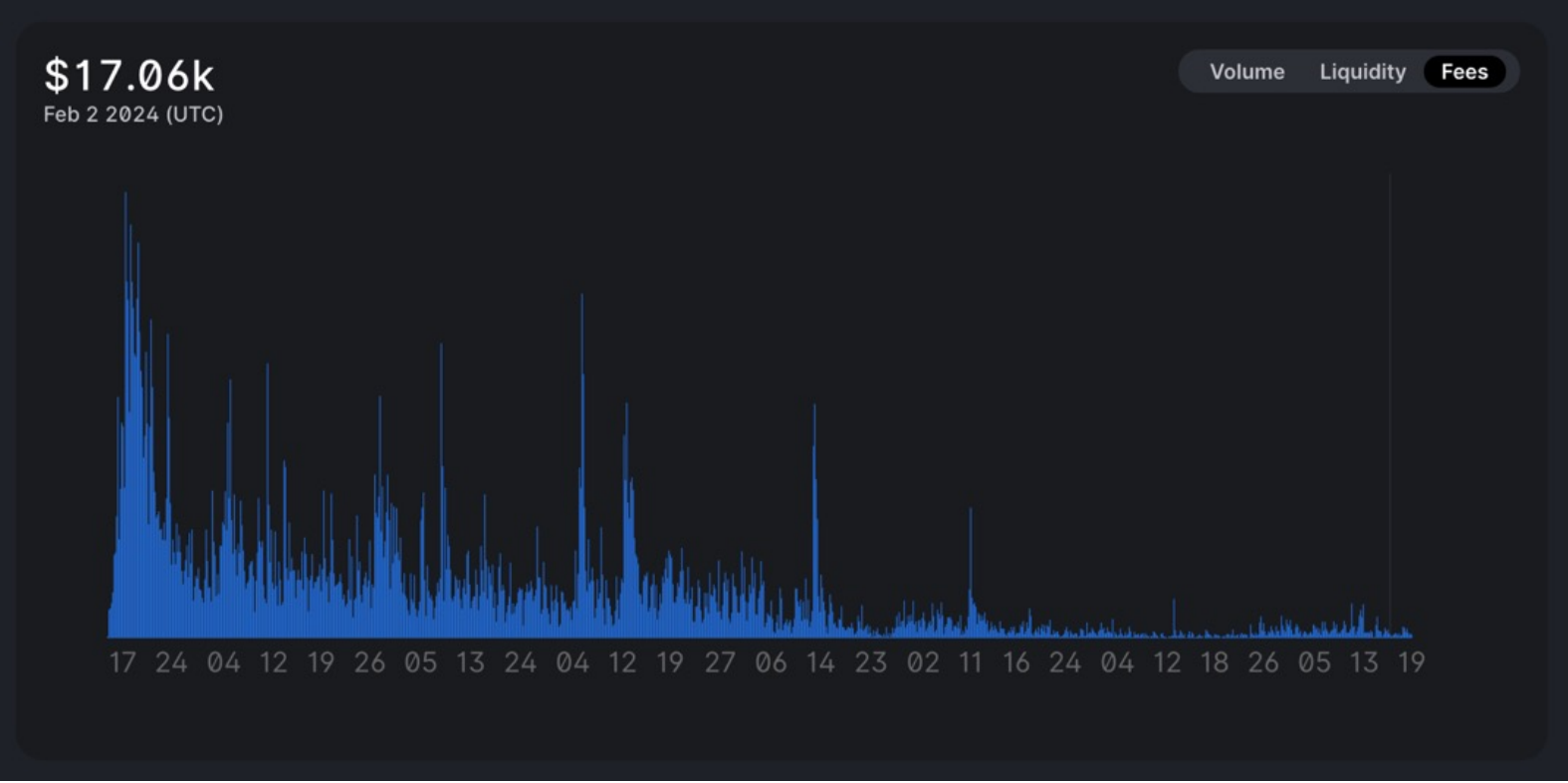}
\label{figure:Ap11}
\end{figure*}

\subsection{Actual level of pool volumes as of 02/02/2024.}\label{App12}

\begin{figure*}[H]
\centering
\includegraphics[width=0.8\textwidth]{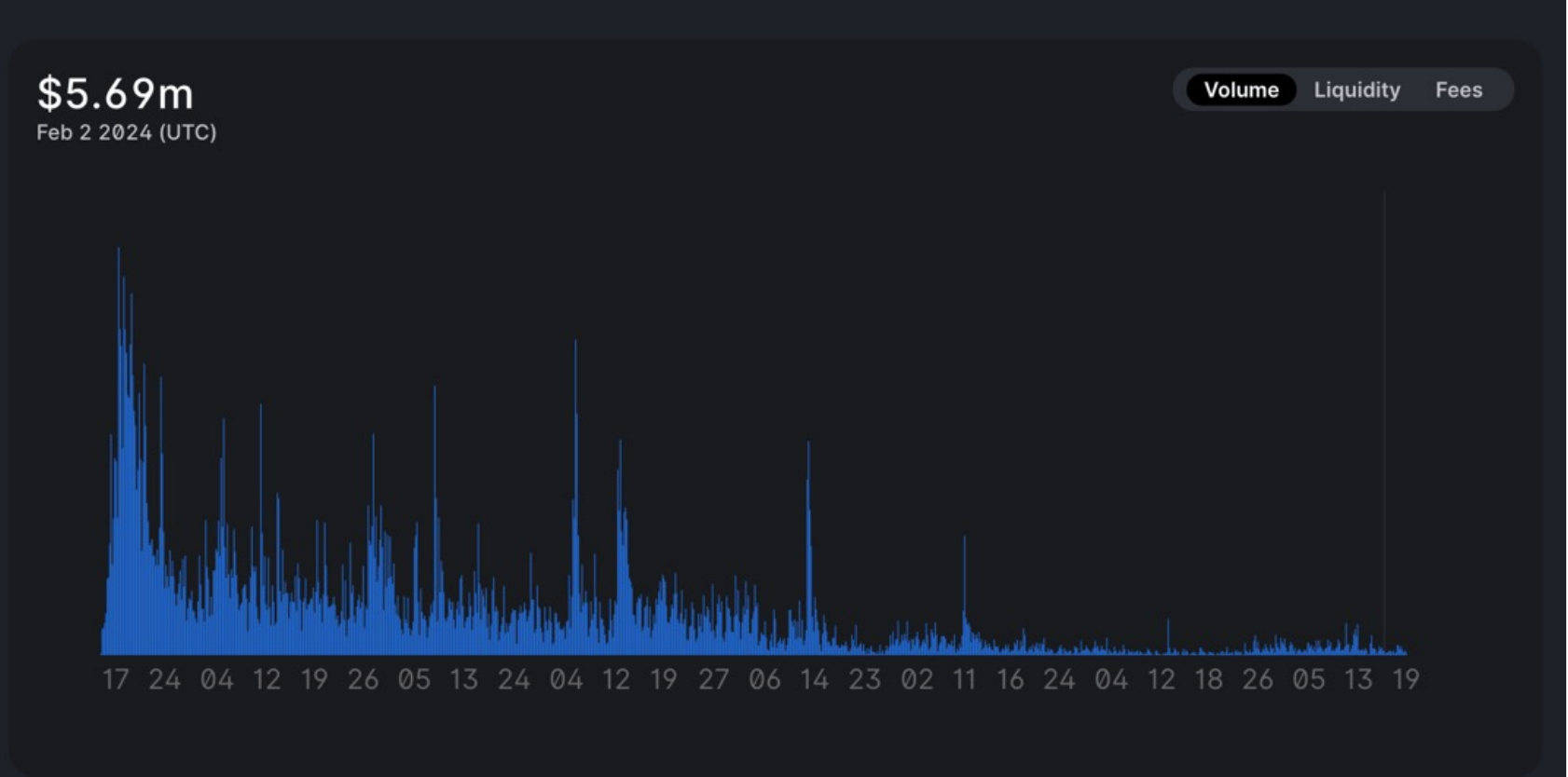}
\label{figure:Ap12}
\end{figure*}

\subsection{Pool liquidity profile as of 02/02/2024.}\label{App13}

\begin{figure*}[!h]
\centering
\includegraphics[width=0.8\textwidth]{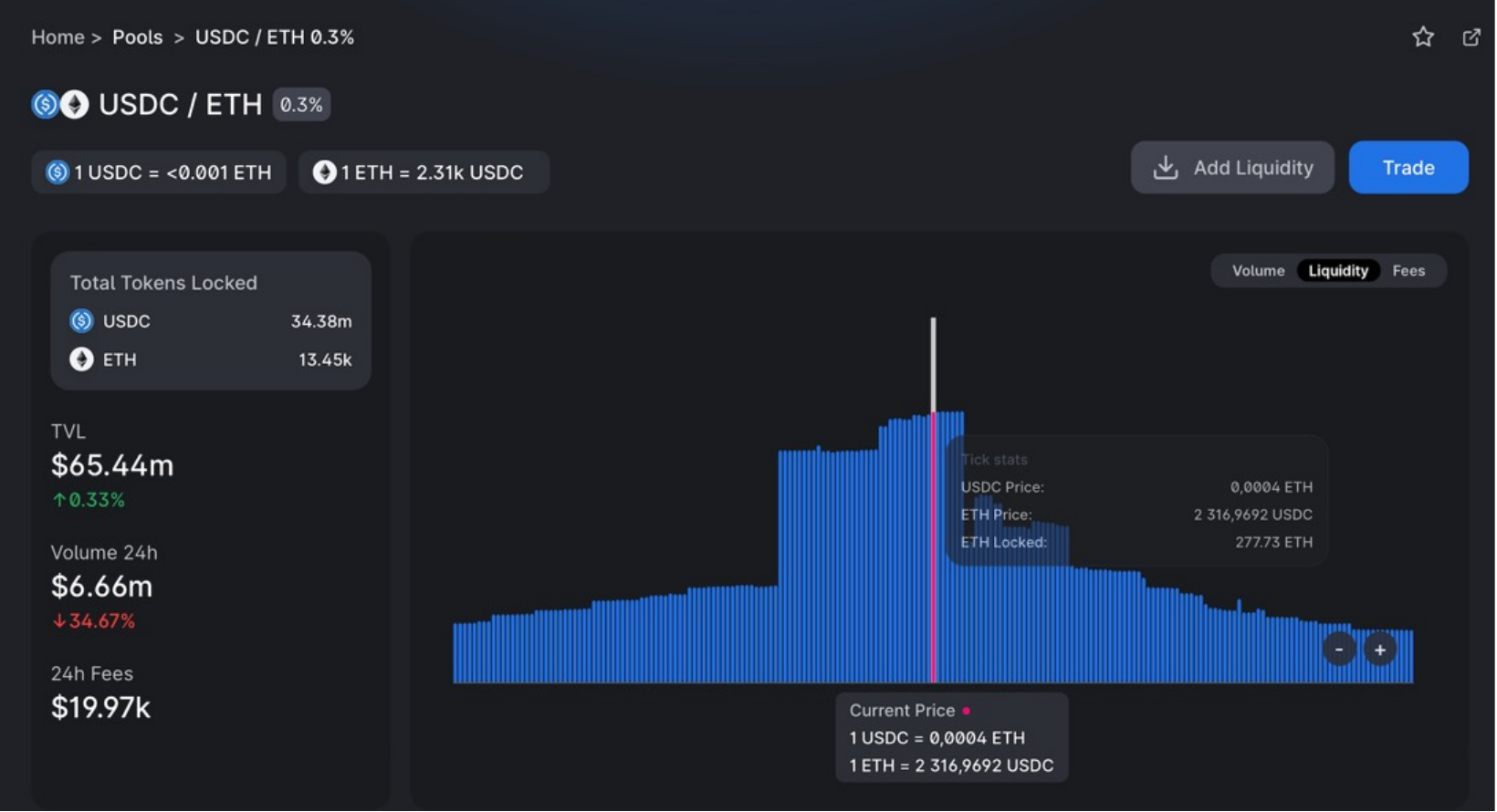}
\label{figure:Ap13}
\end{figure*}
\newpage

\section{Liquidity Curve Profile for Artificial Data}\label{App2}
Profile of the liquidity curve with fixed parameters of $\mu$ = 0.875 and variance = 1 at different boundaries of the distribution under consideration.

\begin{figure*}[!h]
\centering
\includegraphics[width=\textwidth]{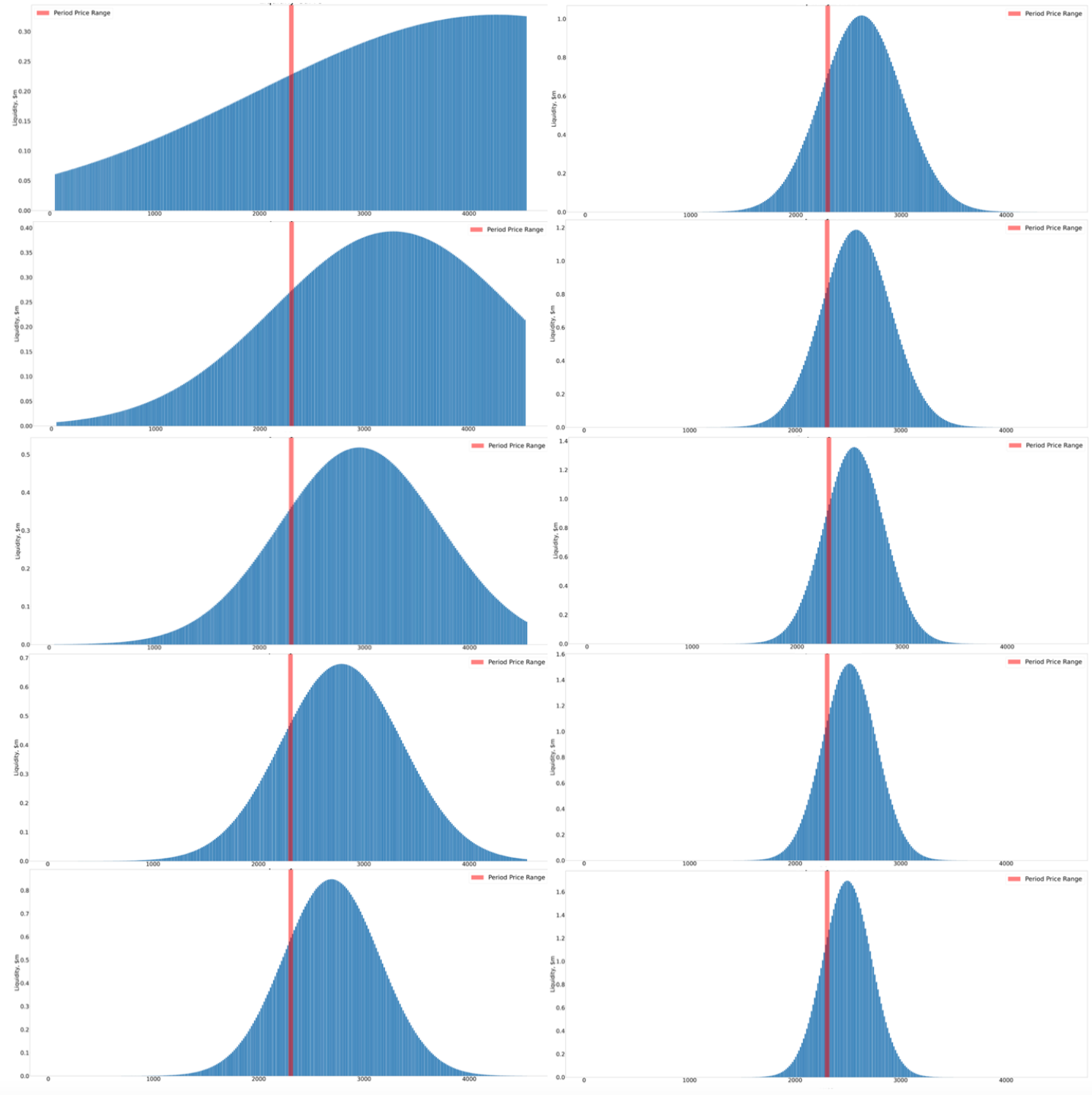}
\label{figure:Ap2}
\end{figure*}
\newpage

\section{Liquidity Curve Profile for Real Data}\label{App3}

For this study, we used the TVL value from Etherscan (source 1) on the date of the study (May 2, 2024) as the average TVL for the period. This decision was made because historical data obtained from The Graph and Uniswap (source 2) raised some questions, and there were no alternative sources.

\subsection{Contract 0x8ad599c3A0ff1De082011EFDDc58f1908eb6e6D8}

\begin{figure*}[!h]
\centering
\includegraphics[width=\textwidth]{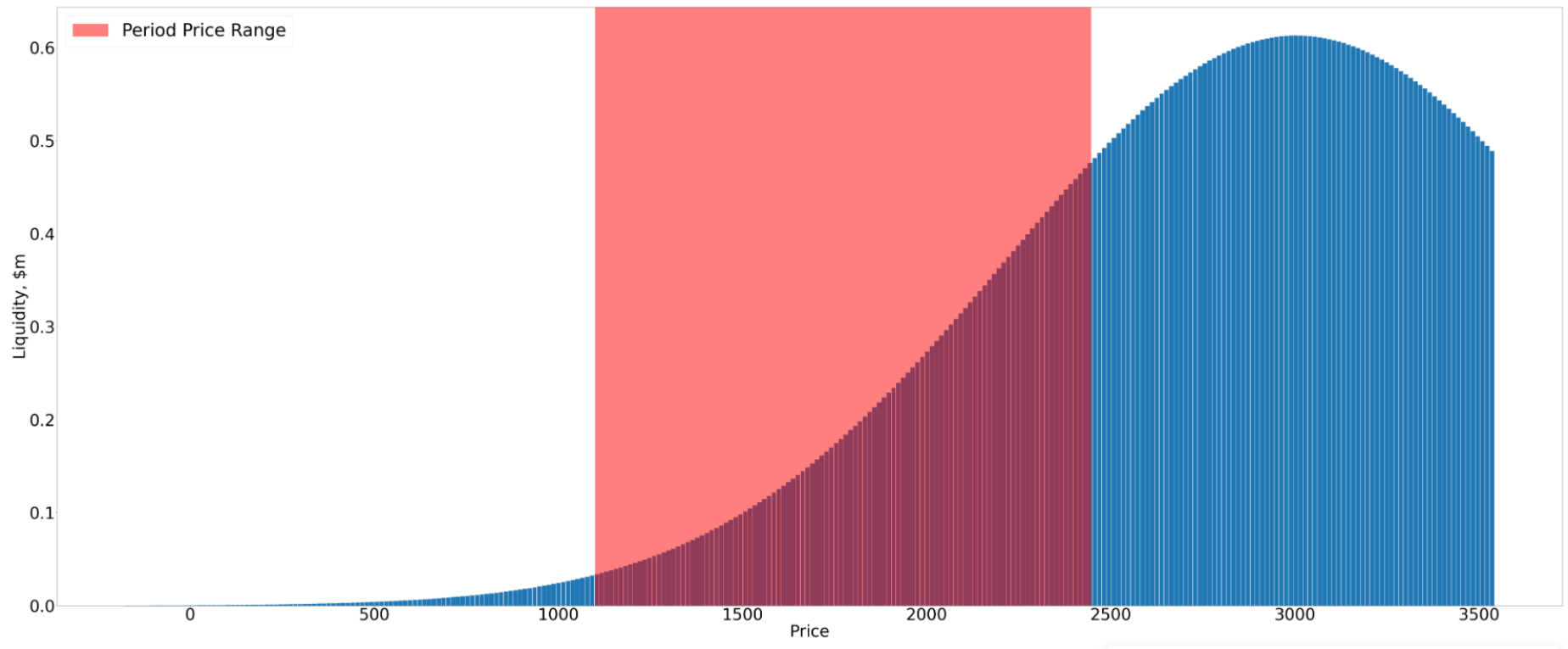}
\end{figure*}

\begin{table}[h]
\centering
\caption{TVL by sources.}
 \begin{tabular}{|c | c|} 
 \hline
  Source, №& {TVL, \$m} \\ [0.5ex] 
 \hline\hline
 1 & 76.4 \\ 
 \hline
 2 & 336.6 \\ [1ex] 
 \hline
 \end{tabular}
 \label{table:tap1}
\end{table}


\newpage

\subsection{Contract 0x88e6A0c2dDD26FEEb64F039a2c41296FcB3f5640}

\begin{figure*}[!h]
\centering
\includegraphics[width=\textwidth]{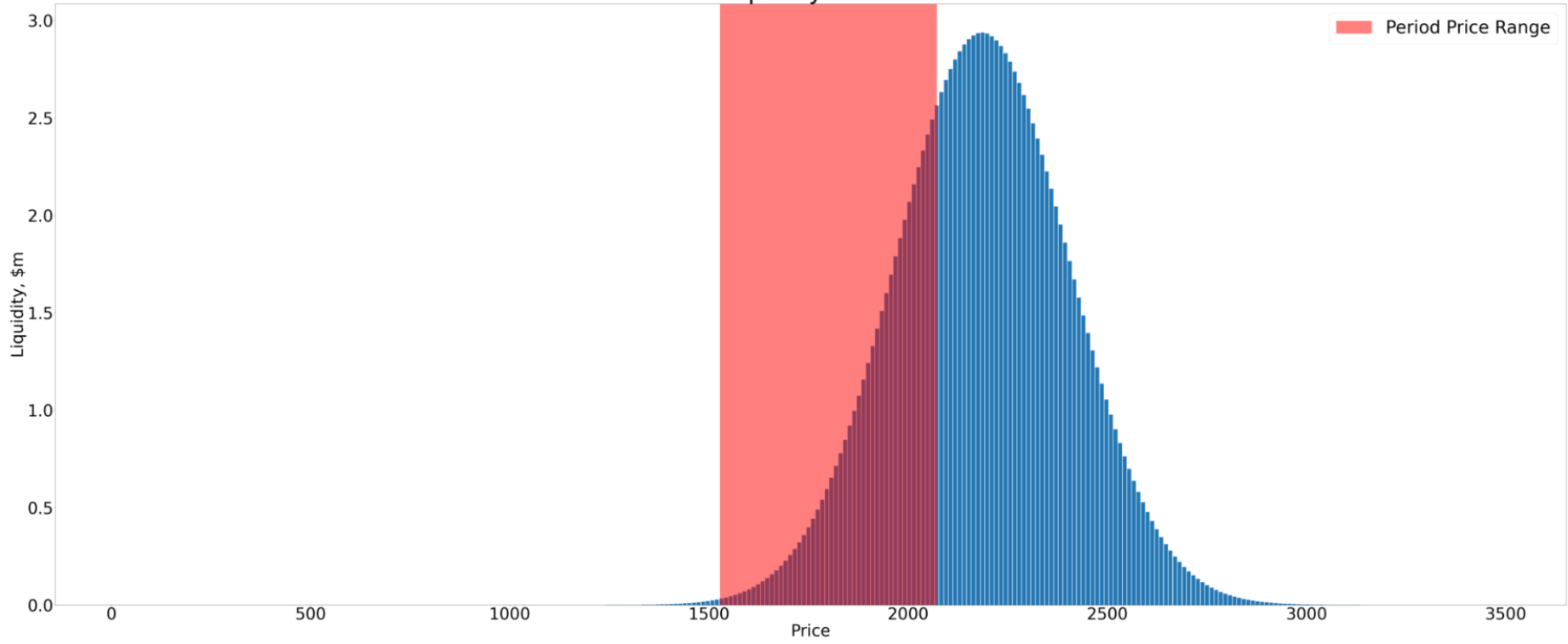}
\end{figure*}

\begin{table}[h]
\centering
\caption{TVL by sources.}
 \begin{tabular}{|c | c|} 
 \hline
  Source, №& {TVL, \$m} \\ [0.5ex] 
 \hline\hline
 1 & 139.6 \\ 
 \hline
 2 & 409.7 \\ [1ex] 
 \hline
 \end{tabular}
 \label{table:tap1}
\end{table}



\subsection{Contract 0xCBCdF9626bC03E24f779434178A73a0B4bad62eD}

\begin{figure*}[!h]
\centering
\includegraphics[width=\textwidth]{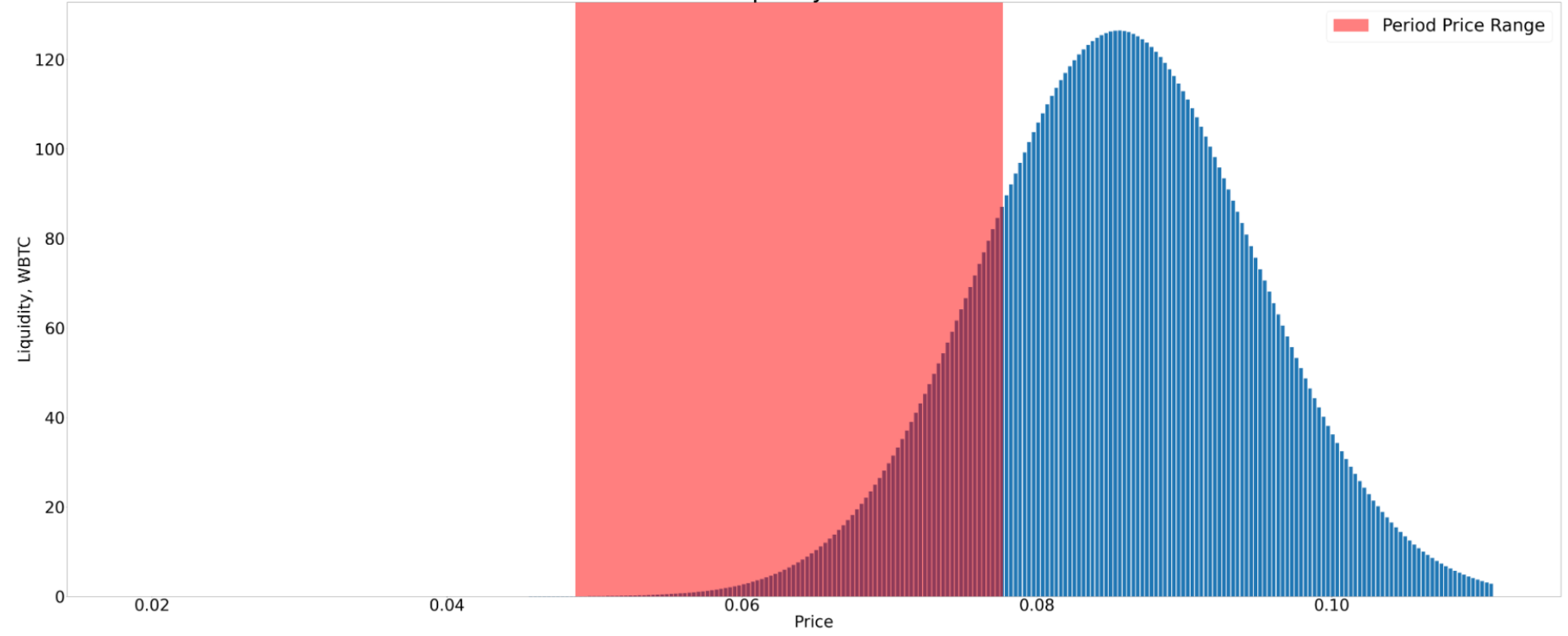}
\end{figure*}
\newpage

\begin{table}[h]
\centering
\caption{TVL by sources.}
 \begin{tabular}{|c | c|} 
 \hline
  Source, №& {TVL, \$m} \\ [0.5ex] 
 \hline\hline
 1 & 282.7 \\ 
 \hline
 2 & 407.5 \\ [1ex] 
 \hline
 \end{tabular}
 \label{table:tap1}
\end{table}



\subsection{Contract 0x3416cF6C708Da44DB2624D63ea0AAef7113527C6}

\begin{figure*}[!h]
\centering
\includegraphics[width=\textwidth]{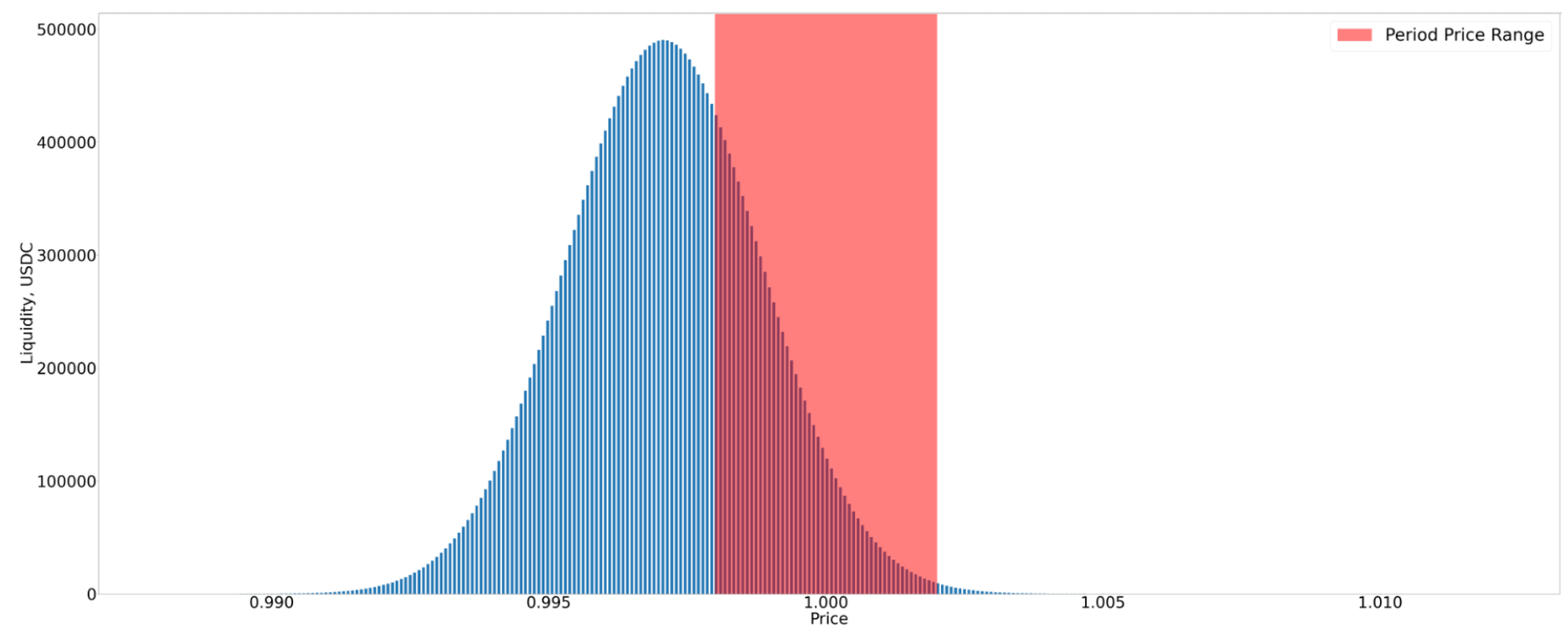}
\end{figure*}

\begin{table}[h]
\centering
\caption{TVL by sources.}
 \begin{tabular}{|c | c|} 
 \hline
  Source, №& {TVL, \$m} \\ [0.5ex] 
 \hline\hline
 1 & 26.9 \\ 
 \hline
 2 & 34.4 \\ [1ex] 
 \hline
 \end{tabular}
 \label{table:tap1}
\end{table}


\newpage

\subsection{Contract 0x4b62Fa30Fea125e43780DC425C2BE5acb4BA743b}

\begin{figure*}[!h]
\centering
\includegraphics[width=\textwidth]{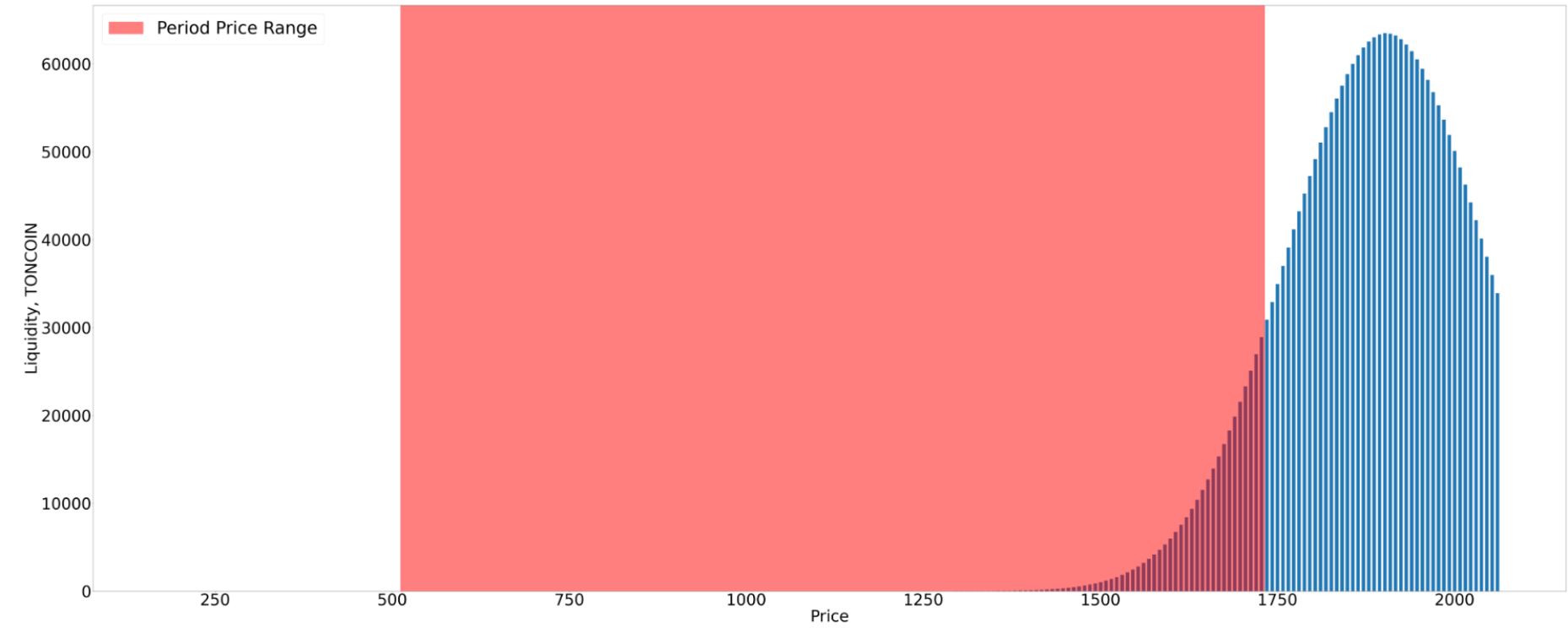}
\end{figure*}

\begin{table}[h]
\centering
\caption{TVL by sources.}
 \begin{tabular}{|c | c|} 
 \hline
  Source, №& {TVL, \$m} \\ [0.5ex] 
 \hline\hline
 1 & 5.2 \\ 
 \hline
 2 & 8.7 \\ [1ex] 
 \hline
 \end{tabular}
 \label{table:tap1}
\end{table}







\end{document}